\documentstyle{cjaa}                    
\input{epsf.sty}                        
\input{psfig.sty}                       

\begin{document}

   \title{Gamma-Ray Bursts in the Swift Era}

   \author{Bing Zhang\inst{}\mailto{bzhang@physics.unlv.edu}} 
   \offprints{B. Zhang}                   

  \institute{Department of Physics \& Astronomy, University of Nevada,
	Las Vegas, NV 89154-4002, USA \\
            \email{bzhang@physics.unlv.edu}
          }

   \date{Received~~2007 xxx; accepted~~2007~~xxx }

   \abstract{Since the successful launch of NASA's dedicated gamma-ray burst 
(GRB) mission, Swift, the study of cosmological GRBs has entered a new era. Here I 
review the rapid observational and theoretical progress in this dynamical research 
field during the first two-year of the Swift mission, focusing on how observational 
breakthroughs have revolutionized our understanding of the physical origins of 
GRBs. Besides summarizing how Swift helps to solve some pre-Swift mysteries, I also 
list some outstanding problems raised by the Swift observations. An outlook 
of GRB science in the future, especially in the GLAST era, is briefly discussed.
   }

   \authorrunning{B. Zhang}            
   \titlerunning{Gamma-Ray Bursts in the Swift era}  


   \maketitle
%
%
\section{Introduction}           
\label{sect:intro}

Gamma-ray bursts (GRBs) are fascinating celestial objects. These short, 
energetic bursts of gamma-rays mark the most violent, cataclysmic explosions 
in the universe, likely associated with the births of stellar-size black holes 
or rapidly spinning, highly magnetized neutron stars. Since the detections 
of their long-wavelength afterglows (Costa et al. 1997; van Paradijs et al.
1997; Frail et al. 1997), GRBs are observationally accessible 
in essentially all electromagnetic wavelengths. They are also potential 
emission sources of ultra-high energy cosmic rays, high-energy neutrinos, 
and gravitational waves. As stellar scale events located at cosmological 
distances, GRBs open a unique window to connect together the branches 
of stellar, interstellar, galactic, and intergalactic astronomy as well as 
cosmology. The study of 
GRBs has been prolific over the past several years. New discoveries on GRBs 
have been ranked several times as one of the ``top-ten scientific 
breakthroughs of the year'' by Science magazine (e.g. \#6 in 2003 and 
\#4 in 2005). The topic of GRBs has been extensively reviewed over the
years (e.g. Fishman \& Meegan 1995; Piran 1999; van Paradijs et al. 2000; 
M\'esz\'aros 2002; Lu et al. 2004; Zhang \& M\'esz\'aros 2004; 
Piran 2005; M\'esz\'aros 2006).

The launch of the NASA's dedicated GRB mission, Swift (Gehrels et al. 2004),
has opened a new era for GRB study. Carrying three 
instruments (Burst Alert Telescope [BAT], Barthelmy et al. 2005a; X-Ray
Telescope [XRT], Burrows et al. 2005a; and UV-Optical Telescope [UVOT],
Roming et al. 2005), Swift is a multi-wavelength observatory that can 
``swiftly'' catch the unpredictable bursts of gamma-rays in the random
directions of the sky within less than 100 seconds with all three 
instruments on target. It allows for the first time detections of 
multi-wavelength GRB early afterglows in a time domain previously
unexplored. 
In slightly over two years of operation, Swift has fulfilled most of its 
pre-mission scientific goals in GRB study, and more importantly, brings 
new surprises and challenges to our understanding of these nature's most 
violent and mysterious explosions. 
The Swift revolution has been summarized in several 
recent reviews (e.g. M\'esz\'aros 2006; O'Brien et al. 2006a; Fox \& 
M\'esz\'aros 2006). 

The plan of this review is the following. Since more extended reviews on
Swift observational data are being written (e.g. N. Gehrels et al.
2007, in preparation), I will not invest great efforts to summarize 
Swift observations in a systematic manner. Rather, I will highlight the most
important Swift observational results, and put more emphasis on 
discussing how new data revolutionize our understanding on the 
nature/physics of the GRB phenomenon. 
I do not intend to discuss GRB basics, which has been
covered in an earlier review (Zhang \& M\'esz\'aros 2004), and I refer
the latest full GRB review by M\'esz\'aros (2006) to those readers who are
interested in both GRB basics and the latest developments in the field. The
basic theme of this review is similar to Zhang \& M\'esz\'aros (2004),
which includes the progress, problems and prospects in the field. However,
by comparing the two reviews, it is encouraging to see that many items
discussed as ``problems'' in the previous review are now included 
as part of ``progress'' (\S2-4). On the other hand, 
the list of ``problems'' (\S5) is not shortened, mainly because new 
observations reveal new puzzles that were not expected before. The 
``prospects'' part (\S6) is as bright as before, in particular in view
of the upcoming high-energy era of GRB study led by the launch of GLAST. 
Due to page limitation, I will make no effort to include all the 
important papers published in the pre-Swift era (my apologies), but 
will try to include most recent papers. Following an earlier 
ChJAA review (Cheng \& Lu 2001), I will also pay special attention
to the latest contributions of the Chinese astronomers in the GRB field.

I'd like to finish the introduction with a time table of major (GRB-related)
events in the first two years of operation of Swift.

\begin{itemize}
\item Nov. 20, 2004: the Swift satellite was successfully launched from 
the Cape Canaveral Air Force Base, Florida, USA;
\item Dec. 27, 2004: Swift BAT detected the brightest gamma-ray events
ever detected by the mankind, a giant flare from the Galactic
Soft Gamma-ray Repeater source SGR 1806-20 (Palmer et al. 2005). This event
was also detected by many other high energy detectors 
(e.g. Hurley et al. 2005; Terasawa et al. 2005).
The event triggered the possibility that a good fraction of short GRBs may
be simply extragalactic SGR giant flares (Hurley et al. 2005, cf. Nakar
et al. 2006b);
\item Jan. 26, 2005 and Feb. 19, 2005: Swift detected two bursts that
show very steep decay in early X-ray afterglows (Tagliaferri et al. 2005;
Goad et al. 2006). The steep decay component is later found to be norm
of most early X-ray afterglows;
\item Apr. 6, 2005: Swift detected its first complete X-ray flare following
a soft GRB. May 2, 2005, the second burst detected on this day by Swift showed
a giant X-ray flare with fluence comparable to that of the prompt gamma-rays.
The results were reported in Burrows et al. (2005b), Romano et al. (2006a), 
Falcone et al. (2006);
\item May 9, 2005: Swift detected the first X-ray afterglow following a
short duration GRB (Gehrels et al. 2005). The XRT error box overlaps 
with a giant elliptical galaxy in a galactic 
cluster at a low redshift ($z=0.225$), giving 
the first evidence of the compact star merger origin of short GRBs (Gehrels
et al. 2005; Bloom et al. 2006a);
\item Two months later on Jul. 9, 2005, HETE-2 triggered another short GRB
(Villasenor et al. 2005), leading to the discovery of the first optical 
afterglow of short GRBs (Fox et al. 2005; Hjorth et al. 2005);
\item Half month later on Jul. 24, 2005, another short GRB was captured
by Swift, whose coordinates are firmly located inside an elliptical
galaxy (but off-center) 
(Barthelmy et al. 2005b; Berger et al. 2005a). This conclusively suggests
that short GRBs have a distinct origin from traditional long GRBs, probably
associated with compact star mergers. The extended X-ray flares following
GRB 050724 (Barthelmy et al. 2005b), on the other hand, pose a great 
challenge to the traditional compact star merger models;
\item By mid 2005, a canonical XRT lightcurve emerged from the early XRT
afterglow data of a sample of bursts (Nousek et al. 2006, see also 
Chincarini et al. 2005), which includes five distinct components (Zhang et al. 
2006). Interpreting these components require new additions to the standard 
fireball model (Zhang et al. 2006; Nousek et al. 2006; Panaitescu et al. 2006a);
\item It became clear in mid-2005 that most GRBs have very dim early optical
afterglows. Most of them are not detectable by Swift UVOT (Roming et al. 2006a);
\item Sep. 4, 2005: Swift detected a GRB with the highest redshift (as of the
end of 2006). The detection of the burst (Cusumano et al. 2006a) prompted the 
IR follow-up observations which led to the identification of its redshift 
$z=6.29$ (Haislip et al. 2006; Kawai et al. 2006; Antonelli et al. 2005);
\item Feb. 18, 2006: Swift detected an extremely long, faint, low-luminosity
GRB (Campana et al. 2006a) at redshift $z=0.0331$ (Mirabal et al 2006),
which is clearly associated with a Type Ic supernova SN 2006aj (Pian et al.
2006). More intriguingly, a distinct thermal X-ray emission component was
detected in the XRT prompt emission spectrum, which may be related to the
shock breakout of the underlying supernova (Campana et al. 2006a);
\item Jun. 14, 2006: Swift detected a peculiar nearby long-duration burst
(Gehrels et al. 2006), which was not associated with a supernova (Gal-Yam
 et al. 2006; Fynbo et al. 2006a; Della Valle et al. 2006b). This peculiar
event calls for reconsideration of the GRB classification scheme;
\item Oct. 7, 2006: Swift detected a very bright GRB (Schady et al. 2006a),
whose early optical flux peaked around 10th magnitude, very close to
the previous record-holder GRB 990123 (Alkerlof et al. 1999). The decay
behavior is however rather different from GRB 990123, likely dominated
by the forward shock emission (Mundell et al. 2006; Schady et al. 2006a).
\end{itemize}


\section{Classes of GRBs}
\label{sect:Classes}

One fundamental question related to GRBs is how many intrinsically different
categories they have, which correspond to intrinsically
different types of progenitor and possibly different types of central 
engine as well. This section is dedicated to this important topic.

\subsection{Short vs. long; Type I vs. Type II}

From the GRB sample collected by Burst And Transient Source Experiment (BATSE) 
on board the Compton Gamma-Ray Observatory (CGRO), a clear bimodal distribution
of bursts was identified (Kouveliotou et al. 1993). Two criteria have been
used to classify the bursts. The primary criterion is duration. A separation
line of 2 seconds was adopted to separate the double-hump duration distribution
of the BATSE bursts. The bimodal distribution is supported by hardness-duration
correlations (Qin et al. 2000).
The supplementary criterion is the hardness - usually denoted as the hardness 
ratio within the two energy bands of the detector. On average,
short GRBs are harder, while long GRBs are softer. So the two distinct
populations of bursts discussed in the literature have been long-soft GRBs 
and short-hard GRBs. Based on duration distribution, a third class of
GRBs with intermediate duration has been proposed (e.g. 
Mukherjee et al. 1998; Horv\'ath 1998; Horv\'ath et al. 2006). The case 
is however not conclusive.

Several prompt emission data analyses regarding the differences between long
and short GRBs have revealed interesting conclusions. Ghirlanda et al. (2004a)
discovered that the short GRBs are hard mainly because of a harder low-energy 
spectral index of the GRB spectral function (the Band function, Band et al.
1993). More interestingly, short GRB spectra are broadly similar to 
those of long GRBs if only the first 2 seconds of data of long GRBs are 
taken into account. Nakar \& Piran (2002) found that temporal properties of
short GRBs are also similar to those of long GRBs in the first
1-2 seconds, with highly variable temporal structures. Liang et al. (2002),
on the other hand, found that the variability time scales of short GRBs are
much shorter than those of long GRBs. 
Dong \& Qin (2005) and Qin \& Dong (2005) present the arguments that the
properties of short GRBs are different from the first two seconds of long 
GRBs. Cui et al. (2005) discovered that long and short GRBs follow two 
distinct sequences in the $E_p$ - hardness ratio sequence. Spectral lag
(the lag of arrival time between softer band emission with respect to harder 
band emission) analyses indicate that the lag in short GRBs is much 
smaller than that in long GRBs (Yi et al. 2006; Norris \& Bonnel 2006;
Gehrels et al. 2006), 
consistent with being zero. In both long and short GRBs, on the other hand, 
the ratios between lags and pulse widths are comparable.
(Yi et al. 2006). This generally explains the much smaller lags in short 
GRBs since their pulses are much narrower.

Afterglow observations shed light onto the nature of these two distinct
classes of bursts. Since 1997 and by Nov. 20, 2006, the afterglows of 
over 200 long GRBs have been detected (Greiner 2006). 
Several cases of solid associations between GRBs
and Type Ib/c Supernovae have been established, which include 
GRB 980425/SN 1998bw at $z=0.0085$ (Galama et al. 1998; Kulkarni et al. 
1998), GRB 030329/SN 2003dh at $z=0.168$ (Stanek et al. 2003; Hjorth et al. 
2003), GRB 031203/SN 2003lw at $z=0.105$ (Malesani et al. 2004), 
GRB 060218/SN 2006aj at $z=0.0331$ (Modjaz et al. 2006; Pian et al. 2006; 
Sollerman et al. 2006; Mirabal et al. 2006; Cobb et al. 2006), and
GRB 050525A/SN 2005nc at $z=0.606$ (Della Valle et al. 2006a). In
some other cases, red SNe bumps have been observed in the late optical
afterglow light curves (Bloom et al. 1999a, 2002; Reichart 1999; Della
Valle et al. 2003; Fynbo et al. 2004; see a comprehensive sample in Zeh 
et al. 2004 and references therein). The host galaxies of long GRBs are 
exclusively star-forming galaxies, predominantly irregular dwarf
galaxies (Fruchter et al. 2006). All these strongly suggest that
most, if not all, long GRBs are produced during the core-collapses of
massive stars, dubbed ``collapsars'', as has been suggested theoretically
(Woosley 1993; Paczy\'nski 1998; MacFadyen \& Woosley 1999; Colgate 1974).
Not long ago, it has been suggested that both observations and theories
are consistent with the hypothesis that every long GRB has an underlying
supernova associated with it (Woosley \& Bloom 2006).

\begin{figure}[t]
   \begin{center}
   \centerline{\psfig{figure=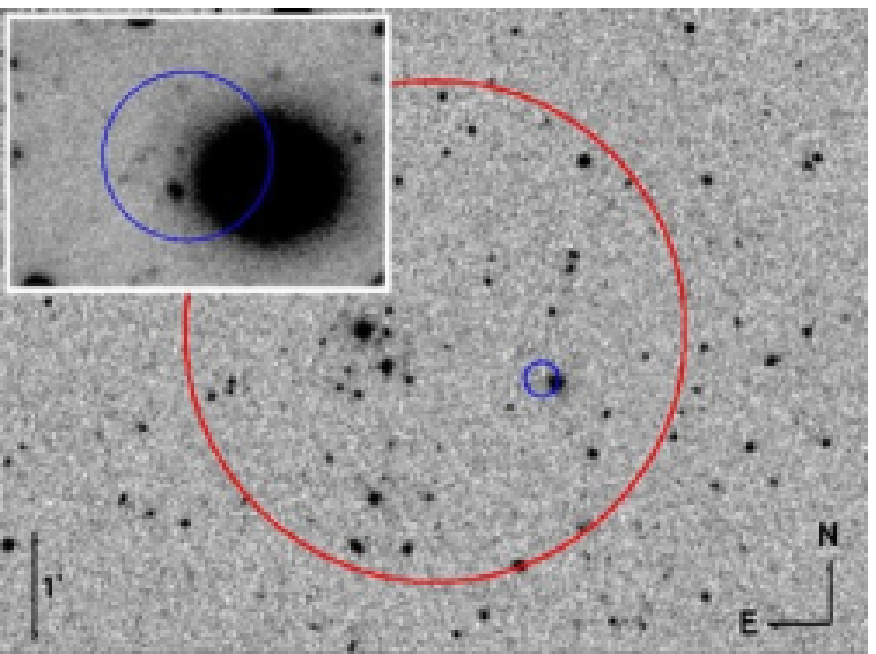,height=6cm}
	\psfig{figure=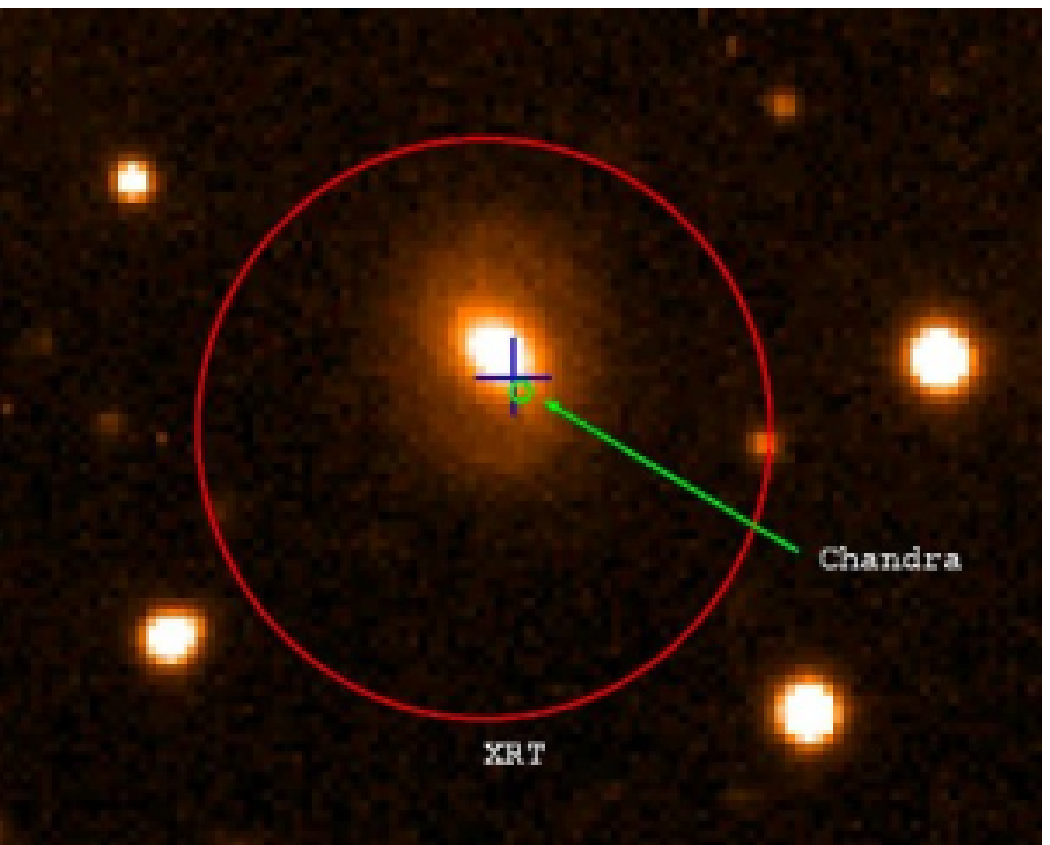,height=6cm}}
   \caption{Two Swift short GRBs associated with elliptical galaxies.
{\em left:} GRB 050509B (Gehrels et al. 2005; Bloom et al. 2006a), the
red and blue circles are BAT and XRT error boxes, respectively;
{\em Right:} GRB 050724 (Barthelmy et al. 2005b; Berger et al. 2005a).}
   \label{shortGRBs}
   \end{center}
\end{figure}

The observations led by Swift (and in small number by HETE-2) have 
revealed a completely different picture for ``short'' GRBs. Since the 
watershed discoveries of the first three short GRB afterglows 
(GRB 050509B at $z=0.226$, Gehrels et al. 2005, Bloom et al. 2006a; 
GRB 050709 at $z=0.1606$, Villasenor et al. 2005, Fox et al. 2005, 
Hjorth et al. 2005; and GRB 050724 at $z=0.258$, Barthelmy et al.
2005b, Berger et al. 2005a), by Nov. 20, 2006, a total number 12
``short'' GRB (with duration shorter than 5 seconds, Donaghy et al.
2006, see Table 1) afterglows have been discovered. The general message 
collected from these observations is that they are intrinsically different 
from long GRBs. GRB 050509B (Gehrels et al. 2005; Bloom et al. 2006)
and GRB 050724 (Barthelmy et al. 2005b; Berger et al. 2005a) are found to 
be at the outskirts of elliptical galaxies, in which star forming rate is 
very low (Fig.\ref{shortGRBs}). 
It is rather unlikely that these two events are associated with 
deaths of massive stars. GRB 050709 (Fox et al. 2005) and GRB 051221A 
(Soderberg et al. 2006b) are associated with star-forming galaxies, but 
they are usually far away from the star forming regions. There are several
other cases for which a robust host galaxy was not identified, but the
host galaxy candidates are of early type (e.g. GRB 060121 Levan
et al. 2006a; GRB 060502B, Bloom et al. 2006b). Deep supernova searches
have been performed, but with negative results (e.g. for GRB 050509B,
Bloom et al. 2006a; GRB 050709, Fox et al. 2005; GRB 050724, Berger et 
al. 2005; GRB 050813, Ferrero et al. 2006; GRB 060505, Fynbo et al.
2006). All these are consistent with the long-held speculation that
some cosmological GRBs are associated with mergers of compact objects,
such as neutron star - neutron star (NS-NS) mergers, neutron star -
black hole (NS-BH) mergers, white dwarf - black hole (WD-BH) mergers,
WD-NS mergers, and even WD-WD mergers (e.g. Pacz\'ynski 1986; 
Goodman 1986; Eichler et al. 1989; Pacz\'ynski 1991; Narayan et al. 1992; 
M\'esz\'aros \& Rees 1992; Ruffert \& Janka 1999; Fryer et al. 1999; 
Rosswog et al. 2003; Aloy et al. 2005; Dermer \& Atoyan 2006; King et al. 
2007; Levan et al. 2006c).
These mergers only involve evolved compact stars, and can happen in
early type galaxies (such as elliptical galaxies). On the other hand,
population studies reveal some novel channels to form compact star
mergers in a relatively short time scale (Belczynski et al. 2002, 2006).
This allows some merger events to happen in star-forming galaxies.
In any case, since there 
is a significant delay in time since the birth of the two compact stars 
before a coalescence happens (due to the loss of orbital angular momentum 
via gravitational radiation), the merger events tend to happen in the
outskirts of the host galaxy since NSs usually receive a large ``kick''
velocity at birth (Bloom et al. 1999b, cf. Grindlay et al. 2006). 
Although some weak nuclear
radioactivity signals would accompany the merger events (e.g. Li \&
Paczy\'nski 1998; Kulkarni 2005), they are nonetheless much fainter than
the typical Type Ib/c supernovae that accompany long GRBs. All these
suggest that the observations of ``short'' GRBs are
consistent with the compact star merger scenario. 

\begin{table}[]
  \caption[]{Durations, redshifts, host galaxies of Type I (``short''-hard) 
GRBs with afterglow detections before Nov. 20, 2006. Bursts marked with `*'
have durations longer than 5 seconds. Several other short GRBs without
afterglow detections include GRBs 050906, 050925, 051105A, 051114, and 
051211A, which are not listed in the table.}
  \label{Tab:tab1}
  \begin{center}\begin{tabular}{lcccccc}
  \hline\noalign{\smallskip}
GRB & Mission  & $T_{90}({\rm s})$ & $z$ & Host galaxy & Location & Refs         \\
  \hline\noalign{\smallskip}
050509B  & Swift  & $0.04\pm 0.004$ & 0.226 & elliptical & outskirts? & [1,2]  \\
050709 & HETE & $0.07\pm 0.01$ & 0.1606 & irregular & outskirts & [3-5] \\
050724 & Swift & $3.0\pm 1.0$ & 0.257 & elliptical & outskirts & [6-9] \\
050813 & Swift & $0.6\pm 0.1$ & -- & -- & -- & [10] \\
050911* & Swift & $\sim 16$ & 0.1646? & galaxy cluster? & -- & [11,12] \\
051210 & Swift & $1.4\pm 0.2$ & -- & -- & -- & [13] \\
051221A & Swift & $1.4\pm 0.2$ & 0.5465 & star forming galaxy & 
slightly off-center & [14,15] \\
051227$*$ & Swift & $8.0\pm 0.2$ & -- & -- & -- & [16,17] \\
060121 & HETE & $4.25 \pm 0.56$ & 1.7? or 4.6? & early-type? & outskirts? & [18-20] \\
060313 & Swift & $0.7 \pm 0.1$ & -- & -- & -- & [21] \\
060502B & Swift & $0.09\pm 0.02$ & 0.287? & early-type?  & outskirts? & [22,23] \\
060505 & Swift& $4.0\pm 1.0$  & 0.089? & star-forming galaxy & -- & [24-26]\\
060614* & Swift & $102\pm 5$ & 0.125 & star-forming galaxy & off-center & [27,28] \\
060801 & Swift & $\sim 0.50$ & 1.1304?? & -- & -- & [29,30] \\
061006 & Swift & $\sim 0.42$ & -- & -- & -- & [31,30] \\
  \noalign{\smallskip}\hline
  \end{tabular}\end{center}
References: [1] Gehrels et al. (2005); [2] Bloom et al. (2006a); [3] Villasenor
et al. (2005); [4] Fox et al. (2005); [5] Hjorth et al. (2005); [6] Barthelmy
et al. (2005b); [7] Berger et al. (2005); [8] Campana et al. (2006b); [9]
Grupe et al. 2006a; [10] Retter et al. (2005); [11] Page et al. (2006a);
[12] Berger et al. (2006a); [13] La Parola et al. (2006); 
[14] Soderberg et al. (2006b); [15] Burrows et al. (2006); [16] Barbier et al.
(2006); [17] Barthelmy et al. (2006); [18] Donaghy et al. (2006);
[19] de Ugarte Postigo et al. (2006); [20] Levan et al. (2006a);
[21] Roming et al. (2006b); [22] Troja et al. (2006); [23] Bloom et al. (2006b);
[24] Palmer et al. (2006); [25] Ofek et al. (2006); [26] Fynbo et al. (2006); 
[27] Gehrels et al. (2006); [28] Mangano et al. (2007b); [29] Racusin et al.
(2006); [30] Berger et al. (2006b); [31] Krimm et al. (2006a).
\end{table}

One important fact from the recent short GRB observations is that they are
not necessarily short. Extended emission following short GRBs has been
seen in about 1/3 of the sample in Table 1 (Norris \& Bonnell 2006).
GRB 050724 (and likely also GRB 050709) was followed by erratic X-ray flares 
that have properties similar to the prompt emission and require restart of 
the central engine (Barthelmy et al. 2005b; Zhang et al. 2006). There has
been evidence of extended emission following short GRBs in the pre-Swift
era (e.g. Lazzati et al. 2001; Connaughton 2002). A closer investigation
reveals that a larger (than 1/3) fraction of BATSE short GRBs actually
harbor observable extended emission (Norris \& Gehrels 2007). All these
greatly challenge the standard merger paradigm. Donaghy et al. (2006)
suggested to increase the separation line between short and long GRBs 
to 5 seconds.

\begin{figure}[t]
   \begin{center}
   \centerline{\psfig{figure=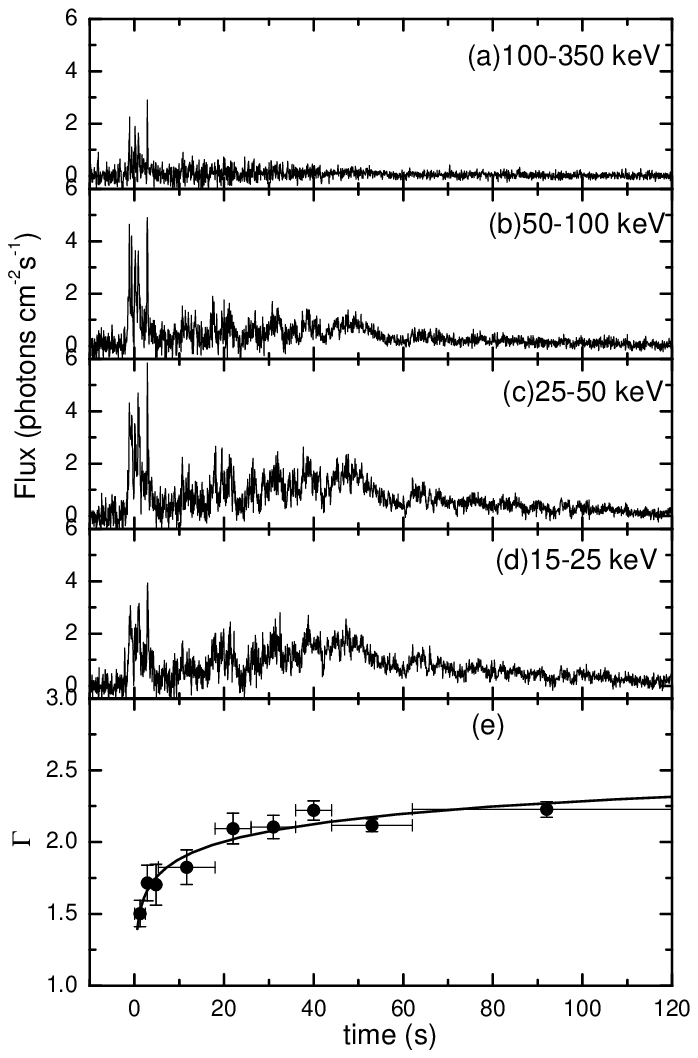,height=7cm}
	\psfig{figure=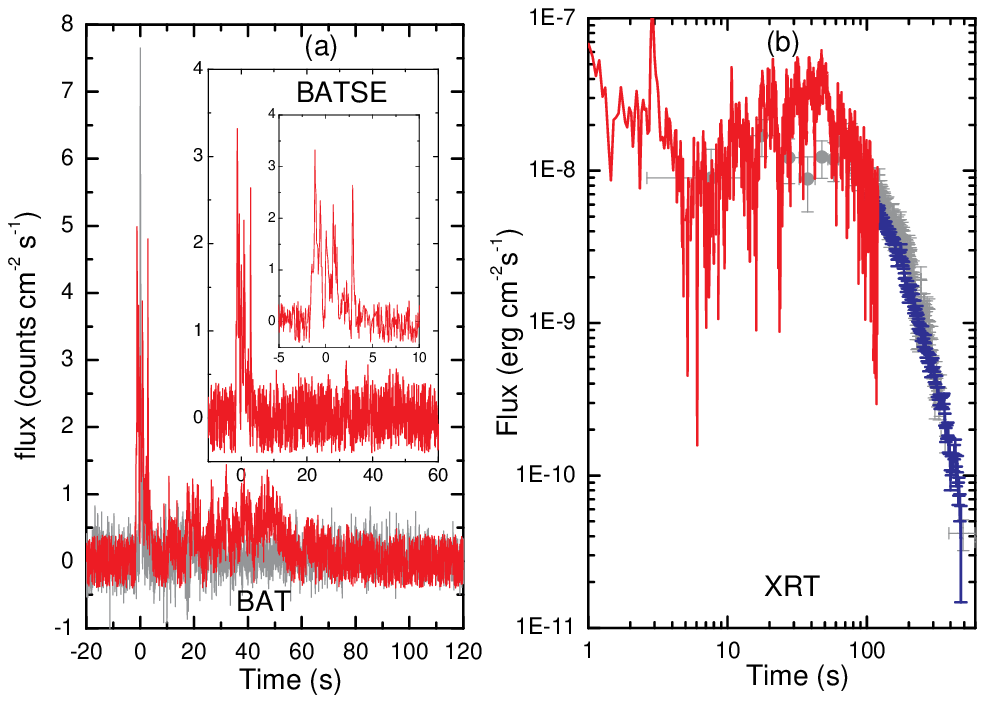,height=7cm}}
   \caption{The peculiar burst GRB 060614 (Gehrels et al. 2006; Zhang et al.
(2007a). {\em left:} Multi-wavelength lightcurves;
{\em Right:} The gamma-ray and X-ray properties of the ``pseudo'' burst
appears similar to GRB 050724 (Zhang et al. 2007a).}
   \label{060614}
   \end{center}
\end{figure}

The discovery of GRB 060614 at $z=0.125$ (Gehrels et al. 2006; Mangano 
et al. 2007b) pushes this issue to the extreme, and breaks
the clean dichotomy of the long vs. short classification regime\footnote{There
are concerns about whether the association of GRB 060614 with the nearby host galaxy
is due to a chance coincidence, e.g. Schaefer \& Xiao 2006; Cobb et al. 2006b. However,
Swift UVOT observation of the burst sets an upper limit to the burst redshift 
to be lower than 1, Gehrels et al. 2006, which rules out the higher redshifts
suggested by those authors.}. With a 
duration of $\sim 100$ s (which securely places it to the ``long'' category),
deep searches of an underlying supernova associated with this burst
came up empty-handed - any underlying
supernova is more than 100 times fainter than other SNe associated with long
GRBs, and is fainter than any SN ever observed (Gal-Yam et al. 
2006; Fynbo et al. 2006a; Della Valle et al. 2006b). More intriguingly, the
spectral lag of the burst is very short - consistent with being a short GRB
(Gehrels et al. 2006). The host galaxy has a relatively low star forming 
rate with respect to other hosts of long GRBs (Gal-Yam et al. 2006; Della 
Valle et al. 2006b; Fynbo et al. 2006a), and the afterglow is located in a 
region far away from the center of the star forming region (Gal-Yam et al.
2006). These aspects seem to be consistent with the properties of short GRBs.
Although the duration is long, a closer look at the lightcurve reveals early
hard spikes (about 5 seconds) followed by a softer emission tail with spectrum
rapidly softening with time (Gehrels et al. 2006; Zhang et al. 2007a;  Mangano 
et al. 2007b). More interestingly, the total energy of GRB 060614 is about 8
times of that of GRB 050724, the only ``short'' GRB that is robustly associated
with an elliptical galaxy. Assuming an empirical relation between isotropic 
energy (or luminosity) and spectral peak energy (the so-called Amati-relation, 
$E_p \propto E_{\rm iso}^{1/2}$, see \S3.4 for more discussion), 
which is found generally valid among bursts 
(Amati et al. 2002; Amati 2006) and also within a same burst (Liang et al. 2004),
Zhang et al. (2007a) generated a pseudo-burst that is about 8 times less
energetic. They found that this synthetic burst is ``short'' with $T_{90}\sim
4.4$ s in the BATSE band. The late soft gamma-ray tails are shifted to the
X-ray band as X-ray flares. This is essentially a carbon-copy of GRB 050724.
This suggests that GRB 060614 is likely simply a more energetic version of
GRB 050724, and should belong to merger-type GRBs (Fig.\ref{060614}). 
Another point is that the beaming-corrected gamma-ray energy of GRB 060614
is comparable to those of other short GRBs, but is about an order of magnitude
less than long ones (Mangano et al. 2007).
Although the possibility that GRB 060614 stands for a third type of GRBs 
is not ruled out (e.g. collapsars without supernova signature, Woolsey 1993), 
it appears that GRB 060614 is a close relative of GRB 050724. It is worth
commenting that some BATSE bursts (e.g. trigger 2703) has the similar 
properties as GRB 060614 (e.g. Norris \& Bonneli 2006). These bursts have
a larger intensity ratio between the extended emission and the prompt 
emission than most other short GRBs (Norris \& Gehrels 2007).

With such a connection, the traditional ``short vs. long'' classification
regime breaks down, and some new terminologies involving multiple criteria 
are needed to define the two GRB categories. Zhang et al. (2007a) suggest the
terms ``Type I' and ``Type II'', by analogy with the supernovae nomenclature
(e.g. Filippenko 1997). As summarized in Zhang et al. (2007a) and Zhang (2006), 
Type I (the previous short-hard) GRBs are usually short (but may have long 
soft tails) and hard (but the tail may be soft), with very short spectral lags 
and with no supernova associations. Like Type Ia supernova, Type I GRBs are 
associated with old stellar population and can be found in all types of host 
galaxies including elliptical galaxies, and are typically in regions with low 
star forming rate, which is usually outskirts of the host galaxy. The most 
likely progenitor candidates are compact star mergers, which involve binary 
systems, again similar to Type Ia SNe. On the other hand, Type II (the previous
long-soft) GRBs are usually long and soft, with long spectral lags and supernova
associations. Like Type II supernovae, they involve core collapses of massive 
stars, which belong to the young stellar population. Their host galaxies are
late type, predominantly irregular, dwarf galaxies. The location is usually 
near the center of the star-forming core of the host galaxy. According to
this new definition, GRB 060614 belongs to Type I. In fact, two other
Swift bursts have been suggested to belong to the ``short''-hard category
even though their durations are longer than 5 seconds. They are GRB 050911 with
$T_{90} \sim 16$s (Page et al. 2006a; Berger et al. 2006a) and GRB 051227 
with $T_{90} \sim 8$s (Barthelmy et al. 2006). These three GRBs are also 
listed in Table 1 as Type I GRBs (marked with `*' symbol), making the total 
number in the sample 15. In the rest of the paper, I will interchangeably use
``Type I / Type II'' and ``short (hard) / long (soft)'' in the text.

It is worth commenting that afterglow modeling also lends indirect support to 
the merger scenario of Type I GRBs. The immediate environment of Type I GRBs
should be tenuous with low ISM density (Panaitescu et al. 2001; Perna \&
Belczynski 2002; Fan et al. 2005a). As a result, afterglow observations
(especially multi-wavelength) may potentially lead to constraints on the
density and thereby shed light onto the nature of the bursts. Afterglow
modeling has been indeed carried out for several short GRBs, and the results
are generally consistent with a low ambient medium density (e.g. Fan et al. 
2005a; Panaitescu 2006a; Roming et al. 2006b; Burrows et al. 2006). Some
short GRBs appear ``naked'' (i.e. no external shock afterglow component)
or completely with no afterglow detection. They are also consistent being
born in a low-density medium. On the other hand,
abnormal afterglow behaviors have been observed. For example, 
GRB 060313 exhibited complex structure with different 
decay indices and flaring (Roming et al. 2006b). The
optical flux fluctuation may be related to density fluctuation of the ambient
medium, or to weak central engine activities.
Another caveat is that a low density medium may not be solely associated with
Type I GRBs. When analyzing GRB radiative efficiency of a sample of Swift 
bursts (most are Type II bursts), Zhang et al. (2007b) found that 
in about 1/3 of the bursts the X-ray band is below the cooling frequency for 
a very long period of time. This suggests a very small $\epsilon_B$ or a very
low ambient density for long GRBs as well.

As for cosmological setting, luminosity and redshift distributions of known
short-hard bursts have been used to constrain the progenitor 
lifetime of compact star binaries. With a small sample of short GRBs
with known redshift, Nakar et al. (2006a) and Guetta \& Piran
(2006) found that in order to reproduce the observed redshift distribution,
the typical progenitor lifetime is typically longer than previously 
believed, and the local burst rate is also
higher than previous believed. On the other hand, Belczynski et al. (2006)
argue a bimodal distribution of the merger times corresponding to two
distinct evolutionary tracks of compact star binaries. They argue that there
exist a population of mergers whose merger time scale is short, so that they
could be found in star-forming galaxies. Zheng \& Ramirez-Ruiz (2006) study
the merger rate in both early-type and late-type galaxies and argue a large
merger time for at least half of short GRBs. The existence of
some possible high-$z$ short GRBs (e.g. Levan et al. 2006a; Berger et al.
2006b) may suggest that there exist some fast evolutionary channels
such as those proposed by Belczynski et al. (2002, 2006). Nakar et al.
(2006a) have tested several possible function forms for the lifetime 
distribution. The high-$z$ short GRBs suggest that the narrow lognormal
distribution tested by Nakar et al. is inconsistent with the data, while
the wide lognormal or a single power law function forms may be still
consistent with the data. 

Another interesting question regarding short-hard GRBs is what fraction is
produced by SGR giant flares in nearby galaxies. The Dec.27, 2004 giant
flare event from SGR 1800-20 has a luminosity $(3.7\pm 0.9)\times 
10^{46} d_{15}^2~{\rm erg ~s^{-1}}$ (where $d_{15}$ is the distance of the
source in unit of 15 kpc), suggesting a high detectability of similar 
events up to $30 d_{15}$ Mpc, which would contribute to a significant fraction 
of BATSE short-hard bursts (Hurley et al. 2005). A search for associations
of well-localized short GRBs with nearby galaxies (Nakar et al. 2006b),
however, sets an upper limit of this fraction of $\sim 15\%$. Schaefer (2006) 
found that most BATSE short hard GRBs were not associated with moderately 
bright nearby galaxies. Tanvir et al.
(2005) on the other hand, report a correlation between the locations of
previously observed short bursts and the positions of galaxies in the local
universe, indicating that about $10\%-25\%$ short GRBs originate at 
$z<0.025$. 

\subsection{GRBs vs. XRFs}

X-ray flashes (XRFs) are the extension of typical long GRBs to the softer,
and fainter regime. They were first identified with the Beppo-SAX satellite
(Heise et al. 2001; Kippen et al. 2002), and studied more extensively with
HETE-2 in the pre-Swift era (e.g. Sakamoto et al. 2005; Lamb et al. 2005a;
D'Alessio et al. 2006). The lightcurves of XRFs are similar to those of 
long GRBs, with rapid temporal variability in many cases 
(Heise et al. 2001; D'Alessio et al. 2006). 
The spectral properties of XRFs are similar to those of GRBs, except that 
the values of the peak energy $E_p$ of the burst $\nu F_\nu$ spectrum are 
much smaller (Cui et al. 2005). 
The peak flux and the total energy fluence of XRFs are also 
correspondingly smaller (Sakamoto et al. 2005; D'Alessio et al. 2006). 
There is no clear separation between GRBs and XRFs. Bursts in the grey zone
are sometimes called ``X-ray rich GRBs'' (XRRs). Besides the traditional 
$E_p$-distribution peak around 200 keV (Preece et al. 2000), there is 
tentative evidence of a second distribution peak of $E_p$ around 15 keV
(Liang \& Dai 2004). The poor statistics however does not
allow a robust claim of two distinct components in $E_p$-distribution. It
is possible that the GRB-XRF $E_p$ distribution forms a broad peak around
100 keV. 

Since the identification of XRFs, many suggestions have been proposed to 
interpret XRFs and their relation with GRBs. In general these models fall
into two broad categories: i.e. XRFs differ from GRBs extrinsically 
(different distances or different viewing angles) or intrinsically 
(different physical parameters, different radiation mechanisms, or even 
different progenitors and central engines). The following is a list of
models of XRFs proposed previously. The first four are ``extrinsic''
models, while the latter five are ``intrinsic'' ones. Swift has detected 
and extensively monitored a handful of XRFs. These observations 
significantly constrained the possible models of XRFs.

\begin{itemize}
\item {\bf High redshift GRBs}. One early speculation (Heise et al.
2001) is that XRFs are distant GRBs so that the redshift effect makes
them softer and fainter. Redshift measurements of several GRBs (e.g.
$z=0.251$ for XRF 020903, Soderberg et al. 2004; $z=0.21$ for XRF 040701,
Soderberg et al. 2005) suggest that at least some of them are nearby events.
Analyses of the $E_p$ predictions in various GRB prompt emission models
suggest that $E_p$ are usually functions of many parameters (including $z$),
and that the dependences on some other parameters (e.g. the bulk Lorentz 
factor $\Gamma$) are more sensitive than the dependence on $z$ (e.g. Table
1 of Zhang \& M\'esz\'aros 2002c). This suggests that redshift should not
be the sole factor to define an XRF. A systematic study of the redshift
distribution of XRFs rules out the suggestion that XRFs are high-$z$ GRBs 
(Gendre et al. 2006a).
\item {\bf Off-beam viewing geometry for a uniform GRB jets.} The energy
budget requirement and the temporal breaks in some afterglows have led to
the suggestion that GRB ejecta are beamed (Rhoads 1997, 1999; Sari et al.
1999; Frail et al. 2001)\footnote{It is worth emphasizing that the 
salient-feature of the jet model, i.e. an achromatic temporal break in 
multi-wavelength afterglows, has not been generally confirmed after two years 
of Swift operation (see e.g. Willingale et al. 2006; Burrows 2006; Zhang 
2007; Covino et al. 2006, and \S3.3, \S5 for more discussion).}. The 
simplest model (not necessarily the most realistic model) suggests that 
the jet forms a uniform conical structure with sharp edge in energy.
Within this scenario, it has been suggested that GRBs correspond to
on-beam geometry while XRFs correspond to off-beam geometry (e.g. Yamazaki
et al. 2002, 2004a). A direct prediction of such a scenario is that the
early lightcurve should rise initially due to the gradual entrance of the
main ultra-relativistic cone into the observer's field of view (Granot et al. 
2002). Recent Swift observations suggest that the afterglows are decaying 
from the very early epoch of the observation (e.g. Schady et al. 2006b;
Mangano et al. 2007a). This essentially rules out a sharp-edge off-beam
geometry of XRFs. In some XRFs, the early decay slope is shallow. However,
an early shallow decay is a common feature of most Swift GRB X-ray afterglows
(see \S3.1 for more discussion). This model may be amended by introducing
a smoothed edge, which is effectively a structured jet as discussed below.
\item {\bf Off-axis viewing geometry for a (one-component) structured jet.}
GRB jets may have significant structure, with an angle-dependent energy
per solid angle and possibly Lorentz factor as well (M\'esz\'aros et al.
1998). An on-axis geometry of a structured jet would modify the afterglow
temporal decay rate (M\'esz\'aros et al. 1998; Dai \& Gou 2001; Panaitescu
2005a), while an off-axis geometry would mimic a jet-break-like lightcurve
as the Lorentz factor along the line-of-sight is reduced to be comparable
to the viewing angle from the jet axis (Rossi et al. 2002; Zhang \& 
M\'esz\'aros 2002b; Wei \& Jin 2003; Kumar \& Granot 2003; Granot \& Kumar 
2003; Panaitescu \& Kumar 2003; Salmonson 2003; Rossi et al. 2004). Within
such a picture, energy per solid angle decreases with viewing angle with
respect to the jet axis. Depending on the unknown jet structure, at
certain viewing angles, an otherwise detected GRB (if viewed near the jet 
axis) would be observed 
as an XRF. The jet angular structure is unknown, and in reality it may not
follow any simple analytical function. For the purpose of modeling, usually
power law jets ($\epsilon(\theta) \propto \theta^{-k}$, and in particular 
$k\sim 2$, Rossi et al. 2002; Zhang \& M\'esz\'aros 2002b) and Gaussian 
jets ($\epsilon (\theta) \propto \epsilon_0 \exp (-\theta^2/2\theta_0^2)$,
Zhang \& M\'esz\'aros 2002b) have been widely discussed. Both models have
been suggested to interpreted XRFs (e.g. Zhang et al. [2004a] for Gaussian
jets and Jin \& Wei [2004] and D'Alesio et al. [2006]
for power law jets). Lamb et al. (2005a) pointed out that
an $\epsilon(\theta) \propto \theta^{-2}$ structured jet tends to over
predict the number of XRFs, which is inconsistent with the rough 1:1:1
number ratio for GRBs, XRRs and XRFs. The Gaussian jets can easily pass
this and several other observational constraints (Zhang et al. 2004a,
X. Dai \& Zhang 2005). Since there are relativistically moving materials 
(though with a smaller energy) along the line of sight, the lightcurve
in this model decays from the very beginning (Kumar \& Granot 2003;
Salmonson 2003), not inconsistent with the observational constraint
from X-ray lightcurves (Mangano et al. 2007a). Yamazaki et al. (2004b)
introduced ``patches'' or ``mini-jets'' in a Gaussian-like structured
jets to present a unified model of long, short GRBs as well as XRRs
and XRFs. As discussed in \S2.1, it is now clear that short GRBs form
a distinct new population from long GRBs, XRRs and XRFs, which cannot
be unified within this model.
\item {\bf Two-component jets.} Another widely discussed model is the
two-component jet model, a special type of structured 
jets. Because this model is motivated physically by progenitor 
models, it receives broad attention. In the collapsar model of 
Type II GRBs, it is natural to expect a hot cocoon surrounding the
central relativistic jet that penetrates from the star (Woosley et al. 1999;
Zhang et al. 2003b, 2004b; M\'esz\'aros \& Rees 2001; Ramirez-Ruiz et al. 
2002a; Mizuta et al. 2006; Morsony et al. 2006). 
The cocoon would form a distinct second less energetic jet component.
Even in the naked GRB models, a neutron-rich MHD outflow would be
naturally separated into a narrow, high-$\Gamma$ proton jet and a wide, 
low-$\Gamma$ neutron jet (Vlahakis et al. 2003; Peng et al. 2005). In
both types of model, if the line of sight sweeps the less-energetic
wide beam, one may observe an XRF. Phenomenologically, the two-component
jet model has been introduced in several other contexts (Lipunov et al.
2001; Berger et al. 2003a). In particular, within the XRF context, Huang
et al. (2004) have interpreted the rising bump of the optical lightcurve
of XRF 030723 within the framework of the two-component jet model. An 
alternative interpretation of the bump is the supernova component 
(Tominaga et al. 2004). The tentative bi-modal distribution of $E_p$ of 
GRBs and XRFs was also suggested as a support to the two-component jet 
model (Liang \& Dai 2004). Swift has followed some XRFs extensively to
very late epochs. The lightcurve of XRF 050416A (Mangano et al. 2007a) 
keeps decaying with a constant slope to very late times. This greatly
constrains the two-component jet model for XRFs: the narrow bright
jet component must not be prominent enough to leave a rebrightening
signature in the lightcurve. This suggests that at least two components
are not required to interpret XRFs. Similar conclusions are drawn from the 
observations of other XRFs (e.g. Levan et al. 2006b).
\item {\bf Intrinsically faint, less collimated jets.} Lamb et al. (2005a)
suggested a toy model invoking varying opening angle of jets. While GRBs
are bright, narrow jets, XRFs are much fainter, wider jets. Starting with
the assumption of constant energy reservoir of all events (as derived 
by Frail et al. 2001; Bloom et al. 2003 - but not confirmed by more
systematic observations by Swift, e.g. Willingale et al. 2006; Burrows
2006; Zhang 2007), they drew the conclusion that GRBs have a typical
opening angle of about 1 degree, while XRFs are essentially isotropic
events. These narrow opening angles for GRBs are inconsistent with
the typical angle derived from the afterglow jet break data (typically 5 
degrees, see Zhang et al. 2004a for more discussion). On the other hand,
if the standard energy reservoir assumption is dropped (as is suggested
by the recent Swift data), this very narrow jet inference is no longer
valid. The constant-slope power law decay lightcurve of XRF 050416A 
(Mangano et al. 2007a) is consistent with a wide-beam jet model. It is 
however worth commenting that the constant-slope power law decay of
X-ray lightcurves seem to be a common feature of some other normal
GRBs as well (e.g. 100 day lightcurve of GRB 060729, Grupe et al. 2006b), 
so that the jet opening angle may not be the crucial criterion to
define whether a burst is a GRB or an XRF. 
\item {\bf Dirty fireballs.} A naive expectation is that bursts with lower
Lorentz factors would receive smaller Doppler boost and therefore give
softer, fainter emission. These dirty fireballs (e.g. Dermer et al. 1999;
Huang et al. 2002) have been suggested as the origin of XRFs. A detailed
study of $E_p$ models (Zhang \& M\'esz\'aros 2002c) suggest that 
depending on prompt emission models, $E_p$ depends on $\Gamma$ in a
non-trivial way. In particular, the popular internal shock model predicts
high $E_p$'s for dirty fireballs. The dirty fireball suggestion is in
any case relevant for the external shock model and models invoking 
internal magnetic field dissipation or photosphere dissipation (Table 1 
of Zhang \& M\'esz\'aros 2002c). Current Swift XRT observations strongly 
suggest an internal origin of GRB prompt emission (Zhang et al. 2006, see 
\S3.1 for more discussion). So the dirty fireball suggestion for XRFs, if 
proven true, may suggest internal dissipation models other than the 
conventional internal shock models. It is interesting to note that the 
latest extreme XRF 060218 discovered by Swift (Campana et al. 2006a) shows
an extremely long and smooth lightcurve with very low $E_p \sim 5$ keV.
Along with other intermediate XRFs (e.g. Lamb et al. 2005a; Sakamoto et
al. 2005, 2006a), XRF 060218 also satisfies the $E_p \propto E_{\rm iso}^{1/2}$
Amati-relation (Amati et al. 2002; Amati et al. 2006). Furthermore, a
study of the multi-band temporal profiles and spectral lags of XRF 060218
suggests that its spectral lags are extremely long (Liang et al. 2006b), 
roughly consistent with the luminosity - lag relation discovered in 
BATSE long GRBs (Norris et al. 2000) and confirmed by Swift Type II GRBs
(Gehrels et al. 2006). All these seem to be consistent
with the straightforward intuition that XRF 060218 is a less-Lorentz-boosted
burst comparing with canonical GRBs, suggesting a dirty fireball. The low
Lorentz factor for XRF 060218 is also inferred from radio observations 
(Soderberg et al. 2006a) and is required in theoretical models to interpret
this peculiar event (Dai et al. 2006b; Wang et al. 2006a; Toma et al. 2007). 
Another comment
is that the dirty fireball suggestion does not exclude the structured jet
and wide opening angle jet model discussed above, since in those models
the Lorentz factors along the line of sight could be also low. On the 
other hand, there is no evidence for off-axis emission for XRF 060218.
Another low-luminosity nearby GRB 980425 has been extensively monitored
in radio at late times, which significantly constrained the off-axis
model for the burst (Waxman 2004a).
\item {\bf Intrinsically inefficient GRBs from clean fireballs.} This is
a specific XRF model within the framework of the internal shock model (Barraud
et al. 2005), since in this model $E_p \propto \theta_p^4 L^{1/2} r^{-1}$ 
(e.g. Zhang \& M\'esz\'aros 2002c), where $\theta_p$ is the internal energy 
of the protons in the internal shocks, $L$ is the wind luminosity, and $r 
\sim\Gamma^2 c \delta t$ is the internal shock radius, and $\delta t$ is 
the variability time scale. A clean fireball (large $\Gamma$) tends to give
a larger internal shock radius, at which magnetic field strength is 
smaller so that the typical synchrotron frequency ($E_p$) is lower. An 
inefficient internal shock reduces $\theta_p$, and also helps to lower 
$E_p$. Indeed in the pre-Swift era it has been found that XRFs have lower
radiative efficiency than GRBs by comparing their prompt emission energy
with the late time kinetic energy inferred from the X-ray afterglow data
(e.g. Soderberg et al. 2004; Lloyd-Ronning \& Zhang 2004). However, using
the earliest Swift data, it is found that XRFs are generally as efficient
as GRBs (Schady et al. 2006a; Zhang et al. 2007b), suggesting that XRFs
are not intrinsically inefficient GRBs. The apparent low efficiency derived 
from the late time X-ray data may be caused by a prolonged energy injection 
epoch in the early phase. This seems to be consistent with the expectation 
of the off-axis structured (e.g. Gaussian-like) jet model of XRFs (Zhang et 
al. 2004a).
\item {\bf Photosphere-dominated emission models.} In GRB fireballs, there
are in principle three emission regions that could potentially contribute 
to the observed prompt gamma-ray emission. Besides the traditional external
shock (Rees \& M\'esz\'aros 1992; M\'esz\'aros \& Rees 1993) and internal
shocks (Rees \& M\'esz\'aros 1994), baryonic and pair photospheres 
(Thompson 1994; M\'esz\'aros \& Rees 2000; Kobayashi et al. 2002;
M\'esz\'aros et al. 2002; 
Rees \& M\'esz\'ros 2005; Ryde 2005; Ramirez-Ruiz 2005; Ryde et al. 2006; 
Thompson et al. 2006) are another important emission site. The domination of 
photosphere emission, under certain conditions, could give rise to soft 
emission that characterizes XRFs. The data generally suggest that the $E_p$ 
distribution of GRBs and XRFs forms a broad distribution peak. As a result, 
to accept the photosphere interpretation of XRFs
(M\'esz\'aros et al. 2002), one should expect that the prompt emission 
of GRBs is also from the photosphere. Such a model is being advocated 
recently (Rees \& M\'esz\'aros 2005; Ryde et al. 2006; Thompson 
et al. 2006).
\item {\bf Completely different origin with respect to GRBs.} The last
possibility is that XRFs are different from GRBs. They may originate from
different progenitors, may have different central engines, and different
radiation mechanisms as well. As discussed above, XRFs seem to be a
natural extension of GRBs to softer and fainter regime. If the difference
is not due to the viewing angle effect, the variation of progenitor and
central engine properties must be gradual and smooth, and probably without
abrupt transition. The recent observations of XRF 060218, however, raise 
new discussion on the topic. The radio afterglow observation of XRF 060218 
(Soderberg et al. 2006a) suggests that the central engine may be a neutron 
star rather than a black hole. A similar conclusion was reached independently 
by modeling of the supernova associated with the XRF (Mazzali et al. 2006). 
A population study suggests that the low-luminosity GRBs such as XRF 060218
may require a distinct new component in the GRB luminosity function (Liang
et al. 2006c). The existence of the thermal X-ray component in the prompt
emission spectrum (Campana et al. 2006a) may require a novel radiation
mechanism different from that for canonical GRBs (e.g. Wang
et al. 2006). The facts that XRF 060218 satisfy the Amati-relation (Amati
et al. 2006) and the lag-luminosity relation (Liang et al. 2006b), on the
other hand, suggest that the radiation mechanism for XRFs should not be
much different from that for GRBs (For more discussion on the two
correlations, see \S3.4). It is worth emphasizing that bursts 
with different progenitor systems and central engines could well share 
the same radiation mechanism, since the fireball properties are generic 
and independent of the unknown central engine.
\end{itemize}

In summary, Swift observations have significantly narrowed down the
possible models of XRFs. The high-redshift scenario is essentially ruled
out. The sharp edge off-beam jet model is disfavored by the early
afterglow data of XRFs. Observations of XRF 060218 (the softest XRF)
suggest that its radiation mechanism should not be very different from
that of GRBs. The long-term constant-slope decay of XRF 050416A 
(Mangano et al. 2007a) disfavors jet models other than the wide-beam
uniform jet model and a large viewing angle Gaussian-like structured
jet model. Regarding the Lorentz factor, in view of the properties of
XRF 060218 (long pulse, long lag, soft spectrum), the dirty (rather 
than clean) fireball scenario is favored. Combining with the fact 
that GRB prompt emission is likely of internal origin (Zhang et al. 
2006), mechanisms other than internal shocks are preferred. Photosphere
or magnetic dissipation mechanisms may be good candidates.

An interesting question is how XRFs are related to the X-ray flares
observed following some GRBs and XRFs (Burrows et al. 2005b). Could it 
be possible that XRFs are simply X-ray flares without prompt emission 
detection? As discussed later (\S3.1.2), X-ray flares generally require
reactivation of the central engine, and therefore would have the same
energy dissipation mechanism as the prompt emission. On the other hand,
observationally the lightcurves of XRFs are more variable than X-ray
flares, which are similar to those of prompt GRBs (Heise et al. 2001;
D'Alessio et al. 2006). This may suggest that prompt XRF emission is
related to the prompt accretion phase at the central engine, while
X-ray flares are related to accretion at late epochs, which generally
predicts a smoother lightcurve consistent with the viscous disk evolution
at large radii (e.g. Perna et al. 2006; Proga \& Zhang 2006).

\subsection{HL-GRBs vs. LL-GRBs}

\begin{figure}[t]
   \begin{center}
   \centerline{\psfig{figure=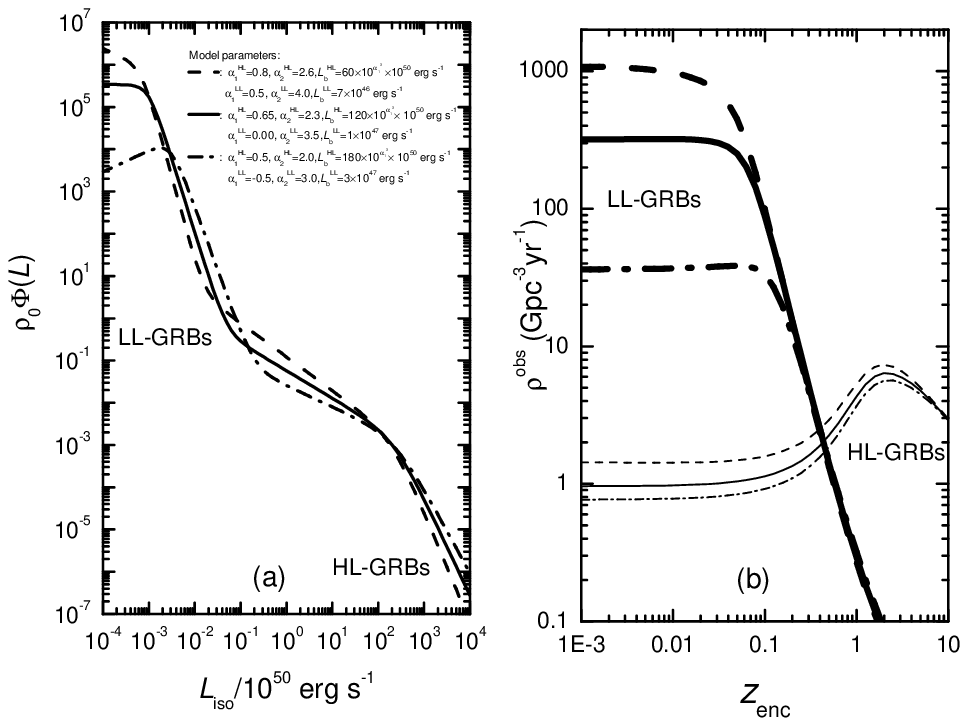,height=8cm}}
   \centerline{\psfig{figure=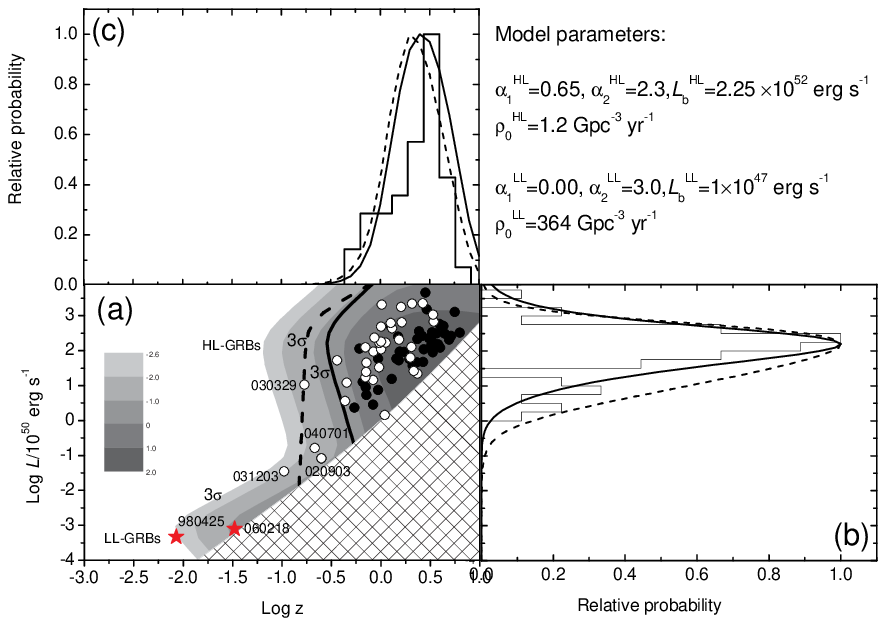,height=7cm}}
   \caption{The HL and LL populations of GRBs. {\em Upper:} The suggested
two-component luminosity function and predicted observed local event rate
as a function of redshift;
{\em Lower:} The model can interpret the observed 2-D luminosity and
redshift distributions of HL and LL GRBs (from Liang et al. 2006c).}
   \label{HL-LL}
   \end{center}
\end{figure}

The detection of XRF 060218 at $z=0.0331$ (Mirabal et al. 2006) within 
1.5 years of operation of Swift (Campana et al. 2006a), together
with the previous detection of GRB 980425 at $z=0.0085$ by BeppoSAX
(Galama et al. 1998), suggest that the local event rate of low-luminosity
(LL) GRBs is very high. The volume enclosed by $z<0.033$ is very small,
$V_{z<0.033} \sim 0.01~{\rm Gpc}^3$. One can naively estimate the
local event rate of these low-luminosity (LL) GRBs ($\rho_0^{\rm LL}$)
by $\rho_0^{\rm LL} V_{z<0.033} (T^{Bepp}\Omega^{Bepp}/4\pi +
T^{Swift} \Omega^{Swift}/4\pi) \sim 2$, where $T^{Bepp} \sim 6$ yr and
$T^{Swift} \sim 1.5$ yr are the operation times for the BeppoSAX and
Swift missions, respectively, and $\Omega^{Bepp} \sim 0.123$ and 
$\Omega^{Swift} \sim 1.33$ are the solid angles covered by the two
missions, respectively. This rough estimate gives $\rho_0^{\rm LL}  
\sim 800 ~{\rm Gpc^{-3}~yr^{-1}}$, which is much greater than the
local event rate of the conventional high-luminosity (HL) GRBs of
$1 ~{\rm Gpc^{-3}~yr^{-1}}$ (e.g. Schmidt 2001) and its simple
extrapolation to low luminosities, i.e. $\leq 10 ~{\rm Gpc^{-3}~
yr^{-1}}$ (Guetta et al. 2004). Such a high event rate for LL-GRBs 
has been independently derived by
several groups (e.g. Cobb et al. 2006a; Pian et al. 2006; Soderberg
et al. 2006a; Liang et al. 2006c). By investigating the 1-D and 2-D
distributions of luminosity and redshift for a sample of GRBs with
known redshifts, Liang et al. (2006c) found that the current sample
is not compatible with a single luminosity function component.
Rather, data require a distinct LL-GRB component other than the
HL-component (Fig.\ref{HL-LL}). The former component has a much higher 
event rate than the latter. In view that GRB 060218's central
engine may be a neutron star rather than a black hole (Mazzali et 
al. 2006; Soderberg et al. 2006a), one would speculate that 
the apparent bimodal distribution in the luminosity function may
be related to the two distinct types of central engines involved, 
e.g. HL-GRBs involve black holes, while LL-GRBs involve neutron stars.

Although these individual LL-GRBs are less energetic and 
under-luminous, due to their much higher event rate, they could 
give an interesting contribution to various diffuse emission backgrounds. 
For example, assuming LL-GRBs produce gamma-rays in internal shocks
similar to HL-GRBs, the protons in LL-GRBs would produce high energy 
neutrinos through photo-meson interaction at the $\Delta$-resonance, 
which have the dominant contribution at energies above $10^{16}$ eV
(Gupta \& Zhang 2007a; Murase et al. 2006).

\subsection{Optically bright vs. dark, optically luminous vs. dim}

In the previous optical follow up observations, GRBs are generally
divided into two categories, optically bright and optically dark
ones (e.g. Jakobsson et al. 2004; Rol et al. 2005). 
The latter typically account for $\sim 50\%$ of the total 
population\footnote{Swift UVOT does not detect optical afterglows
for $\sim 67\%$ of the Swift bursts. Combining with ground-based
follow ups, the non-detection rate is $\sim 45\%$ (P. Roming, 2006,
private communication).}.
The discovery of the early optical flash of GRB 021211 (Fox et al.
2003b; Li et al. 2003a) in the HETE-2 era had led to the speculation
that as long as observations are performed early enough, most dark
bursts are not dark. However, Swift UVOT did not detect a large
number of bursts even at very early epochs (Roming et al.
2006a). It is not straightforward to define an optically ``dark'' 
burst. Jakobsson et al. (2004) and Rol et al. (2005) use the criterion
that a burst is ``darker'' than it is expected to be (using spectral
extrapolation from the X-ray band for example) to define a dark
burst. This might be the most quantitative method to define late
darkness. At early times, it is somewhat ``expected'' theoretically
to observe optical emission originated from the reverse shock
(which is not a strong contributer to the X-ray band). In such
a case, the X-ray band and the optical band are not from a same
emission component, rendering the quantitative definition 
inconclusive. The existence of X-ray flares (and possible other
internal-related emission in the X-ray band) makes the case
even more complicated (see e.g. the completely different early 
X-ray/optical lightcurves in GRB 060418 and GRB 060607A,
Molinari et al. 2006). The non-detection of a large fraction of 
Swift bursts by UVOT (Roming et al. 2006a) 
at least suggest that the reverse shock
component is insignificant. Among the other possible reasons of 
optical darkness, foreground extinction, circumburst absorption, 
and high redshift are the best candidates.

Among the optically bright GRBs, it is intriguing to discover that
there are two sub-classes, namely optically luminous and optically dim 
categories (Liang \& Zhang 2006a; Nardini et al. 2006; Kann et al.
2006). The rest-frame lightcurves of GRBs with known redshifts
are found to follow two ``universal'' tracks. The rest-frame 10-hour
luminosities of the bursts with known redshifts 
show a clear bimodal distribution. The optically dim
bursts all appear to be located at redshifts lower than $\sim 1$ and
their lightcurve tends to be smooth and single-pulsed, while the
optically luminous bursts have a wider redshift distribution and
the lightcurves are more complex (Liang \& Zhang 
2006a). A related dichotomy in the prompt emission properties
(lags and internal luminosity functions) was identified by 
Hakkila \& Giblin (2006, see also Norris 2002). The origin of the 
dichotomy is unknown. The two universal tracks of afterglow 
lightcurves may be related to the different total explosion 
energies involved in the two groups of bursts. It is worth
commenting that Gendre \& Boer (2005) have reported two groups
of X-ray afterglow lightcurves. However, this is not confirmed
by the Swift data (O'Brien et al. 2006; Willingale et al. 2006).
The lack of evidence in X-rays to support the bimodal optical 
luminosity distribution is puzzling. On the other hand, there is
growing evidence that some early X-ray afterglow emission may
be more related to the GRB central engine, and hence, is a different
component from the optical one (which is likely from the external
shock). This might be the reason of the discrepancy.

\section{Physics of GRBs and afterglows}
\label{sect:Res}

The standard GRB fireball model has been extensively reviewed (Piran 1999; 
M\'esz\'aros 2002; Zhang \& M\'esz\'aros 2004; Lu et al. 2004; Piran 2005; 
M\'esz\'aros 2006). Regardless of the nature of the explosion, 
the generic fireball shock model invokes a relativistically expanding
ejecta. According to this model, the ejecta is 
intrinsically intermittent and unsteady, and is composed of many 
mini-shells with a wide range of bulk Lorentz factors. 
Internal shocks (Rees \& M\'esz\'aros 1994) are likely developed before 
the global fireball is decelerated by the ambient medium, which are
generally believed to be the emission sites of the observed prompt GRB
emission. Alternatively, magnetic dissipation may be responsible for
the prompt gamma-ray emission even without internal shocks.
The fireball is decelerated at a larger distance after 
sweeping enough interstellar medium whose inertia becomes noticeable, 
and the blastwave enters a self-similar deceleration regime at later
times (Blandford \& McKee 1976). Upon deceleration, a pair of shocks
forms. A long-lived forward shock propagating into the ambient medium
gives rise to the long-term broad band afterglow (M\'esz\'aros \& Rees
1997a; Sari et al. 1998); and a short-lived reverse shock propagating
into the ejecta itself gives rise to a possible optical/IR
flash and a radio flare (M\'esz\'aros \& Rees 1997a, 1999; 
Sari \& Piran 1999a,b). The relativistic ejecta are likely collimated 
(Rhoads 1999; Sari et al. 1999), and the jets may have substantial 
angular structures (Zhang \& M\'esz\'aros 2002b; Rossi et al. 2002). 
This general theoretical framework has been successful to interpret 
most of the observational data in the pre-Swift era. 
With the successful launch and operation of Swift, we now have 
unprecedented information about GRB afterglows, which sheds 
light on many outstanding problems in the pre-Swift era (Zhang 
\& M\'esz\'aros 2004 for a summary): e.g. central engine, composition 
and geometric configuration of the GRB fireball, and its interaction 
with the ambient medium.

It is informative to clarify the definitions of ``prompt emission''
and ``afterglow'' at this point. Traditionally, ``prompt emission'' 
refers to the emission component detected by the gamma-ray detector
(sometimes also optical emission simultaneously detected during the
gamma-ray emission phase); while all the emissions detected by other 
instruments at later times are termed as ``afterglow''. 
On the other hand, Swift observations strongly
suggest that such a definition scheme is not physical. X-ray flares,
if strong and hard enough, would be included as part of prompt 
emission (e.g. by comparing GRB 050724 and GRB 060614, Zhang et al.
2007a). Physically, it is more meaningful to define emission 
components as of ``internal'' (central engine) or ``external'' 
(medium) origins. In such a scheme, the central engine related emission 
likely extends to much later epochs and can no longer be defined as ``prompt''.
In the following discussion, I will still stick to the conventional
terminology, but will discuss the distinct physical meanings of various 
emission components.

\subsection{A canonical X-ray afterglow lightcurve}

\begin{figure}[t]
   \begin{center}
   \centerline{\psfig{figure=noflare-lc.ps,angle=-90,width=7cm}
\psfig{figure=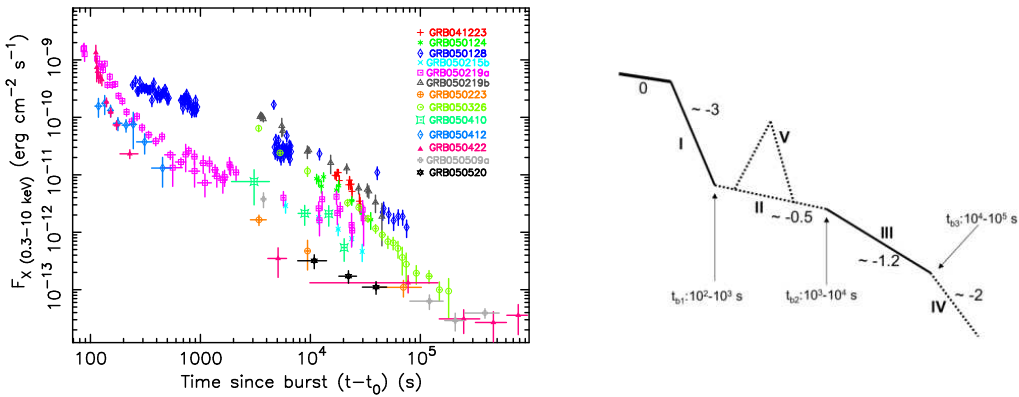,angle=-90,width=7cm}}
   \caption{A canonical X-ray afterglow lightcurve revealed by 
Swift XRT observations. {\em Left:} Data (from Nousek et al. 2006);
{\em Right:} A cartoon picture (from Zhang et al. 2006).}
   \label{XRT-lc}
   \end{center}
\end{figure}

One of the major discoveries of Swift is the identification of a
canonical X-ray afterglow behavior (Nousek et al. 2006; Zhang et al.
2006; O'Brien et al. 2006b; Chincarini et al. 2005, 
see Fig.\ref{XRT-lc}). Besides the prompt
emission phase (denoted as `0'), there are a total of five components in 
the X-ray lightcurves. Not every burst has all five components, so that
their lightcurves may vary from one another. In any case, their afterglow
lightcurve components could be generally fit into this generic picture.
The five components are:
\begin{itemize}
\item {\bf Steep decay phase (I):} Typically smoothly connected to the 
prompt emission (Tagliaferri et al. 2005; Barthelmy et al. 2005c), with a 
temporal decay slope $\sim -3$ or steeper (sometimes up to $\sim -10$, e.g.
Vaughan et al. 2006; Cusumano et al. 2006b; O'Brien et al. 2006b) extending
to $\sim (10^2-10^3)$s. Usually with a different spectral slope from 
the later afterglow phases\footnote{The steep decay component has been
observed in two BeppoSAX bursts: GRB 990510, Pian et al. 2001 and GRB 
010222, in't Zand et al. 2001. It was Swift that reveals that this is a 
common feature among bursts.}. 
\item {\bf Shallow decay phase (II):} Typically with a temporal decay slope
$\sim -0.5$ or flatter extending to $\sim (10^3-10^4)$s, at which a 
temporal break is observed before the normal decay phase (e.g. Campana et al.
2005; De Pasquale et al. 2006a). There is no spectral evolution across the
break.
\item {\bf Normal decay phase (III):} Usually with a decay slope $\sim -1.2$,
and usually follows the predictions of the standard afterglow model
(M\'esz\'aros \& Rees 1997a; Sari et al. 1998; Chevalier \& Li 2000).
A systematic test of the afterglow closure-relations (e.g. Table 1 of
Zhang \& M\'esz\'aros 2004) suggests, however, that a fraction of bursts
do not satisfy any afterglow model (Willingale et al. 2006).
\item {\bf Post Jet break phase (IV):} Occasionally observed following the 
normal decay phase, typically with a decay slope $\sim -2$, satisfying 
the predictions of the jet model.
\item {\bf X-ray flares (V):} Appear in nearly half of GRB afterglows. Sometimes
multiple flares appear in one GRB. Typically have very steep rising and
decaying slopes (Burrows et al. 2005b; Falcone et al. 2006; Romano et al.
2006a) with $\delta t/t \ll 1$. Appear in both long-duration 
(Falcone et al. 2006) and short-duration GRBs (Barthelmy et al. 2005b;
Campana et al. 2006b), and both GRBs and XRFs (Romano et al. 2006a).
\end{itemize}
Except for the normal decay and the jet-break phases, all the other 
three components were not straightforwardly expected in the pre-Swift
era\footnote{The flare-like signature was seen by Beppo-SAX, but it
was interpreted as the onset of the afterglow (Piro et al. 2005). More
detailed theoretical calculations (Lazzati \& Begelman 2006; Kobayashi
\& Zhang 2007) suggest that the onset of afterglow cannot produce a
sharp lightcurve feature to interpret X-ray flares.}.
As of the time of writing, the steep decay phase and X-ray flares
are better understood, while the shallow decay phase is still
a mystery. 

\subsubsection{Steep decay phase: tail of the prompt emission}
 \label{sec:steep}

The generally accepted interpretation of the steep decay phase is the
tail emission due to the so-called ``curvature effect'' (Fenimore et
al. 1996; Kumar \& Panaitescu 2000; Dermer 2004;
Zhang et al. 2006; Panaitescu et al. 2006a; Dyks et al. 2005,
for discussion of the curvature effect in the prompt gamma-ray
phase, see e.g. Kocevski et al. 2003; Shen et al. 2005;
Qin \& Lu 2005; Qin et al. 2006a). The basic assumption of this
interpretation is that the GRB emission region is disconnected from
the afterglow region (the external shock), and that the emission from
the GRB emission region ceases abruptly. This is consistent with the
conjecture of internal shocks or other internal dissipation mechanisms
(e.g. photosphere dissipation, magnetic field reconnection, etc). 
Since it is generally
assumed that the ejecta has a conical geometry, the curvature of the
radiation front causes a propagation delay for high-latitude emission
from the line of sight. Combining with the variation of the Doppler
factor at different latitudes, one gets a simple prediction 
$\alpha=2+\beta$ for the emission outside the $\Gamma^{-1}$ emission
cone, where the convention $F_\nu \propto t^{-\alpha} \nu^{-\beta}$
is adopted. The salient feature of this interpretation is that 
it could be directly tested since both $\alpha$ and $\beta$ could be
measured directly from the observational data, given that two
complications are treated properly (Zhang et al. 2006): First, 
for internal emissions, every time when the central engine restarts,
the clock should be re-set to zero\footnote{We notice that for 
external shock related emissions, taking the GRB trigger time as
the time zero point is generally required (Lazzati \& Begelman
2006; Kobayashi \& Zhang 2007).}. In a $\log - \log$ lightcurve,
this usually introduces an ``artificial'' very steep decay if the 
GRB trigger time (which is usually taken as $t=0$) significantly 
leads the time zero point ($t_0$) of the corresponding
emission episode. Second, the observed decay is the superposition 
of the curvature effect decay and the underlying afterglow decay 
from the external shock. One needs to
subtract the underlying afterglow contribution before performing the
test. The credibility of the curvature effect interpretation 
is that by properly taking into account the two
effects mentioned above, the steep decay is consistent
with $\alpha=2+\beta$ with $t_0$ shifted to the beginning of the
last pulse of prompt emission (Liang et al. 2006a) at least in some
cases.

Besides the standard curvature effect model, other 
interpretations for the steep decay phase 
have been discussed in the literature. 
\begin{itemize}
\item In some cases, the steep-decay slope may be shallower than 
the expectation of the curvature effect\footnote{It is worth noticing
that generally a decay slope steeper than the curvature effect prediction
is not allowed, unless the jet is very narrow. Usually, even if
the intrinsic temporal decay slope is steeper than $2+\beta$,
the curvature effect nonetheless takes over to define the decay slope.}.
This would suggest that the emission in the shock region may not cease 
abruptly, but rather decay (cool) with time gradually, leading to a 
decaying internal shock afterglow (Fan \& Wei 2005; Zhang et al. 2006). 
\item Yamazaki et al. (2006) study the curvature effect of an
inhomogeneous fireball (mini-jets). They found that the decay tail
is generally smooth, but sometimes could have structures, which may
interpret the small-scale structure in some of the decay tails.
\item Pe'er et al. (2006b) suggest that the emission from the
relativistically expanding hot plasma ``cocoon'' associated with 
the GRB jet could also give rise to the steep decay phase
observed by Swift.
\end{itemize}

Motivated by the discovery of the spectrally evolving tails
in GRB 050724 (Campana et al. 2006a) and GRB 060614 (Gehrels et al.
2006; Zhang et al. 2007a; Mangano et al. 2007b), recently Zhang et al. 
(2007c) performed a systematic time-dependent spectral analysis of 17 
bright steep decay tails. They found that while 7 tails show no
apparent spectral evolution, the other 10 do. A simple curvature
effect model invoking an angle-dependent spectral index cannot 
interpret the data. This suggests that the curvature effect is not 
the sole factor to control the steep decay tail phase at least in 
some bursts. Zhang et al. (2007c) show that some of the spectrally 
evolving tails might be interpreted
as superposition of the curvature effect tail and an underlying 
``central engine afterglow'', which is soft but decays ``normally''.
Such a component has been seen in GRB 060218 (Campana et al. 2006a),
which cannot be interpreted by the standard external shock afterglow
model and may be from a decaying central engine (Fan et al. 2006). 
The strong spectral evolutions in GRB 050724,
GRB 060218, and GRB 060614, however, cannot be interpreted with such
a model. They may be interpreted as cooling of the internal-shocked
region (Zhang et al. (2007c). 

It is interesting to notice that in some bursts (e.g. GRB 050421,
Godet et al. 2006; GRB 050911, Page et al. 2006a) the X-ray afterglow 
is dominated by the steep decay component 
(with overlapping X-ray flares). Such naked GRBs may be 
surrounded by a very tenuous medium so that the external shock 
component is very faint.

\subsubsection{X-ray flares: restarting the central engine}
 \label{sec:flares}

\begin{figure}[t]
   \begin{center}
   \centerline{\psfig{figure=flare-lc.ps,angle=-90,width=7cm}
\psfig{figure=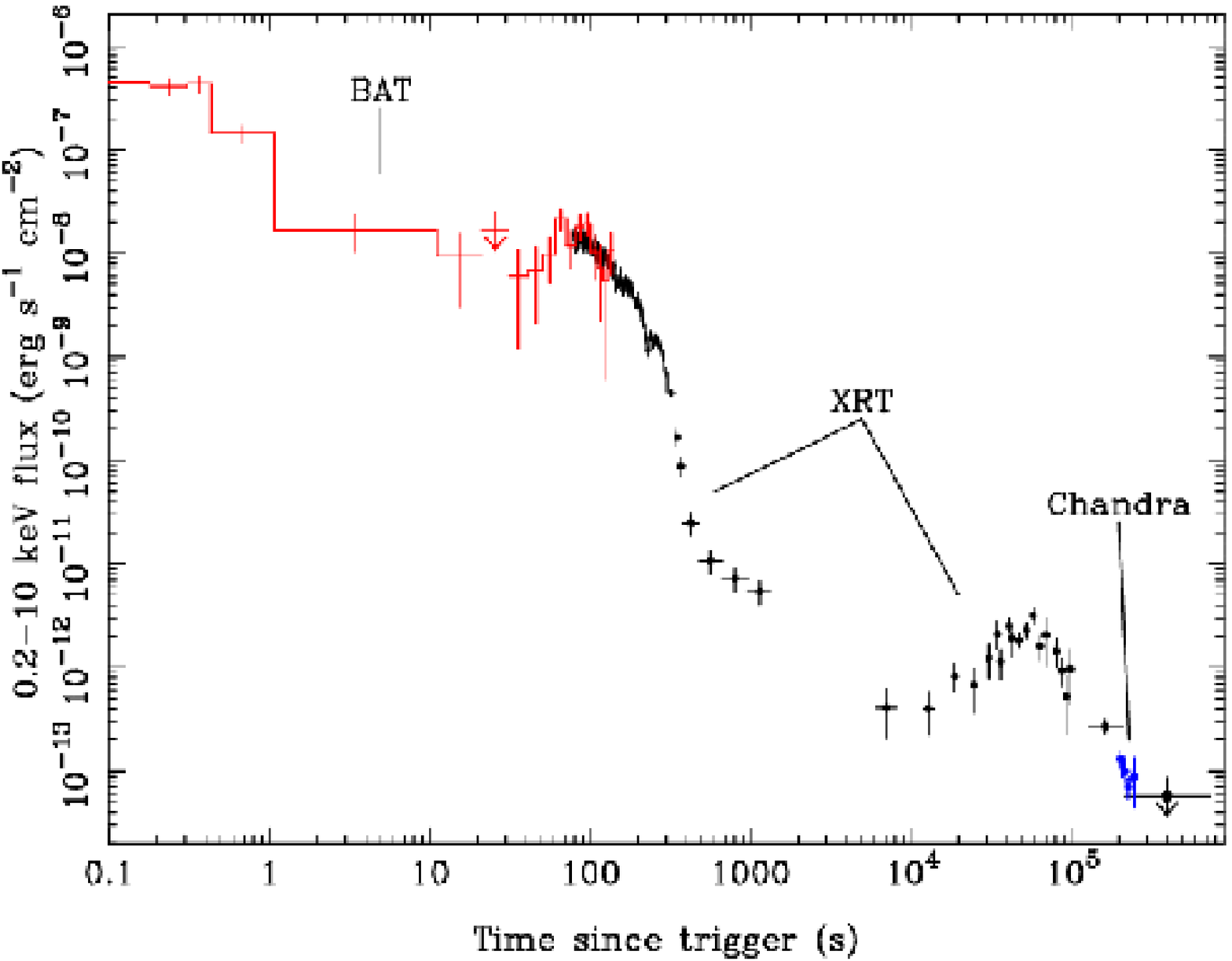,width=7cm}}
   \caption{Lightcurves show erratic X-ray flares. {\em Left:} Several 
long GRBs including the giant flare of 050502B (from Nousek et al. 2006);
{\em Right:} Flares following the short GRB 050724 
(from Barthelmy et al. 2005b).}
   \label{flares}
   \end{center}
\end{figure}

The X-ray flares have the following observational properties 
(Burrows et al. 2005b; Falcone et al. 2006; Romano et al. 2006a;
Chincarini et al. 2007, Burrows et al. 2007; see Fig.\ref{flares}): 
Rapid rise and fall times with $\delta t/t_{peak}
\ll 1$; many light curves have evidence for the same decaying afterglow
component before and after the flare; multiple flares are observed
in some bursts with similar properties; large flux increases at the
flares; typically degrading fluence of flares with time, but in rare
cases (e.g. GRB 050502B) the flare fluence could be comparable with 
that of the prompt emission; 
flares soften as they progress; and later flares are less
energetic and more broadened than early flares. These properties 
generally favor the interpretation that most of them are not associated 
with external-shock related events. Rather they are the manifestations
of internal dissipations at later times, which requires restarting
the GRB central engine (Burrows et al. 2005b; Zhang et al. 2006; 
Fan \& Wei 2005; Ioka et al. 2005; Wu et al. 2005a; Falcone et al.
2006; Romano et al. 2006a; Lazzati \& Perna 2007)\footnote{It had been
questioned whether well-separated gamma-ray pulses are due to restarting
of the central engine or inhomogeneity within the central engine
outflow in the pre-Swift era (e.g. Ramirez-Ruiz et al. 2001), but the
case was inconclusive.}. Compared with the 
external shock related models, the late 
internal dissipation models have the following two major advantages
(Zhang et al. 2006): First, since the clock needs to be re-set each 
time when the central engine restarts, it is very natural to explain 
the very sharp rising and falling lightcurves of the flares. Second,
energetically the late internal dissipation model is very economical.
While in the refreshed external shock models a large energy budget is
needed (the injection energy has to be at least comparable to that
already in the blastwave in order to have any significant injection
signature, Zhang \& M\'esz\'aros 2002a), the internal model
only demands a small fraction of the prompt emission energy to
account for the distinct flares. 

The leading candidate of the late internal dissipation model is the
late internal shock model. In such a model, the collisions could be
between the fast shells injected later and the slow shells injected
earlier during the prompt phase (e.g. Zou et al. 2006; Staff et al.
2006) or between two shells injected at later times (see Wu et al.
2005a for a categorization of different types of collisions). One
concern is whether later collisions between two slow shells injected
during the prompt phase could give rise to the observed X-ray flares.
This is generally not possible.  In order to produce late
internal shocks, the two slow shells must both have a low
enough Lorentz factor so that at the time of collision they do not
collide with the decelerating blastwave. Also in order not to
collide with each other earlier, their relative Lorentz factor
$\Delta \Gamma$ must be very small. When they collide,
the internal energy is usually too small to give rise to 
significant emission. Should such a collision occur, most 
likely it has no interesting observational effect (Lazzati \&
Perna 2007; Zhang 2007). Generally,
in the internal shock model the observed time sequence reflects the
time sequence in the central engine (Kobayashi et al. 1997). As a 
result, the observed X-ray flares $(10^2-10^5)$s after the prompt
emission must imply that the central engine restarts during this
time span, say, typically thousands of seconds but could be as late 
as days after the prompt emission is over.

The late internal dissipation model of X-ray flares is also tested
by Liang et al. (2006a). The same logic of testing the steep decay
component is used. The starting assumption is that the decay of 
X-ray flares are controlled by the curvature effect after the
abrupt cessation of the internal dissipation, so that $\alpha=
2+\beta$ is assumed to be valid. After subtracting the underlying 
forward shock afterglow contribution, Liang et al. (2006a) searched for 
the valid zero time points ($t_0$) for each flare to allow the decay
slope to satisfy the requirement of the curvature effect model.
If the hypothesis that flares are of internal origin is correct, 
$t_0$ should be generally before
the rising segment of each flare. The testing results are impressive: 
Most of the flares indeed have their $t_0$ at the beginning of the 
flares. This suggests that the internal dissipation model 
is robust for most of the flares. It is worth emphasizing that even 
the late slow bump at around 1 day following the short
GRB 050724 (Barthelmy et al. 2005b; Campana et al. 2006b) satisfies
the curvature effect model, suggesting that the central engine is 
still active even at 1 day after the trigger. 
This is also consistent with the
late Chandra observation of this burst (Grupe et al. 2006a) that
indicates that the afterglow resumes to the pre-flare decay slope 
after the flare.

Having identified the correct model for the flare phenomenology,
one is asked about a fundamental question: how to restart the
central engine. No central engine models in the pre-Swift era have
specifically predicted extended activities far after the prompt
emission phase. Prompted by the X-ray flare observations, the
following suggestions have been made recently, and none is proven
by robust numerical simulations yet at the moment.
\begin{itemize}
\item {\bf Fragmentation or gravitational instabilities in the massive
star envelopes.} King et al. (2005) argued that the collapse of
a rapidly rotating stellar core leads to fragmentation. The delayed
accretion of some fragmented debris after the major accretion 
event leads to X-ray flares following collapsar-related GRBs. 
\item {\bf Fragmentation or gravitational instabilities in the accretion
disk.} Observations of the short GRB 050724 (Barthelmy et al. 2005b; 
Campana et al. 2006b; Grupe et al. 2006a) reveal that it is also 
followed by several X-ray flares starting
from 10s of seconds all the way to $\sim 10^5$s. The properties of
these X-ray flares are similar to those in long GRBs. The 
requirement that both long and short GRBs should produce X-ray
flares with similar properties prompted Perna et al. (2006) to
suggest that fragmentation in the accretion disk, the common
ingredient in both long and short GRB models, may be the agent for
episodic accretion that powers the flares.
\item {\bf Magnetic barrier around the accretor.} Based on 
MHD numerical simulations in other contexts (e.g. Proga \&
Begelman 2003) and theoretical arguments, Proga \& Zhang (2006) 
argued that a magnetic barrier near the black hole may act as an 
effective modulator of the accretion flow. The accretion flow can be 
intermittent in nature due to the role of magnetic fields. This
model does not require the flow being chopped (e.g. due to 
fragmentation or gravitational instabilities) at larger radii before
accretion, although in reality both processes may occur altogether.
The magnetic barrier model is in accordance with the magnetic origin
of X-ray flares based on the energetics argument (Fan et al. 2005d).
\item {\bf NS-BH merger.} Flares in GRB 050724 (Barthelmy et al. 2005b)
pose a great challenge to the previous compact star merger models.
Numerical simulation of NS-NS mergers typically gives a short
central engine time scale (0.01-0.1)s, if the final product is a
BH-torus system (Aloy et al. 2005). In order to account for the late 
time flares in 050724, Barthelmy et al. (2005b) suggest a possible NS-BH 
merger progenitor system. Numerical simulations of BH-NS merger systems 
have been performed. Although X-ray flares at 100s of seconds 
or later still challenge the model, extended accretion over several
seconds could be produced (Faber et al. 2006; Shibata \& Uryu 2006;
cf. Rosswog 2005).
Lately Rosswog (2006) shows that if materials are launched into eccentric
orbits during a compact binary coalescence, the fallback of these materials
would last long enough to power X-ray flares
hours after the coalescence. 
\item {\bf NS-NS merger with a postmerger millisecond pulsar.} Dai et al.
(2006a) argued for a possible solution of the extended X-ray flares 
following NS-NS merger GRBs. Numerical simulations have shown that
the product of a NS-NS merger may not be a BH (Shibata et al.
2005), if the NS equation-of-state is stiff. Instead, the final
product may be a differentially-rotating massive neutron star.
If the initial magnetic fields of the NS is not strong, 
the $\alpha-\Omega$ dynamo action would induce magnetic explosions 
that give rise to late internal shocks to produce 
X-ray flares thousands seconds after the trigger
(Dai et al. 2006a). Earlier discussion on $\alpha-\Omega$
dynamo within the GRB context can be found in (Thompson \& Duncan 1993;
Kluzniak \& Ruderman 1998; Ruderman et al. 2000; Rosswog et al. 2003).
Price \& Rosswog (2006) suggest transient superstrong magnetic fields
during mergers through numerical simulations.
\item {\bf Multi-stage central engine.} Gao \& Fan (2006) and Staff et al.
(2006) proposed multi-stage central engine models to interpret 
X-ray flares. 
\item {\bf White dwarf - neutron star mergers.} A related model is the
WD-NS merger scenario revoked by King et al. (2007) as an effort to 
interpret ``long'' GRB 060614 without supernova associations. In view of
the close analogy between GRB 060614 and GRB 050724 (Zhang et al. 2007a),
this model can be relevant for X-ray flares following short GRBs.
\end{itemize}
Some other flare models have been discussed in the literature. The models
that can only interpret one flare (e.g. the synchrotron self-inverse 
Compton in reverse shock, Kobayashi et al. 2007, and the companion model,
MacFadyen et al. 2005) are found unattractive in view that multiple 
flares within a same burst seem to be common, and that the 
properties of the single flares are essentially the same as those of
multiple flares. The suggestion that flares result from collisions of
density clumps by the external shock (Dermer 2006) is so far not 
supported by numerical calculations. At least for clump angular sizes 
larger than $1/\Gamma$, numerical calculations show very smooth features
incompatible with the X-ray flare data (e.g. Zhang et al. 2006; Huang
et al. 2006, 2007; Nakar \& Granot 2006). For 
smaller clumps (Dermer 2006), fast variability is possible, but current
calculations fail to produce giant flares such as that in GRB 050502B
(Falcone et al. 2006). Giannios (2006) interprets multiple X-ray
flares as delayed magnetic dissipation in a decelerating Poynting-flux
dominated jet without introducing reviving the central engine. It is
however not clear how to interpret the clear $t_0$-resetting of flares
as discovered by Liang et al. (2006a).

\subsubsection{Shallow decay phase: still a mystery}

The shallow decay phase could follow the steep decay phase or 
immediately follow the prompt emission (O'Brien et al. 2006b;
Willingale et al. 2006). In most cases, a same shallow decay phase
is detected in the optical band as well (e.g. Mason et al. 2006).
This component is very likely related to the 
external shock. However, the very origin of this shallow decay 
phase is more difficult to identify,
since there exist several different possibilities that are not easy 
to differentiate among each other from the X-ray observations.  
Generally the spectral index does not change across the temporal
break from the shallow decay phase to the normal decay phase 
(but some slight change occurs in some bursts, 
see Willingale et al. 2006). This essentially rules out the
models that invoke crossing of a spectral break across the band.
The nature of the break should be then either hydrodynamical or 
geometrical.

The following models have been discussed in the literature.

\begin{itemize}
\item {\bf Energy injection invoking a long-term central engine.} The most
straightforward interpretation of the ``shallower-than-normal'' phase
is that the total energy in the external shock continuously increases
with time. This requires substantial energy injection into the 
fireball during the phase (Zhang et al. 2006; Nousek et al. 2006; 
Panaitescu et al. 2006a). There are two possible energy injection
schemes (Zhang et al. 2006; Nousek et al. 2006). The first one is to 
simply invoke a long-lasting central
engine, with a smoothly varying luminosity, e.g. $L\propto t^{-q}$
(e.g. Zhang \& M\'esz\'aros 2001a). In order to give interesting 
injection signature $q<1$ is required; otherwise the increase of the 
total energy in the blastwave is negligible. Such a 
possibility is valid for the central engines invoking a spinning-down
pulsar (Dai \& Lu 1998a,b; Zhang \& M\'esz\'aros 2001a)
or a long-lasting BH-torus system (MacFadyen et al. 2001). One
possible signature of this scenario that differentiates it from the
varying-$\Gamma$ model discussed below is a strong relativistic
reverse shock, if at the shock interacting region
the $\sigma$-parameter (the ratio between the Poynting flux and the
kinetic flux) is degraded to below unity (Dai 2004; Yu \& Dai 2006).
Alternatively, if $\sigma$ is still high at the shock region, 
the reverse shock may be
initially weak, but would still become relativistic if the engine
lasts long enough (i.e. this is effectively 
a rather thick shell, Zhang \& Kobayashi
2005). The observational data suggest a range of $q$ values with
a typical value $q\sim 0.5$. This is different from the requirement
of the analytical pulsar model ($q=0$). However, numerical calculations
suggest that a pulsar model can fit some of the XRT lightcurves
(Fan \& Xu 2006; De Pasquale et al. 2006b; Yu \& Dai 2006).
\item {\bf Energy injection from ejecta with a wide $\Gamma$-distribution.}
This model invokes a distribution of Lorentz factor of the ejecta 
with the low-$\Gamma$ ejecta lagging behind the high-$\Gamma$ ones,
which pile up onto the blastwave when the high-$\Gamma$ part is 
decelerated (Rees \& M\'esz\'aros 1998). In order to produce a 
smooth power law decay, the $\Gamma$-distribution needs to be close
to a power law with $M(>\Gamma) \propto \Gamma^{-s}$. A significant
energy injection requires $s>1$. The temporal break around $(10^3
-10^4)$ s suggests a cutoff of Lorentz factor around several tens,
below which $s$ becomes shallower than unity (Zhang et al. 2006).
Granot \& Kumar (2006) have used this property to constrain the
ejecta Lorentz factor distribution of GRBs within the
framework of this model.
The reverse shock of this scenario is typically non-relativistic
(Sari \& M\'esz\'aros 2000), since the relative Lorentz factor
between the injection shell and the blastwave is always low
when the former piles up onto the latter.
\item {\bf Delayed energy transfer to the forward shock.} Analytically,
the onset of afterglow is estimated to be around $t_{dec} =
{\rm max}(t_\gamma, T)$, where $t_\gamma \sim 5~{\rm s}
(E_{\rm K,52}/n)^{1/3} (\Gamma_0/300)^{-8/3}(1+z)$ is the time scale
at which the fireball collects $\Gamma^{-1}$ of the rest mass of
the initial fireball from the ISM, and $T$ is the duration of the 
explosion. The so-called ``thin'' and ``thick'' shell cases
correspond to $t_\gamma > T$ and $t_\gamma < T$, respectively
(Sari \& Piran 1995; Kobayashi et al. 1999). Numerical
calculations suggest that the time scale before entering the
Blandford-McKee self-similar deceleration phase is long, of
order several $10^3$ s (Kobayashi \& Zhang 2007). This suggests
that it takes time for the kinetic energy of the fireball to
be transferred to the medium. In a high-$\sigma$ fireball, 
there is no energy transfer during the propagation of a reverse
shock (Zhang \& Kobayashi 2005). Although
energy transfer could happen after the reverse shock disappears,
this potentially further delays the energy transfer process
Detailed numerical simulations are needed to verify this.
The shallow decay phase may simply reflect the slow energy
transfer process from the ejecta to the ambient medium. This model 
(e.g. Kobayashi \& Zhang 2007) predicts a significant curvature
of the lightcurves. This is consistent with some of the lightcurves
that show an early ``dip'' before the shallow decay phase. For those
cases with a straight shallow decay lightcurve, one needs to 
invoke superposition of the rising lightcurve and the steep decay 
tail to mimic the observations.
\item {\bf Off-beam jet model.} Geometrically one can invoke an
off-beam jet configuration to account for the shallow decay. Eichler
\& Granot (2006) show that if the line of sight is slightly outside
the edge of the jet that generates prominent afterglow emission,
a shallow decay phase can be mimicked by the combination of the
steep decay GRB tail. However, the expectation of this model is 
the correlations among the slow decay slope, hump luminosity and
its epoch, which are not confirmed observationally (Panaitescu 2006b).
Toma et al. (2006) discussed a similar model within the framework of 
the patchy jet models.
\item {\bf Two-component jet model.} A geometric model invoking two
jet components to produce two-component afterglows 
could also fit the shallow-decay data, since additional
free parameters are invoked (Granot et al. 2006; Jin et al. 2006;
Panaitescu 2006b).
\item {\bf Precursor model.} Ioka et al. (2006) suggest that if there 
is a weak precursor leading the main burst, a shallow decay phase 
can be produced as the main fireball sweeps the remnants of the
precursor.
\item {\bf Varying microphysics parameter model.} One could also invoke
evolution of the microphysics shock parameters to reproduce the
shallow decay phase (Ioka et al. 2006; Fan \& Piran 2006a;
Granot et al. 2006; Panaitescu et al. 2006b).
\item {\bf Dust scattering model.} Shao \& Dai (2006) suggest that 
small angle scattering of X-rays by dust could also give rise
to a shallow decay phase under certain conditions.
\item {\bf Cannonball model.} Dado et al. (2006) explains the 
canonical X-ray afterglow lightcurve within the framework of the
cannonball model, which invokes a series of different radiation
mechanisms to explain different segments of the lightcurves.
\item {\bf Central engine afterglow.} Finally, it remains possible that
the shallow decay phase is not related to the external shock, but is
due to a long-lived central engine activity. In such a case, the X-ray
emission and the optical emission may be two distinct components.
\end{itemize}

Can different possibilities be differentiated by more abundant data?
It seems to be a challenging task. I am inclined to the
first three interpretations on the above list. For the two energy 
injection models, one expects different reverse shock signatures 
(i.e. relativistic reverse shock for the long-term central engine 
model and non-relativistic reverse
shock for the varying-$\Gamma$ model). This would give different radio
emission properties at early times. On the other hand, the uncertainty
of the composition of the central engine outflow (e.g. the $\sigma$
parameter) would make the reverse shock signature of the former model
more obscured. The delayed energy transfer model (the third one on
the above list) is the simplest.
If it is correct, the so-called shallow decay phase is nothing but
a manifestation of the onset of afterglow (Kobayashi \& Zhang 2007).
The peak time can be then used to estimate the bulk Lorentz factor of
the fireball (which is $\sim 100$ for standard parameters).
This might be the case for at least some of the bursts.

\subsection{Optical, IR \& radio afterglows}

In the pre-Swift era, the afterglow observations were mainly carried 
out in the optical and radio bands. The late time optical/radio
observations have been focused on identifying temporal breaks in
the lightcurves, which are generally interpreted as the ``jet breaks''
(see Frail et al. 2001; Bloom et al. 2003; Ghirlanda et al. 2004b;
Dai et al. 2004; Friedman \& Bloom 2005; Liang \& Zhang 2005 for
compilations of the jet break data in the pre-Swift era).
Broad-band modeling was carried out for a handful of well observed
bursts (Wijers \& Galama 1999;
Panaitescu \& Kumar 2001, 2002; Yost et al. 2003), and the
data are generally consistent with the standard external shock
afterglow model. In some cases, very early optical flashes have been
discovered (e.g. GRB 990123, Akerlof et al. 1999; GRB 021004,
Fox et al. 2003a; GRB 021211, Fox et al. 2003b; Li et al. 2003a), 
which are generally interpreted as emission from the reverse shock
(Sari \& Piran 1999a; M\'esz\'aros \& Rees 1999; Kobayashi \& Sari
2000; Kobayashi 2000; Wang et al. 2000; Fan et al. 2002; 
Kobayashi \& Zhang 2003a; Zhang et al. 2003; Wei 2003; Kumar \& 
Panaitescu 2003; Panaitescu \& Kumar 2004; Nakar \& Piran 2004). 
Early radio flares have been 
detected in a sample of GRBs (Frail et al. 2003), which are also 
attributed to the reverse shock emission (Sari \& Piran 1999a;
Kobayashi \& Sari 2000; Soderberg \& Ramirez-Ruiz 2003). While
optical robotic telescopes such as ROTSE indeed reported non-detections 
of optical early afterglows of some bursts, the general 
expectation for Swift before the launch has been that the UVOT 
would collect a good sample of early afterglow lightcurves to 
allow a detailed study of GRB reverse shocks.

In the Swift era, UVOT has been regularly collecting optical photons
$\sim 100$s after the burst triggers for most GRBs. Ground-based
robotic telescopes (e.g. ROTSE-III, PAIRITEL, RAPTOR, P60, TAROT, 
Liverpool, Faulkes, KAIT, PROMPT, etc) 
have promptly observed most targets
whenever possible. A good list of early optical detections have
been made. However, the majority of bursts have very dim or
undetectable optical afterglows (Roming et al. 2006a).
This suggests that in most cases the reverse shock, if any, is
not significant.

\begin{figure}[t]
   \begin{center}
   \centerline{\psfig{figure=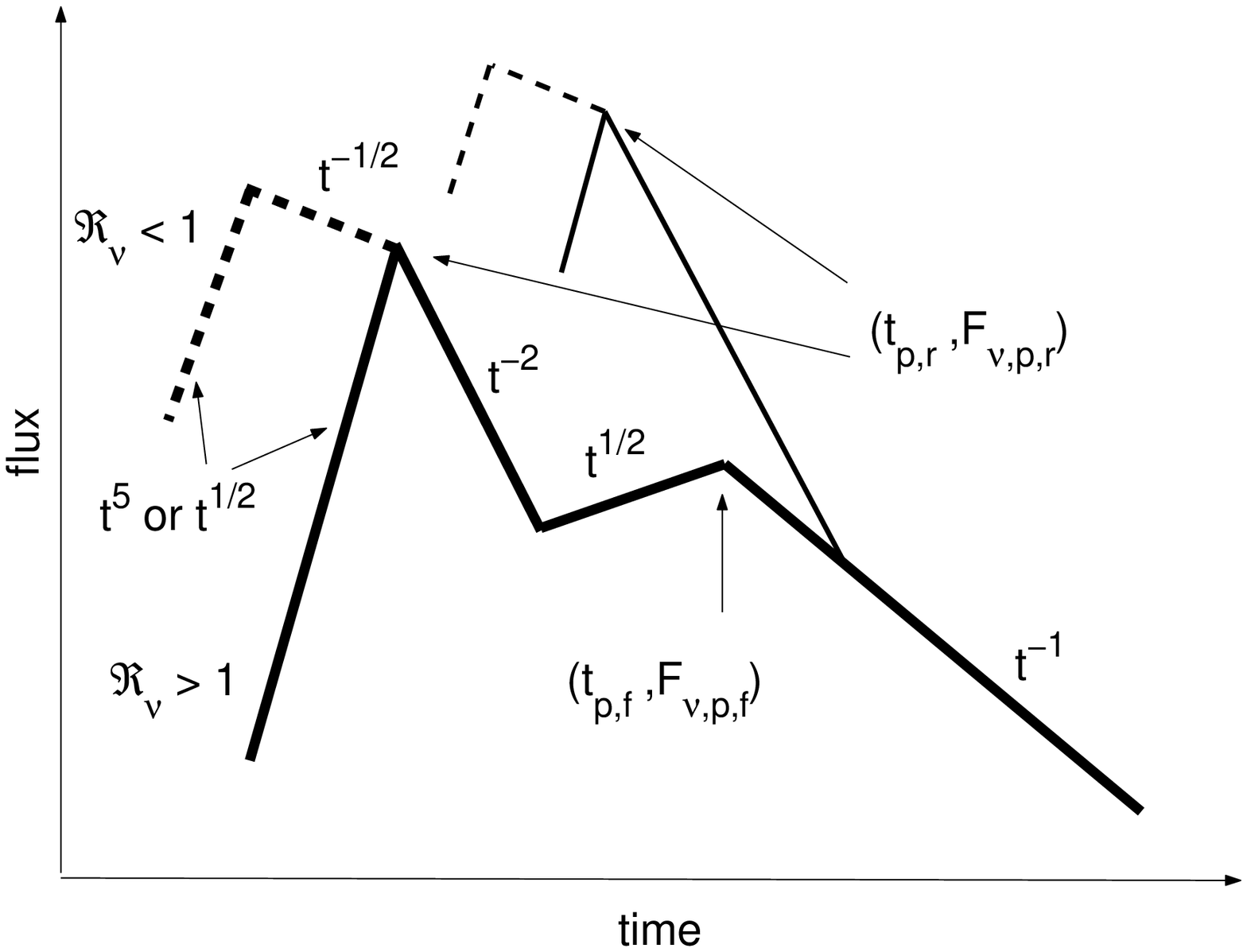,width=6cm}
    \psfig{figure=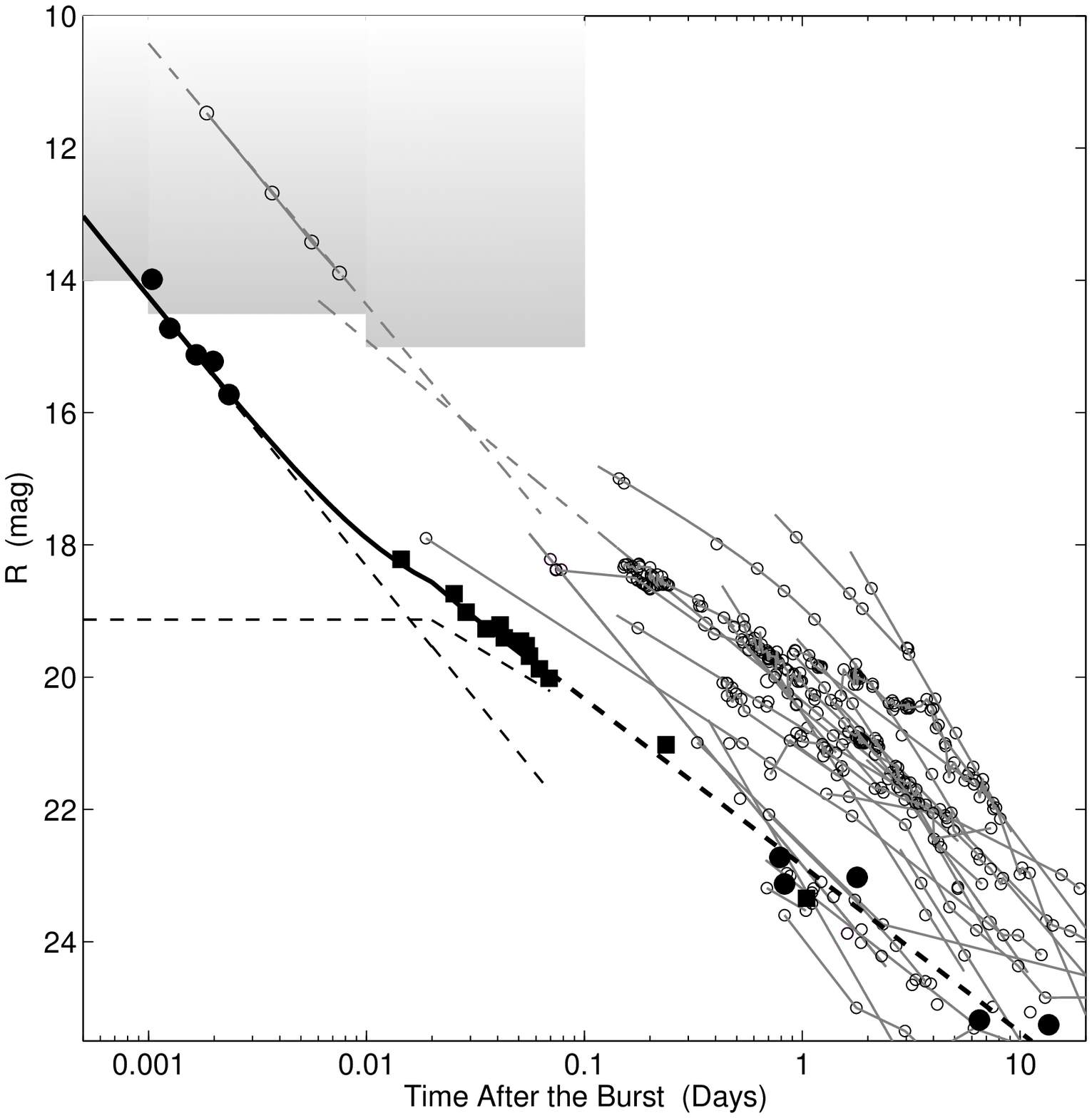,width=6cm}}
   \centerline{\psfig{figure=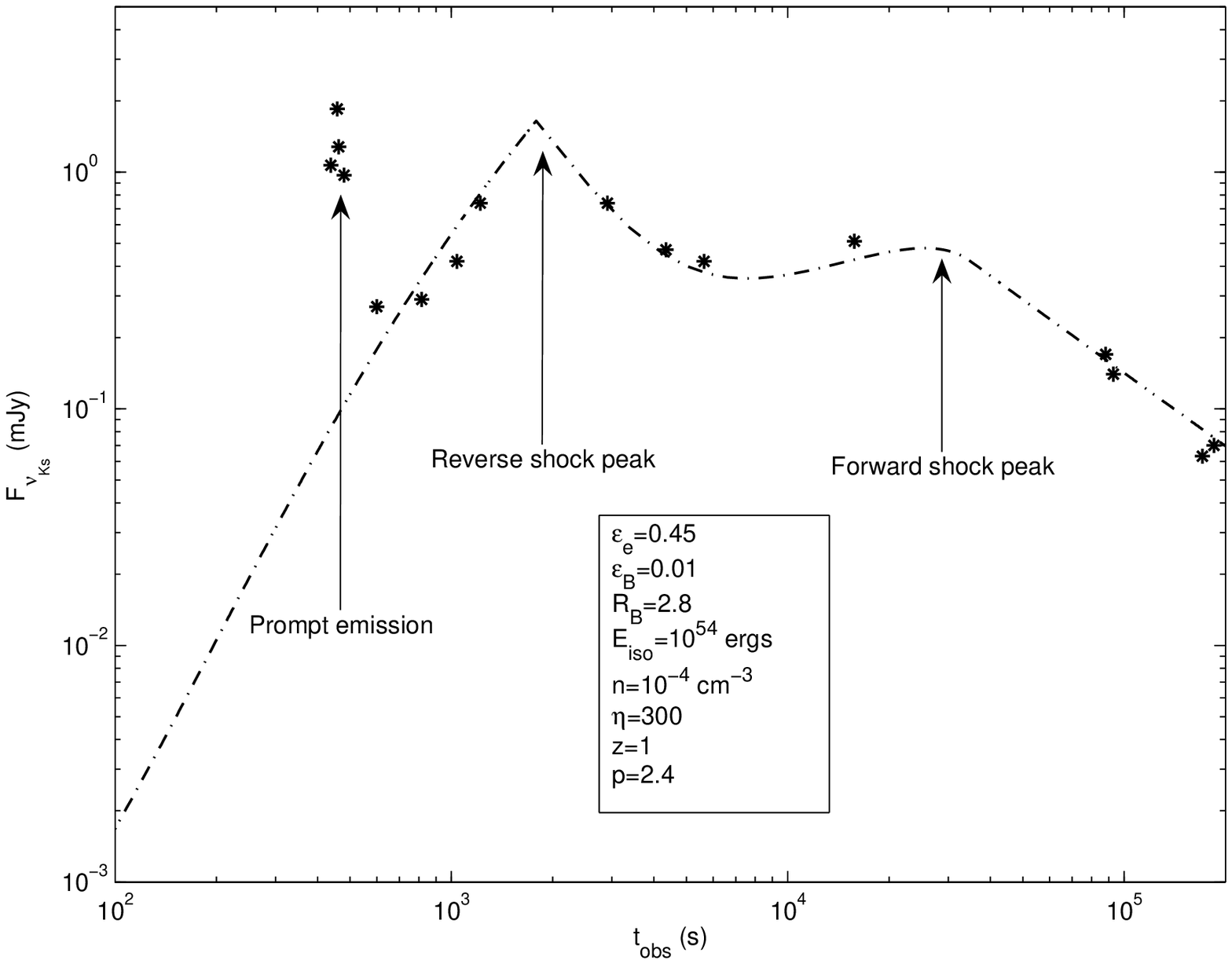,width=6cm}
    \psfig{figure=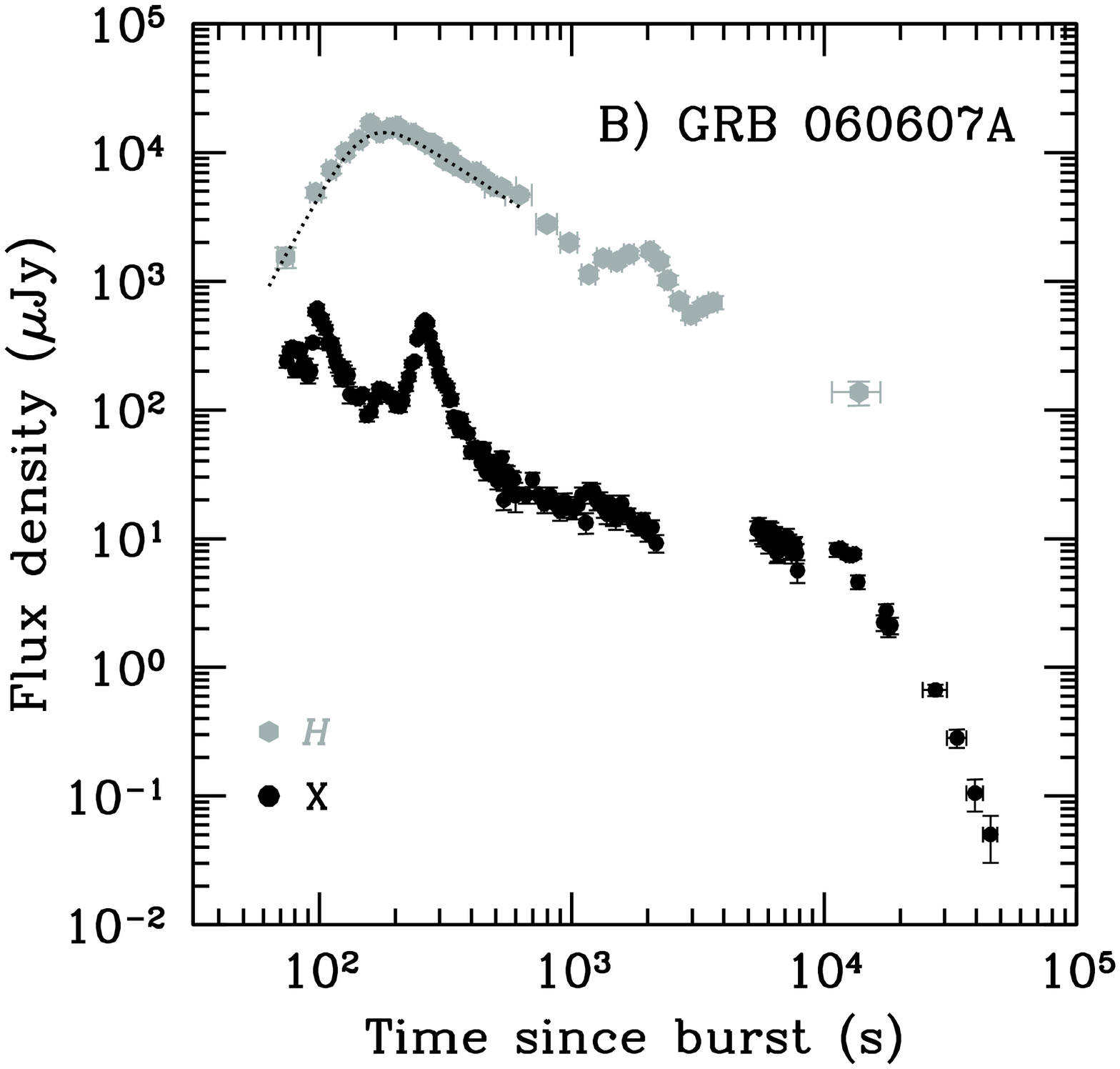,width=6cm}}
   \caption{Early optical afterglow lightcurves related to reverse and forward
shock emission. {\em (a) Top left:}
Theoretically expected early optical lightcurves, which show two types of behavior:
flattening and rebrightening (from Zhang et al. 2003); {\em (b) Top right:}
The flattening lightcurves detected from GRB 990123 and GRB 021211 (from
Fox et al. 2003b); {\em (c) Bottom left:} The rebrightening lightcurve
detected from GRB 041219A (data from Blake et al. 2005, model from Fan et al.
2005c); {\em (d) Bottom right:} The case that there is no evidence of reverse
shock in GRB 060607A (from Molinari et al. (2006)).
}
   \label{opt-lc}
   \end{center}
\end{figure}

Figure \ref{opt-lc}(a) displays the theoretically predicted early
optical afterglow lightcurves (Zhang et al. 2003) in the constant
medium density (ISM) model\footnote{For the model involving a
stratified stellar wind medium, see Chevalier \& Li (2000); 
Wu et al. (2003); Kobayashi \& Zhang (2003b); Kobayashi et al. 
(2004); Zou et al. 2005.}. 
The thick solid line shows two peaks: the first peak 
followed by $\sim t^{-2}$ decay is the reverse shock emission 
peak time, which is typically at the shock crossing
time ($t_{dec}$). The second peak followed by $\sim t^{-1}$
is the forward shock peak, which corresponds to the time 
when the typical synchrotron frequency $\nu_m$ crosses the
optical band. Depending on parameters, the forward shock peak
could be buried below the reverse shock component (the thin
solid line). One therefore has two cases of optical flashes:
rebrightening-type and flattening-type. 
A unified study of both reverse shock and forward shock emission
suggests that the rebrightening lightcurves should be generally 
expected, if the shock microphysics parameters ($\epsilon_e$, 
$\epsilon_B$, $p$, etc) are the same in both shocks. On the other hand, 
these microphysics parameters may not be the same in both shocks.
In particular, if the central engine is strongly magnetized,
as is expected in several progenitor models, the outflow likely
carries a primordial magnetic field, which is likely amplified
at the shocks. It is then possible to have $R_B = 
(\epsilon_{B,r}/\epsilon_{B,f})^{1/2} \gg 1$ in some cases.
This is actually the condition to realize the flattening-type 
lightcurves (Zhang et al. 2003). In order to interpret the bright 
optical flash and the subsequent flattening lightcurves in GRB 990123
and GRB 021211, one typically requires $R_B \sim 10$ or 
more (Fan et al. 2002; Zhang et al. 2003; Kumar \& Panaitescu
2003; Panaitescu \& Kumar 2004). Besides GRB 990123, the
flattening-type lightcurve was also detected for GRB 021211
(Fox et al. 2003b; Li et al. 2003, see Fig.\ref{opt-lc}(b)).

The $\epsilon_B$ parameterization is based on a purely hydrodynamical 
treatment of shocks
with magnetic fields put in by hand. Invoking a strong magnetic 
component in the reverse shock region raises the necessity to 
treat the dynamics more carefully with a dynamically important
magnetic field. Zhang \& Kobayashi (2005) studied the reverse
shock dynamics and emission for an outflow with an arbitrary
$\sigma$ parameter. They found that the most favorable case
for a bright optical flash (e.g. GRB 990123 and GRB 021211) is
$\sigma \sim 1$, i.e. the outflow contains roughly equal amount
of energy in magnetic fields and baryons. This is understandable:
For a smaller $\sigma$, the magnetic field in the reverse shock
region is smaller, and the synchrotron emission is weaker (see also
Fan et al. 2004a). For a larger $\sigma$, the magnetic field is 
dynamically important, whose pressure dominates the outflow region. 
The shock becomes weak or does not exist at all when $\sigma$
is large enough. 

The lack of bright optical flashes such as those observed in GRB 990123
and GRB 021211 is therefore not surprising. In order to have a bright
flattening-type flash, one needs to by chance have an outflow with $\sigma 
\sim 1$, while both larger and smaller $\sigma$'s would lead to not very
significant optical flashes. Even without additional suppression
effects, a non-relativistic shock with $\sigma=0$ would generally
give a reverse shock peak flux below the forward shock peak level
(Kobayashi 2000; Nakar \& Piran 2004; Zhang \& Kobayashi 2005). 
On the other extreme, a high-$\sigma$ flow would lead to very weak
reverse shock emission or no reverse shock at all (Zhang \& Kobayashi 
2005). Thus the tight early UVOT upper limits (Roming et al. 2006a) 
are not completely out of expectation.
Additional mechanisms to suppress optical flashes have been
discussed in the literature. Beloborodov (2005) argued that Compton
cooling of electrons by the prompt MeV photons may be a way to
suppress the optical flashes. Kobayashi et al. (2007) suggested that
a dominant synchrotron-self-Compton process in the reverse shock
region would suppress the synchrotron optical emission. Li et al. 
(2003b) and McMahon et al. (2006) suggested a pair-rich reverse
shock with weak optical emission.

Despite of the general disappointments, several bright optical
flashes have been detected in the Swift era, which could be generally
interpreted within the reverse/forward shock model discussed above.
The IR afterglow of GRB 041219A (Blake et al. 2005) is well modeled
by a rebrightening lightcurve (Fan et al. 2005c) (see Fig.\ref{opt-lc}(c)). 
Another flattening lightcurve was detected from GRB 060111B
(Klotz et al. 2006). Marginal reverse shock signatures may be
present in GRB 050525A (Blustin et al. 2006; Shao \& Dai 2005),
GRB 050904 (Gendre et al. 2006b; Wei et al. 2006),
GRB 060117 (Jelinek et al. 2006) and GRB 060108 (Oates et al. 2006).
Data also suggest a second type of optical flashes, which tracks the
gamma-ray lightcurves (for GRB 041219A, Vestrand et al. 2005).
These optical flashes are likely related to internal shocks
(M\'esz\'aros \& Rees 1999), probably neutron rich (Fan \& Wei
2004; Fan et al. 2005c, cf. Zheng et al. 2006). The time lags
between the prompt gamma-ray and optical emission have been revealed
in GRB 990123 and GRB 041219A (Tang \& Zhang 2006).
In some cases (e.g. GRB 050820A, Vestrand et al. 2006), the 
contributions from both the tracking component and the external shock 
component are detected from the early optical lightcurve.

There are however cases that clearly show no reverse shock component
at all in the early optical afterglows. GRB 061007 (Mundell et al. 2006;
Schady et al. 2006a) is such a case. Reaching a peak magnitude $<11$
(similar to 9th magnitude of GRB 990123), both the X-ray and optical
lightcurves show single power law decaying behavior from the very
beginning ($\sim 80$ s after the trigger). This suggests a strong 
external forward shock emission with enormous kinetic energy 
(Mundell et al. 2006) or a structured jet with very early jet 
break (Schady et al. 2006a). The reverse shock emission in this case 
is believed to peak at the radio band (Mundell et al. 2006).
Molinari et al. (2006) recorded the densely-covered early optical
afterglow lightcurves of GRB 060418 and GRB 060607A, which are both
characterized by a round-shaped single bump that could be interpreted
as the forward shock emission at the onset of afterglow. There is no
evidence of a reverse shock component at all (Fig.\ref{opt-lc}(d)). 
Among other possibilities,
this is consistent with a high-$\sigma$ flow where the reverse shock
is completely suppressed (Zhang \& Kobayashi 2005).

Wiggles and bumps have been observed in several pre-Swift GRB
optical afterglows (e.g. GRB 021004, Holland et al. 2003; GRB 030329, 
Lipkin et al. 2004). Models to interpret these variabilities usually
invoke external shock related processes, such as refreshed shocks,
density fluctuation, inhomogeneous jets, or multiple
component jets (Panaitescu et al. 1998; Zhang \& M\'esz\'aros 2002a;
Lazzati et al. 2002; Heyl \& Perna 2003; 
Nakar et al. 2003; Berger et al. 2003a; Granot et al. 2003; Ioka
et al. 2005). Early optical lightcurves may contain neutron
decay signatures (Beloborodov 2003; Fan et al. 2005b). Kobayashi
\& Zhang (2003a) interpreted the early fluctuations in GRB 021004
as a rebrightening lightcurve by combining both reverse and 
forward shock emission (see also Fan et al. 2005c for GRB 041219A).
Ioka et al. (2005) pointed out that some optical
fluctuations are difficult to interpret within any external shock
related schemes, and they require reactivation of the central 
engine. That erratic X-ray flares generally require 
late central engine activities raises the question whether
some optical flashes/flares are also due to the same origin (but 
softer and even less energetic, e.g. Zhang 2005). Recent optical
afterglow observations reveal that ``anomalous'' 
optical afterglows seem to be the common feature 
(Stanek et al. 2007; Roming et al. 2006c). Although some
of them could be accommodated within the external shock related
models, some optical flares do show similar properties to X-ray
flares (e.g. $\delta t/t < 1$, Roming et al. 2006c), which
demands late central engine activities. For example, the
optical fluctuations detected in the short GRB 060313 optical
afterglows (Roming et al. 2006b) may be better interpreted as 
due to late central engine activities than due to density 
fluctuations (e.g. Nakar \& Granot 2006).
Efforts to model optical flares using the late internal shock model
have been carried out recently (Wei et al. 2006; Wei 2007). The
results suggest that for plausible parameters, even the traditional
reverse shock optical flashes such as those in GRB 990123, GRB 041219A
and GRB 060111B could be interpreted within the late internal shock
model.

Due to a higher mean redshift of Swift bursts than that of pre-Swift
bursts (Berger et al. 2005b; Jakobsson et al. 2006a), 
the efficiency to detect radio afterglows 
is lower in the Swift era. According to GCN Circular statistics
(e.g. Greiner 2006), 17 radio afterglows were detected among about
200 GRBs detected by Swift in the first 2 years. Short-lived radio
transients have been seen in some of these bursts (e.g. Soderberg
et al. 2006c), some of which may be 
related to reverse shock emission (D. Frail, 2006, personal 
communication).

\subsection{Afterglow temporal breaks}

Temporal breaks (usually steepening breaks) have been commonly observed
in broad-band afterglow lightcurves. The origin of these temporal breaks 
in multi-wavelength afterglows is still not well understood. Theoretically 
one expects the following four types of temporal breaks. 
\begin{itemize}
\item {\bf Jet breaks.} This is expected if the GRB outflow is
collimated. The break occurs when the fireball is decelerated enough
so that the relativistic beaming angle $1/\Gamma$ ($\Gamma$ is the
bulk Lorentz factor of the fireball) becomes larger than the geometric
collimation angle. At this time the observer starts to feel the energy
deficit outside the jet cone. In the meantime, the jet starts to
expand sideways due to a horizontally propagating sound wave. 
Both effects tend to steepen the lightcurve, and when combined
together, result in a temporal break from a decay index $\sim -1$ to
$\sim -2$ (Rhoads 1999; Sari, Piran \& Halpern 1999). For a structured
jet, the argument still applies given that the jet opening angle is
replaced by the observer's viewing angle (Zhang \& M\'esz\'aros 2002b;
Rossi et al. 2002). Alternatively, cylindrical jets (Cheng et al. 2001)
have been discussed in the literature.
There are two distinct predictions for the jet
model. (1) Because it is a pure hydrodynamical effect, the break must
be {\em achromatic}, i.e. a temporal break should simultaneously
occur in all wavelengths, including X-ray, IR/optical and radio. (2) 
The hydrodynamical effect should not affect the microscopic shock
physics. The energy index of shock accelerated electrons should
remain the same across the break, so should the photon index. 
\item {\bf Injection breaks.} This is expected during the early phase of
the afterglow when the total energy in the blastwave is still
increasing with time. This could be due to either a long-lived GRB 
central engine (Dai \& Lu 1998a,b; Zhang \& M\'esz\'aros 2001) or
a wide distribution of the ejecta Lorentz factor 
(Rees \& M\'esz\'aros 1998; Sari \& M\'esz\'aros 2000; 
Granot \& Kumar 2006). The break happens upon the sudden
cessation of energy injection. Some ``jet breaks'' requiring $p<2$ may
be also modeled by an injection break (Panaitescu 2005a). The temporal
break separating the shallow decay component and the normal decay
component in the canonical XRT early lightcurves is typically
attributed to such an injection break (Zhang et al. 2006; Nousek et
al. 2006; Panaitescu et al. 2006a). Since the process is again
hydrodynamical, the break should be also achromatic. The spectral
index across the break should remain the same, although a change
of spectral index is not ruled out since the shock acceleration
process may be altered when the injection ceases suddenly. 
\item {\bf Spectral breaks.} This is the temporal break when a spectral
break crosses the band. The typical spectral breaks include the
characteristic synchrotron frequency $\nu_m$ and the cooling frequency
$\nu_c$ (Sari et al. 1998). Alternatively, the accelerated
electrons may have an intrinsic break in their energy spectrum
(Li \& Chevalier 2001; Wei \& Lu 2002a). The
crossing of the corresponding photon spectral break across the band
would give rise to a distinct temporal break. Two predictions
of this model are in distinct contrast to those of the previous two
models. (1) The break must be chromatic, typically rolling from high
energy bands (X-ray) to low energy bands (optical and radio) (but
see otherwise, e.g. $\nu_c$ increases with time for the wind afterglow
models, Dai \& Lu 1998c; Chevalier \& Li 1999; 2000); (2) the
spectral indices before and after the break should be distinctly
different (generally in a predictable way).
\item {\bf Transrelativistic breaks.} A steepening temporal break is 
expected at late times when an isotropic fireball turns from the 
highly relativistic phase to the non-relativistic phase (Wijers et 
al. 1997; Huang et al. 1998, 1999; Dai \& Lu 1999; Huang \& Cheng
2003). If the transition
happens after the jet break, the transrelativistic break would be a
flattening break (Livio \& Waxman 2000). Such a transition may have 
been observed in late radio afterglows (e.g. Frail et al. 2003),
but there is no robust evidence that this break shows up in optical
and X-ray lightcurves. The over 100-day follow-up observations of 
GRB 060729 (Grupe et al. 2006b) show a steady decay of the X-ray
afterglow flux, suggesting that the transrelativistic phase
happens at even later times at least for this burst. 
During the transition phase, the
counter-jet beaming to the opposite direction may be detected
through the observed excess radio emission (Li \& Song 2004).
\end{itemize}

Besides these breaks, more complicated lightcurves breaks may arise
due to collisions between a late shell and the decelerating blast 
wave (Panaitescu et al. 1998; Zhang \& M\'esz\'aros 2002a; Granot 
et al. 2003), collisions between the blastwave and a density jump or 
a wind termination shock (e.g. Dai \& Lu 2002b; Dai \& Wu 2003; 
Ramirez-Ruiz et al. 2005; Pe'er \& Wijers 2006), etc.

The jet break interpretation has been generally accepted in the
pre-Swift era\footnote{It has been doubted whether a jet model can 
indeed interpret the observed breaks, e.g. Wei \& Lu (2000b, 2002b).}. 
This model could alleviate the energy budget
problem encountered by some GRBs (e.g. GRB 990123, Kulkarni et
al. 1999), and it is enhanced by the empirical relation that the
geometrically corrected gamma-ray energy is quasi-standard (Frail et
al. 2001; Bloom et al. 2003). By identifying the temporal break times
as ``jet break times'' when  $\Gamma^{-1} = \theta_j$ is satisfied
(where $\Gamma$ is the bulk Lorentz factor and $\theta_j$ is the
jet opening angle, Rhoads 1999; Sari et al. 1999, see
Frail et al. 2001; Bloom et al. 2003; Ghirlanda et al. 2004b;
Friedman \& Bloom 2005; Liang \& Zhang 2005 for compilations of the
jet break data), one could infer the geometric configuration
and the total energy budget of some bursts. In particular, several 
empirical relations related to afterglow temporal breaks have been 
discussed in the literature.
\begin{itemize}
\item {\bf Frail relation:} Frail et al. (2001) and Bloom et al. (2003)
found that the beaming-corrected gamma-ray energy is essentially constant, 
i.e. $E_{\gamma,iso} 
\theta_j^2 = E_j \sim$ const. Since the standard jet model predicts
$t_j \propto E_{\gamma,iso}^{1/3} \theta_j^{8/3}$ (Sari et al. 1999),
this relation is generally consistent with $E_{\gamma,iso} \propto 
t_{j}^{-1}$.
\item {\bf Ghirlanda relation:} Ghirlanda et al. (2004b) found that the
beaming-corrected gamma-ray energy is not constant, but is related to
the rest-frame spectral peak energy ($E_p$) through $E_p \propto
E_{\gamma,j}^{2/3}$. Again expressing $E_{\gamma,j}$ 
in terms of $E_{\gamma,iso}$
and $t_j$, this relation is effectively $E_p \propto E_{\gamma,iso}^{1/2}
t_j^{1/2}$. Notice that the Ghirlanda relation and the Frail relation
are incompatible with each other.
\item {\bf Liang-Zhang relation:} Liang \& Zhang (2005) took one step back.
They discard the jet model, and only pursue an empirical relation
among three observables, namely $E_p$, $E_{\gamma,iso}$ and the
{\em optical band break time} $t_b$. The relation gives $E_p \propto
E_{\gamma,iso}^{0.52} t_b^{0.64}$. It is evident that if $t_b$ is 
interpreted as the jet break time, the Liang-Zhang relation is rather
similar to the Ghirlanda relation. However, the former has the 
flexibility of invoking chromatic temporal breaks across different
bands. So violating the Ghirlanda relation in other wavelengths
(e.g. in the X-ray band, Sato et al. 2007) 
does not necessarily disfavor the Liang-Zhang relation.
\item {\bf Willingale relation:} Recently Willingale et al. (2006) 
performed a systematic study of the shallow-to-normal decay transition
breaks in the early X-ray afterglows of a sample of Swift GRBs. By
{\em assuming} they are jet breaks (the results actually suggest that
they are not jet breaks), they found a new sequence of
correlation which is parallel to the Ghirlanda relation. This is 
effectively a new series of $E_p-E_{\rm \gamma,iso}-t_b$ relation
as discussed by Liang-Zhang, but by replacing the optical breaks by
X-ray breaks. The fact that the two correlations form a parallel
sequence is intriguing.
\end{itemize}

The growing trend in the Swift era is that some breaks we see in the
broad-band afterglows may not be jet breaks, and that the very origin
of these breaks is still a mystery. This also raises the concern 
whether the pre-Swift ``jet breaks'' are indeed jet breaks. In fact,
the ``smoking-gun'' feature of the jet breaks, i.e. the achromatic 
behavior, was not robustly established in any of the pre-Swift bursts. 
The best case was GRB 990510 (Harrison et al. 1999), in which clear 
multi-color optical breaks were discovered, which are consistent with 
being achromatic. The radio data are also consistent with having a 
break around the same time. However, based on radio data alone, one 
cannot robustly fit a break time that is consistent with the optical 
break time (D. Frail, 2006, private communication). Most of other 
previous jet breaks were claimed using one-band data only, mostly in 
optical, and sometimes in X-ray or radio. 

It has been highly expected that the multi-wavelength 
observatory Swift would clearly detect achromatic breaks 
in some GRBs to verify
the long-invoked GRB jet scenario. The results are however
discouraging. After detecting nearly 200 bursts, few ``textbook''
version jet breaks are detected. The lack of 
detections may be attributed partially to the intrinsic 
faintness of the Swift afterglows, and partially to the very low 
rate of late time optical follow-up observations. A higher average
redshift also pushes required observations of already faint objects
to even later observed times.
Achromatic breaks were indeed observed in some bursts, but few
satisfy the salient features expected in the jet model. For example,
GRB 050801 (Rykoff et al. 2006) and GRB 060729 (Grupe et al. 2006b)
have an early achromatic break covering both the X-ray and optical
bands. However, the break is the transition from the shallow decay
phase to the normal decay phase, which is likely an injection break
rather than a jet break. GRB 050525A (Blustin et al. 2006) has
an achromatic break in X-ray and optical bands, which might be
interpreted as a jet break. However, the post-break temporal indices
in both X-ray and optical bands are too shallow to comply with 
the $\propto t^{-p}$ prediction. An interesting case for a claimed
achromatic jet break was GRB 060526 (X. Dai et al. 2007). However, 
the break indices before and after the break cannot be accommodated
within the simple jet model (cf. Panaitescu 2006b). Maybe the best
case is GRB 060614 (Mangano et al. 2007). An achromatic jet break
around 100 ks was seen by both the XRT in the X-ray band and by 
the VLT in the optical band. 
The post break temporal indices, although not identical,
are similar to each other. Detailed modeling by Panaitescu (2006b) 
suggest that GRB 060124 may be also added to the jet break list.

In most other cases, data seem not to support the existence of jet 
breaks. The data also cast doubts on some of the previous identified 
jet breaks. These pieces of evidence are collected in the following.
\begin{itemize}
\item Optical follow up of GRB 060206 reveals a clear temporal
break that would be regarded as a typical jet break should the X-ray
have not been collected (Monfardini et al. 2006). However, X-ray data 
show a remarkable single power law decay without any evidence of a 
break at the optical break time (Burrows 2006). Notice that Stanek 
et al. (2007) reported a contaminating X-ray source, which may make
the case less conclusive.
\item Many other X-ray afterglows also show remarkable single
power law decays extending to very late times (10 days or later,
Burrows 2006).
The lower limits of the beaming-corrected gamma-ray energy
of many bursts already greatly exceed the standard energy 
reservoir value suggested by Frail et al. (2001) and Bloom et 
al. (2003) (Burrows 2006).
\item Based on the Ghirlanda relation, Sato et al. (2007) have 
searched for expected jet breaks of three Swift bursts in the 
X-ray band with null results. The sample is expanded by Willingale
et al. (2006) to seven. This suggests that Ghirlanda
relation is not a common relation satisfied by most bursts.
This fact however does not disfavor the Liang-Zhang relation,
since an optical break may still exist at the expected time
if the breaks are chromatic. Late time optical observations
are needed to test whether the Liang-Zhang relation is 
generally valid/violated for most bursts.
\item Covino et al. (2006) summarized the search of achromatic breaks
of Swift afterglows using high-quality multi-wavelength data, and 
reported that no convincing case is identified.
\end{itemize}

It is worth mentioning that in several cases, the X-ray data
are consistent with (not robustly suggest) having a jet break. These 
include GRBs 050315, 050814, 050820A, 051221A and 060428A (see Burrows
2006 for a review, also Panaitescu 2006b). In particular, late Chandra 
ToO observations of the short GRB 051221A reveal a possible jet break
(Burrows et al. 2006). This, together with the achromatic jet break 
claimed for GRB 060614 (Mangano et al. 2007), suggest that at least 
some Type I GRBs are collimated. 

The shallow-to-normal transition break in early X-ray afterglow 
lightcurves has been generally interpreted as injection breaks
(see \S3.1.3 for more discussion). However, in some cases, clear
chromatic features have been revealed (e.g. Fan \& Piran 2006a;
Panaitescu et al. 2006b; Huang et al. 2007), which rejects the 
interpretation at least for those cases.

The data seem to suggest that there might exist other types 
of temporal breaks at least for some bursts 
that are not related to jet breaks and injection breaks. 
A very interesting feature of the afterglow breaks is that
the X-ray breaks systematically lead the optical breaks,
which in turn systematically lead the radio breaks. This fact,
along with the chromatic breaks in both X-rays (e.g. 
Panaitescu et al. 2006b) and optical (e.g. Monfardini et al.
2006), drives Zhang (2007) to speculate an ad hoc scenario
to interpret these temporal breaks as well as the Liang-Zhang
and Willingale ($E_{\gamma,iso}-E_p-t_b$) relations. 
In this scenario, the
spectral break in the prompt gamma-ray emission ($E_p$) and
the chromatic temporal breaks in the afterglow lightcurves
may be all related to the same electron energy distribution
break that rolls down from high energy to low energy. 
Initially the break is in the 
gamma-ray band, which defines the $E_p$ in the prompt emission
spectrum. Later this break moves to the X-ray band in $\sim 
(10^3-10^4)$ s, giving rise to the early injection-like breaks
in some bursts. The break keeps rolling down to the optical band
around a day, which can account for the pre-Swift optical
breaks that were interpreted as jet breaks. Later it moves
to the radio band in $\sim 10$ days. Such a scenario gives
a natural link between $E_p$ and the optical and X-ray break times 
$t_b$ in the Liang-Zhang and Willingale relations, which is 
otherwise difficult to explain. (A similar scenario has been adopted
by Wang et al. 2005 and Dai et al. 2005 to interpret the radio 
afterglow of the 2004, Dec.27 giant flare 
of SGR 1806-20.) The scenario has some
difficulties (see Zhang 2007 for more discussion). The most 
severe one is that one expects changes of the spectral index
across the breaks. In the X-ray band, this is not the case for
most bursts, but there are still some cases that might satisfy
the constraint (Willingale et al. 2006). No spectral change has been
established in some optical breaks as well (e.g. Panaitescu 2005b).
On the other hand, the scenario may be still valid for at least
some bursts, and it is testable with broad-band densely-covered afterglow
follow up observations. A hard test of this scenario is to find some 
bursts that have a break crossing through the X-ray, optical and 
radio bands in turn. Although no clear example is available in the 
Swift data sample, the previous GRB 030329 may satisfy the
requirement of this model. It has been claimed that there are
two ``jet breaks'' in this burst (Berger et al. 2003a): an early
optical break and a later radio break. These two breaks were
used to argue a two-component jet model for this burst.
Within the scenario proposed here, the two breaks are simply
the same break rolling over the optical and radio bands at
different times. In view of the sequential relation between the
Liang-Zhang and Willingale relations, a prediction of this 
scenario is that one would observe an ``injection-like'' break
in the X-ray band first (say, thousands of seconds), and then detect
a ``jet-like'' break in the optical band later (say, around a day),
and a radio break at even later times (say, around 10 days).
Whether or not such detections will be made would prove or falsify
this ad hoc scenario (Zhang 2007). In such a scenario, the temporal
breaks do not give us information about collimation and GRB
energetics.

\subsection{Panchromatic observations \& prompt emission models}

The panchromatic, prompt observations of GRBs in the Swift era greatly
advanced our understanding of GRB prompt emission. 

Most of GRBs show an early steep decay tail (Tagliaferri et al. 2005;
Goad et al. 2005; Barthelmy et al. 2005c). Interpreted as the 
curvature effect of high-latitude emission (see \S3.1.1 for discussion),
this component suggests that prompt emission and afterglow are
from distinct emission regions. This finally settles the 
internal vs. external shock debate of the prompt emission site (see
Zhang et al. 2006 for more discussion). A small fraction of bursts
do not show the steep decay phase (O'Brien et al. 2006b; Liang et al.
2006a; Willingale et al. 2006; Mundell et al. 2006; Schady et al. 
2006a), so that the prompt emission and the early afterglow are
smoothly connected together and the prompt emission might be of
external origin as well. This might be due to an early 
deceleration of the fireball, likely due to a very large initial
Lorentz factor and/or a dense medium. 

Within the internal scenarios of the prompt emission, it is still
unclear where the energy dissipation site (internal shocks, magnetic
reconnection region, or baryonic and pair photosphere) is and what the
radiation mechanism (synchrotron or jitter emission, inverse Compton or 
a combination of thermal and non-thermal emission components) would be. 
For discussion of internal prompt emission models, 
see e.g. M\'esz\'aros et al. (1994); Thompson (1994); Daigne \& 
Mochkovitch (1998); Pilla \& Loeb (1998); Medvedev \& Loeb (1999);
Lloyd \& Petrosian (2000); 
Ghisellini et al. (2000); Panaitescu \& M\'esz\'aros (2000); 
Medvedev (2000); M\'esz\'aros \& Rees (2000); Spruit et al. (2001);
Drenkhahn \& Spruit (2002); M\'esz\'aros et al. (2002); Zhang \& 
M\'esz\'aros (2002c); Dai \& Lu (2002a); Pe'er \& Waxman (2004a, 2005); 
Rees \& M\'esz\'aros (2005); Pe'er et al. (2005, 2006a); Ryde (2005);
Ryde et al. (2006); Thompson et al. (2006),
and Zhang \& M\'esz\'aros (2004) for a critical review.  Due to the
uncertainties inherited in the GRB jet composition and the degeneracy 
of models to interpret the limited prompt emission data, it has been
a difficult task to identify the correct scenario for GRB prompt 
emission.

\begin{figure}[t]
   \begin{center}
   \centerline{\psfig{figure=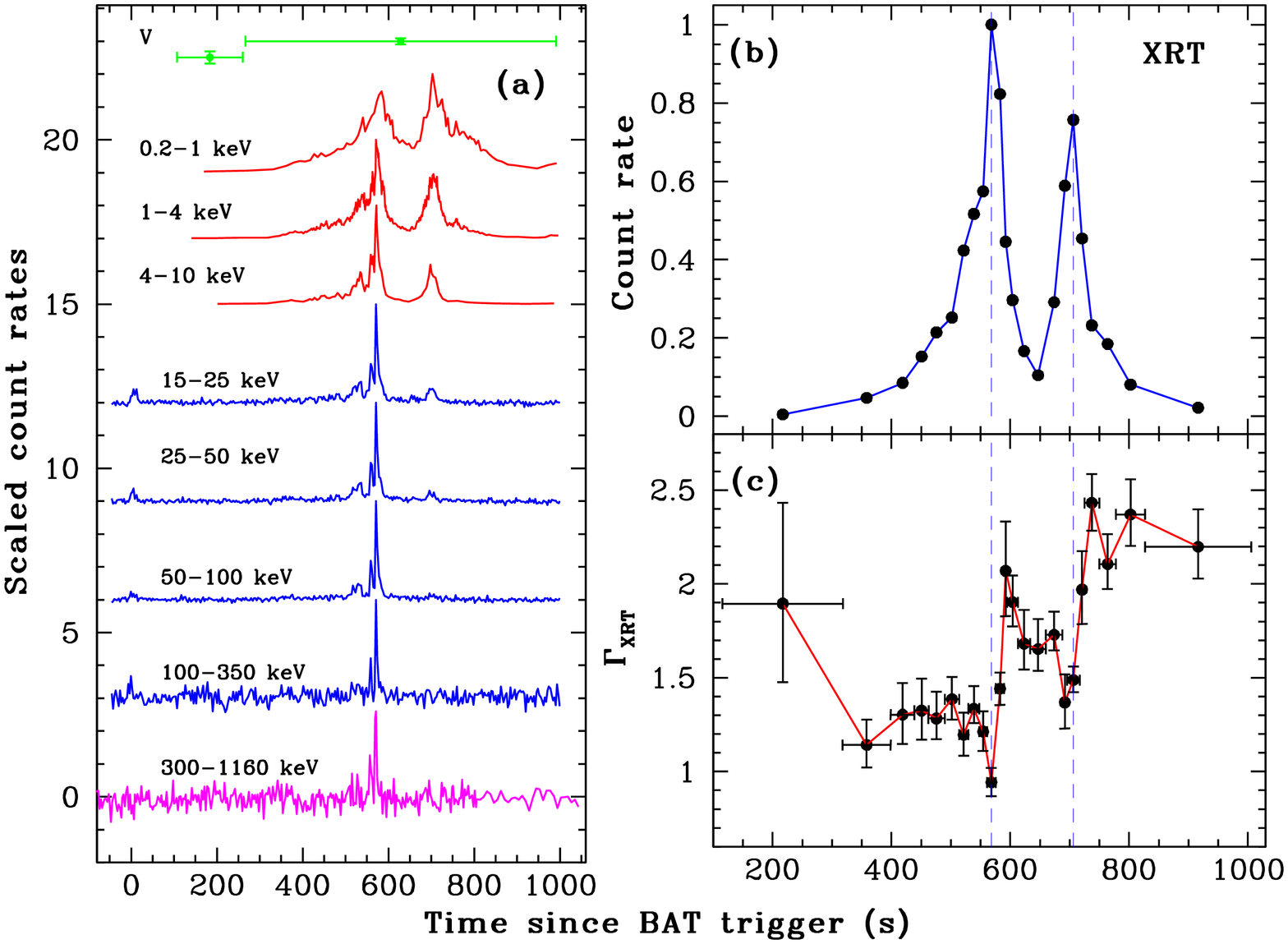,angle=-90,height=8cm}}
   \centerline{\psfig{figure=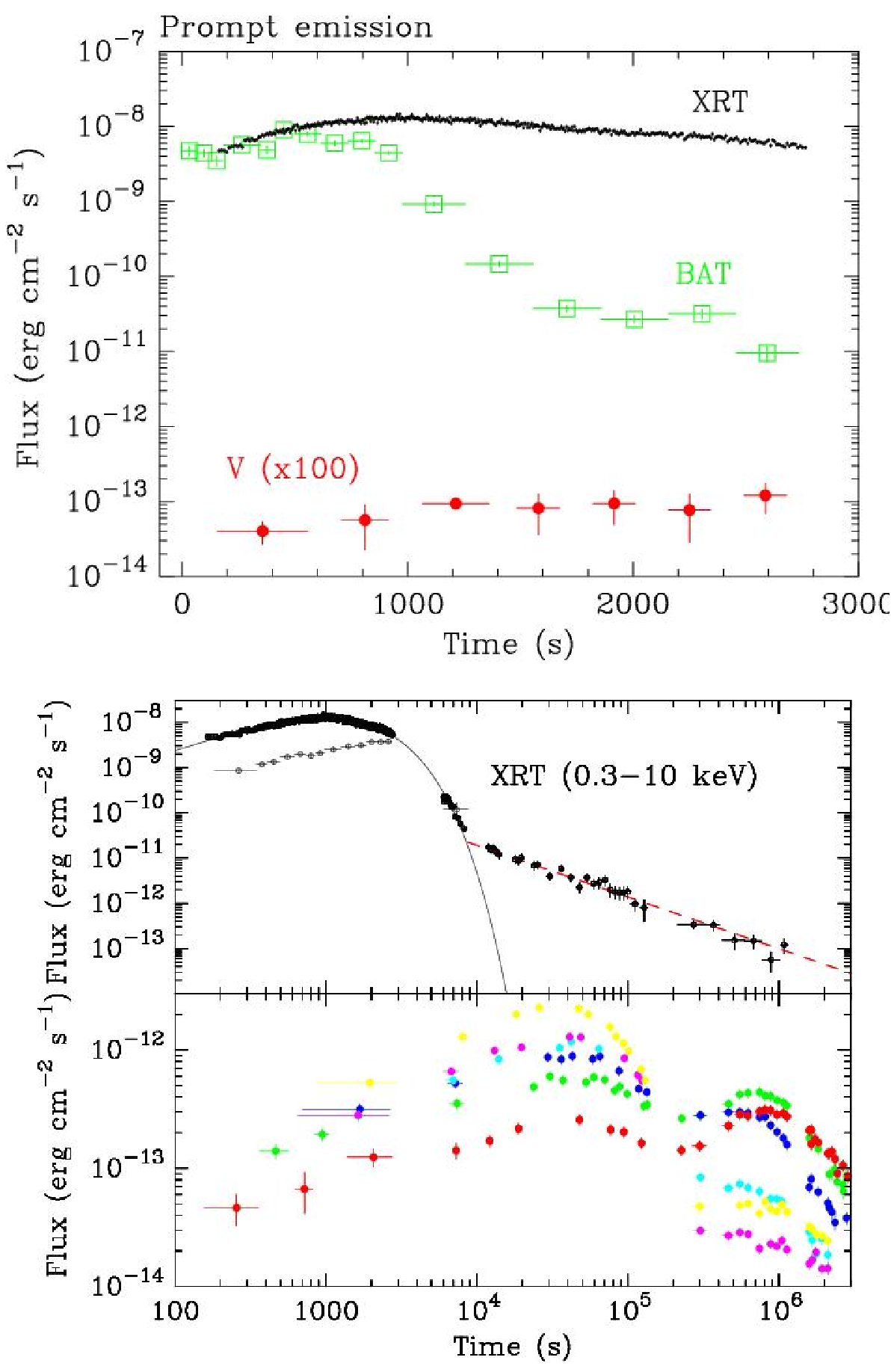,height=9cm}}
   \caption{Panchromatic observations the GRB prompt emission phase.  
{\em Upper:} GRB 060124 (Romano et al. 2006b); {\em Lower:} GRB 060218 
(Campana et al. 2006a).
}
   \label{prompt}
   \end{center}
\end{figure}

BAT is a narrow-band gamma-ray detector. For most of the Swift bursts, 
due to the narrow bandpass, it is difficult to precisely determine the 
prompt emission spectrum, especially the peak energy $E_p$. In most 
cases, the BAT spectrum can be only fitted by a simple power law 
(Zhang et al. 2007b). Nonetheless, by combining hardness ratio 
information (Cui et al. 2005), $E_p$ of the bursts could be estimated, 
which are generally consistent with those derived from joint 
BAT-Konus(WIND) fits (Zhang et al. 2007b). There is a rough correlation 
between the photon index $\Gamma$ and the derived $E_p$ (Zhang et al. 
2007a,b; Sakamoto et al. 2006b), which can be used to roughly estimate 
$E_p$. For most bursts, Swift prompt emission observations do not provide
more information than that gained in the BATSE era. Nonetheless, in rare 
cases Swift was triggered by a weak precursor (e.g. GRB 050117,
Hill et al. 2006; GRB 060124, Romano et al. 2006b; and GRB 061121,
Page et al. 2006b, see Burrows et al. 2007 for a review), so that all 
three instruments were 
targeted on the bursts during the prompt emission. This also happened
for GRB 060218 whose prompt emission was long enough (Campana et al. 
2006a). These panchromatic observations (Fig.\ref{prompt}) unveil 
unprecedented spectral and temporal information of GRB prompt emission.

Statistically, the following empirical relations related to
GRB prompt emission properties have been discovered in the pre-Swift 
era. Most of them are found still valid in the Swift era.

\begin{itemize}
\item {\bf Luminosity - spectral lag (Norris) relation:} This relation
suggests that more luminous bursts have shorter spectral lags ($\tau$). 
For Type II (long-soft) GRBs, the relation reads $L_{\rm iso} \propto 
\tau^{1.2}$ (Norris et al. 2000; Schaefer et al. 2001). 
This relation was confirmed by Swift
bursts (Gehrels et al. 2006) including the peculiar long-soft 
GRB 060218 (Liang et al. 2006b). The interpretation of this relation
is non-trivial from the first-principle prompt emission models. 
If one however assumes a standard radiation unit in the comoving frame,
this relation may be simply related to a varying Doppler-boosting
parameter among bursts (Salmonson 2000; Ioka \& Nakamura 2001; Norris
2002).
\item {\bf Luminosity - variability (Fenimore-Reichart) relation:}
This relation suggests that more luminous bursts tend to have more
variable lightcurves (Fenimore \& Ramirez-Ruiz 2000; Reichart et al.
2001). The scatter of this relation is large, and the index is 
subject to debate (e.g. Guidorzi et al. 2005; Reichart 2005;
Guidorzi et al. 2006; Li \& Paczy\'nski 2006). The definition of 
variability is also 
instrument-dependent. The origin of this relation may have something
to do with the same kinetic effect to interpret the $L-\tau$ relation
(Ioka \& Nakamura 2001) or the screening effect of the pair photosphere 
(Kobayashi et al. 2002; M\'esz\'aros et al. 2002).
\item {\bf Amati and Yonetoku relations:} Amati et al. (2002) discovered 
a simple relation $E_p \propto E_{\gamma,iso}^{1/2}$ with bursts with 
known redshifts (cf. Nakar \& Piran 2005; Band \& Preece 2005). Apparent
outliers of the Amati relation include GRB 980425 and GRB 031203, but 
arguments (e.g. Ghisellini et al. 2006) suggest that they may not be 
intrinsic outliers should these events be detected by instruments like
Swift (i.e. with a wider spectral coverage to detect prompt X-ray 
emission). A similar correlation has been noticed from the BATSE 
sample without redshift information (Lloyd et al. 2000). A related
relation is $E_p \propto L_{p,iso}^{1/2}$ (Yonetoku et al. 2004, see
also Wei \& Gao 2003),
where $L_{p,iso}$ is the isotropic peak luminosity. From the 
first-principle physical models, $E_p$ could be derived as functions of
multiple unknown parameters, including the isotropic energy of the 
emitter and the unknown Lorentz factor (Table 1 of Zhang \& 
M\'esz\'aros 2002c). Thus any model may be adjusted to interpret the
Amati relation given an input $E_{\rm iso}-\Gamma$ relation. For 
example, in the internal shock model, the Amati relation could be
reproduced if $\Gamma$ is insensitive to $E_{\rm iso}$ (Zhang \&
M\'esz\'aros 2002c). For photosphere-dominated prompt emission models,
one needs a different $E_{iso}-\Gamma$ relation (or effectively
$\theta_j-\Gamma$ relation if a $\theta_j - \Gamma$ relation is
established) to interpret the Amati-relation, and such a correlation
was regarded as more natural (Rees \& M\'esz\'aros 2005; 
Thompson et al. 2006). A similar argument was raised within a 
wind-deceleration model (Thompson 2006). On the other hand, if one
assumes a standard emission unit in the comoving frame, 
the Amati-relation could be reproduced by the viewing angle effect 
for some types of jet configurations (e.g. Yamazaki et al. 2004a; 
Eichler \& Levinson 2004; Toma et al. 2005).
\item {\bf Firmani relation:} Firmani et al. (2006a) discovered a 
tight correlation with prompt emission data only, which reads
$L_{p,iso} \propto E_p^{3/2} T_{0,45}^{-1/2}$, where $T_{0,45}$
is the time of the enhanced burst emission. There has been no proposed
interpretation of this relation.
\item {\bf Frail, Ghirlanda, Liang-Zhang \& Willingale relations:}
For completeness, I repeat here the four empirical relations involving
afterglow temporal breaks discussed in \S3.3, but focusing on their
interpretations. The Frail relation suggests a standard energy 
reservoir, a hypothesis which is not confirmed by the Swift data.
The tight Ghirlanda and Liang-Zhang relations connect prompt emission
properties $E_p$ and $E_{\gamma,iso}$ with the afterglow properties
($t_j$ or $t_b$). It has been claimed that these relations
could be understood within the ad hoc annular jet model with the 
assumption of a standard comoving radiation unit (e.g. Levinson 
\& Eichler 2005; Eichler \& Levinson 2006). The photosphere model 
(Thompson et al. 2006) may give a more physical interpretation. 
On the other hand, Swift XRT data now do not support the Ghirlanda 
relation (Sato et al. 2007; Willingale et al. 2006), which renders 
efforts to interpret the relation invoking a jet break not very 
meaningful. If one discards the jet framework, the rolling electron 
spectral break hypothesis (Zhang 2007, see \S3.3 for more discussion) 
may be a possibility to interpret the Liang-Zhang relation. The close 
relationship between the Liang-Zhang relation and the Willingale 
relation seems to lend support to this suggestion. More data are 
needed to test the prediction of this scenario (\S3.3).
\end{itemize}

Besides the above global properties, some Swift observations of GRBs
have shed new light on the prompt emission mechanisms. In the 
following I'll list various pieces of (sometimes controversial) 
inference about GRB prompt emission drawn in the Swift era. 

\begin{itemize}
\item Ryde (2005) argues that the so-called Band-function of GRB spectrum
can be actually decomposed into the combination of a thermal and a
non-thermal component. This model was further enriched by more
physical models involving photosphere thermal emission and Compton
dissipation above it (Ryde et al. 2006; Thompson et al. 2006). Time
dependent modeling with latest Swift data is on-going (Ryde et
al. 2007).
\item Using some general observational constraints (but assuming
shock dissipation and synchrotron/IC radiation mechanisms),
Kumar et al. (2006) constrained the emission properties
of two bursts (GRBs 050126 and 050219A) that display smooth and
single-peaked gamma-ray lightcurves. The conclusion is that the
radiation site is close to the deceleration radius (contrary to
the closer-in photosphere radius derived from other arguments
as discussed above), and that the radiation mechanism is synchrotron
self-Compton. Kumar et al. (2007) used a larger sample to reach
a similar conclusion about the large emission radius, and pointed
out that neither internal shocks nor external shocks seem to 
interpret the data satisfactorily. 
A large emission radius is also independently
estimated by Lyutikov (2006a) using the duration of the steep-decay 
GRB tails within the framework of curvature effect interpretation
and by assuming a standard jet opening angle.

\item A traditional problem of the synchrotron emission model of
GRB prompt emission is the ``fast-cooling'' problem (Ghisellini
et al. 2000). For standard parameters, the cooling frequency is
much lower than the 100 keV range, so that the predicted low-energy
photon index is -3/2, steeper than that of most bursts (but is
satisfied in some bursts, e.g. GRB 060124, Romano et al. 2006b).
Using some general observational constraints, Pe'er \& Zhang (2006)
derived the parameter sets of the internal shock synchrotron
radiation model. They argued that the data could be reproduced if
one assumes that the post-shock magnetic fields decay in a 
length scale much shorter than the comoving width, about $10^5
-10^6$ skin depths. By introducing a synchrotron self-absorption
break, the model can interpret the broad-band data of GRB 050820A.
The suggestion may lead to a slow-cooling synchrotron model for 
prompt emission, which is consistent with the broad-band data of
GRB 061121 (Page et al. 2006b). A similar suggestion has been
proposed for afterglows (Rossi \& Rees 2003). The hypothesis is
probably consistent with the ongoing numerical simulations of 
relativistic collisionless shocks (J. Arons 2006, personal
communication). 
\item The broad-band data of super-long GRB 060218 (Campana et al. 
2006a) during the prompt emission phase allow detailed modeling
of GRB prompt emission for the first time. However, so far no
model can successfully interpret the whole data set. The faint
early UVOT observation severely constrains that the emission
mechanism is likely not synchrotron (the extrapolation of the
observed emission according to the synchrotron model 
predicts a much higher flux, Dai et al. 2006b). Ghisellini et al
(2007) invokes synchrotron self-absorption to accommodate the 
optical flux deficit. The presence of
a thermal X-ray component (probably due to shock breakout,
Campana et al. 2006a, but see Li 2007) provides an extra source
for inverse Compton emission (e.g. Dai et al. 2006b). A bulk
Compton scattering model has been also proposed (Wang et al. 2006a), 
which suggests that the radiation mechanism of this (and probably 
also other) LL-GRBs may be different from that of canonical GRBs. 
On the other hand, the compliance of both Amati and Norris relations
of this burst (Amati et al. 2006; Liang et al. 2006b) seem to
suggest that its radiation physics should not be distinctly
different from that of canonical GRBs. The hitherto most detailed 
prompt emission data of GRB 060218 seem to defy interpretation
and to greatly challenge
our basic understanding about the GRB radiation mechanism.
\end{itemize}

\subsection{Radiative efficiency}

One interesting question is the GRB radiative efficiency, which
is defined as $\eta=E_\gamma/(E_\gamma+E_K)$, where $E_\gamma$ and
$E_K$ are the isotropic gamma-ray energy and kinetic energy of the
afterglow, respectively. The reason why $\eta$ is important to
understand the explosion mechanism is that it is related to the 
energy dissipation mechanism of the prompt emission, which is not
identified. The standard picture is internal shock dissipation,
which typically predicts several percent radiative efficiency
(Kumar 1999; Panaitescu et al. 1999, cf. Beloborodov 2000,
Kobayashi \& Sari 2001). Other mechanisms (e.g. magnetic dissipation)
may have higher efficiencies although a detailed prediction is not
available. It is of great interest to estimate $\eta$ from the data,
which can potentially shed light onto the unknown energy
dissipation process.

In order to estimate $\eta$, reliable measurements of both $E_\gamma$
and $E_K$ are needed. While $E_\gamma$ could be directly measured
from the gamma-ray fluence if the GRB redshift is known, 
measurement of $E_K$ is not trivial, as it requires detailed 
afterglow modeling. In the pre-Swift era, attempts to estimate
$E_K$ and $\eta$ using late time afterglow data have been made
(e.g. Panaitescu \& Kumar 2001, 2002; Freedman \& Waxman 2001;
Berger et al. 2003b; Lloyd-Ronning \& Zhang 2004). The jet sideways
expansion effect (Rhoads 1999; Sari et al. 1999; Huang et al. 2000)
may somewhat affect the estimates of the efficiency (Zhao \& Bai
2006). The presence of an early shallow decay phase in Swift XRT
afterglows suggest that $E_K$ likely increases with time. The 
$\eta$ values measured using the late time data are therefore no 
longer reliable. For a constant energy
fireball, ideally early afterglows may be used to study the radiative
loss of the fireball. However the shallow decay phase due to energy
injection smears the possible signature and makes such a diagnosis
difficult.

A systematic analysis of GRB radiative efficiencies using the first-hand 
Swift data has been carried out by Zhang et al. (2007b). Similar 
analyses using second-hand data for smaller samples of bursts were 
carried out by Fan \& Piran (2006a) and Granot et al. (2006). The 
conclusions emerging from these studies suggest that in most cases 
the efficiency is very high (e.g. $>90\%$) if $E_K$ right after the 
burst is adopted. However, using $E_K$ at a later time 
when the injection phase is
over one typically gets $\eta \sim$ several percent. The nature
of the shallow decay phase is therefore essential to understand
the efficiency. For example, if the shallow decay phase is due to
continuous energy injection, the GRB radiative efficiency must be
very high - causing problems to the internal shock model. If, however,
the shallow decay is simply due to the delay of energy transfer into the
forward shock (Kobayashi \& Zhang 2007), 
the GRB radiative efficiency is just the right one
expected from the internal shock model. The investigation of
Zhang et al. (2007b) also suggests that XRFs may not be 
intrinsically less efficient GRBs, in contrast to the pre-Swift 
expectation (Soderberg et al. 2004; Lloyd-Ronning \& Zhang 2004). 
Also as far as the radiative efficiency 
is concerned, there is no fundamental difference
between Type I and Type II GRBs (see also Bloom et al. 2006a and
Lee et al. 2005). This suggests that both types of GRBs share the 
same radiation physics.

\section{Cosmological setting}

GRBs are cosmological events. The close connection between Type II
GRBs with deaths of massive stars make GRBs potential tracers of 
star forming and probably metallicity history of the Universe.
In view that the history of the Universe during the so-called ``dark
age'' (from cosmic background radiation at $z \sim 1100$ to the epoch 
when first quasars were formed around $z \sim 7$) is still poorly known 
(Loeb \& Barkana 2001 for a review), GRBs, as bright beacons in the deep
Universe, would be the unique tool to illuminate the dark Universe and
allow us to unveil the re-ionization history of the Universe. There
are several reasons to believe that high-$z$ GRBs exist
and are detectable. First, due to a favorable $k$-correction factor and
the time-dilation effect, theoretically high-$z$ GRBs are not much
dimmer than their nearby sisters for both prompt gamma-ray emission
and afterglow in the infrared and radio wavelengths (Lamb \&
Reichart 2000; Ciardi \& Loeb 2000; Gou et al. 2004;
Ioka \& M\'esz\'aros 2005). In fact, the GRB
redshift record holder GRB 050904 (Cusumano et al. 2006a; Haislip et
al. 2006; Kawai et al. 2006; Frail et al. 2006) at $z=6.295$ has very 
bright prompt gamma-ray emission and early infrared and radio afterglows. 
Second, based
on several empirical standard candles (e.g. Fenimore \& Ramirez-Ruiz
2000; Norris et al. 2000; Amati et al. 2002) one could derive the
``pseudo''-redshifts of a large sample of GRBs. The results suggest that 
over $10\%$ of GRBs are at $z > 6$. This is also consistent with the 
theoretical prediction of the GRB rate assuming GRBs tracing the cosmic star 
formation history (Bromm \& Loeb 2002, 2006).  
Third, numerical simulations suggest that first generation stars
form at around $z \sim 20$ (Bromm et al. 1999; Abel et al. 2002),
which is generally also consistent with the conclusion drawn from the 
cosmic microwave background data collected by WMAP (Bennett et al. 2003
Spergel et al. 2006). Finally, a high fraction of high-$z$ bursts is
also inferred from redshift distribution of Swift bursts (Jakobsson et
al. 2006a). It is highly expected that
GRBs would break the current redshift record held by faint galaxies,
which would then bring unprecedented information about the reionization
history of the early Universe.
The first two year of Swift observations have detected at least four
bursts with $z>5$: GRB 050814 at $z=5.3$ (Jakobsson et al. 2006b),
GRB 050904 at $z=6.29$ (Cusumano et al. 2006a; Haislip et al. 2006; 
Kawai et al. 2006), GRB 060522 at $z=5.11$ (Cenko et al. 2006), 
and GRB 060927 at $z=5.6$ (Fynbo et al. 2006b). 
The lower rate than predicted is very likely due
to the challenge of promptly performing IR spectroscopic observations
of the high-$z$ bursts. 
The low UVOT detection rate of Swift GRBs (Roming et al.
2006a) could be partially due to a good fraction of high-$z$ GRBs.
In fact, based on prompt emission data, some high-$z$ GRB candidates
have been suggested (e.g. GRB 050717, Krimm et al. 2006b).

The study of high-$z$ GRBs reveals interesting features. GRB 050505 
(Hurkett et al. 2006) at $z=4.275$ has a host galaxy with a damped
Lyman-alpha system with the highest column density (Berger et al. 2005c).
High resolution spectroscopy reveals fine-structure transition features
which can be used to infer gas densities and diffuse radiative
conditions of the host galaxy (Chen et al. 2005).
The study of the GRB redshift holder 050904 (Cusumano et al. 2006a;
Haislip et al. 2006; Kawai et al. 2006; Watson et al. 2006)
is even more fruitful:
A detailed spectroscopic study (Totani et al. 2006) suggests that 
the Universe is already largely ionized at $z=6.3$. Afterglow
observations (Frail et al. 2006) and modeling (Gou et al. 2006)
reveal a relative high density circumburst medium around the
burst. The detection of a bright optical flare similar to GRB 990123
(Bo\"er et al. 2006) suggests a possible bright reverse shock emission
component (Wei et al. 2006; Gou et al. 2006). The most erratic
flaring activity in X-rays (Cusumano et al. 2006a,c) suggests a
super-long active central engine (e.g. Zou et al. 2006). A
speculation is that this might be related to a more massive
(or probably more rapidly-rotating) progenitor star.

An interesting question is whether GRB properties evolve with
redshift. The possibility has been raised in the literature
based on various different arguments (e.g. Lloyd-Ronning et al. 
2002; Wei \& Gao 2003; Donaghy et al. 2004; Firmani et al. 2004;
Salvaterra \& Chincarini 2006). On the other hand, observational 
selection effects (e.g. only bright GRBs are detectable at high
redshifts), which are difficult to address, tend to mimic
an apparent evolutionary pattern, rendering a robust claim of
evolutionary effect difficult. A statistical study of GRBs with
known redshift (Liang et al. 2006c) suggests that the observed
luminosity and redshift distributions could be well reproduced
without introducing evolutionary effects. More data are needed
to draw firmer conclusions. From the theoretical point view, first
generation stars tend to be massive due to their low metallicity 
(Abel et al. 2002). If these stars also produce GRBs, the bursts
may be more energetic. The deaths of these stars, however, quickly
contaminate the interstellar medium, so that the next generation
stars may not be very different from the stars seen now. The
evolutionary pattern, if any, may be more complicated
than a simple power law dependence on $(1+z)$.

Another interesting question is whether (Type II) GRBs trace the 
cosmic star forming history only (e.g. Totani 1997).  
Tentative evidence that 
metallicity is another important factor to make a GRB has been 
collected (e.g. Ramirez-Ruiz et al. 2002b; Prochaska et al. 2004; 
Stanek et al. 2006; Modjaz et al. 2007, cf. Campana et al. 2007). 
This factor is currently not included in
most GRB population studies (e.g. Perna et al. 2003; 
Lloyd-Ronning et al. 2004; Zhang et al. 2004a; Lamb et al. 2005a; 
X. Dai \& Zhang 2005; Guetta et al. 2005; Lin et al. 2005; 
Xu et al. 2005b; Liang et al. 2006c, etc). It is 
interesting to explore how to incorporate the metallicity factor
in a quantitative way and how different the results would be
with the metallicity factor included. 

The study of Type I GRBs within the cosmological context has 
just started (e.g. Nakar et al. 2006a; Guetta \& Piran 2006;
Belczynski et al. 2006; Berger et al. 2006b). The data are 
consistent with there being a delay of Type I GRBs with respect to 
the star forming history of the Universe. Better understanding of
the cosmological setting of Type I GRBs will be achieved in a few
years when more data become available.

I'd like to finish this section by discussing an exciting but
controversial field: the GRB cosmology. The cosmological setting
of (Type II) GRBs suggests that they can be invaluable tools to
measure the structure of the Universe if GRBs are standard candles.
Since GRBs have higher redshifts than the Type Ia SNe, it is
promising that GRBs would extend the measurement of the Universe
to the high-redshift regime that Type Ia SNe cannot attain.
The fundamental question is whether there exists a physically-understood, 
narrowly-clustered tight correlation that can serve as a
standard candle. Most previously claimed GRB correlations have
been listed in \S3.3 and \S3.4. Earlier attempts to build GRB
Hubble diagrams (e.g. Schaefer 2003; Bloom et al. 2003) have failed
to put meaningful constraints on the cosmological parameters, since
the correlations that were used have very large scatter.
It was after the discovery of the tight Ghirlanda correlation (Ghirlanda
et al. 2004b) when the GRB cosmology started to make progress (Dai et al.
2004; Ghirlanda et al. 2004c; Xu et al. 2005a; Firmani et al. 2005; Xu 
2005). The approach was however criticized by Friedman \& Bloom (2005)
who pointed out several uncertainties inherited in the Ghirlanda
relation. Liang \& Zhang (2005) discarded the jet model and proposed
the model-independent $E_{\rm iso}-E_p-t_b$ correlation, which is
tight enough for the cosmological purpose. Lately, Firmani et al.
(2006a) discovered a tight correlation using prompt emission data
only, and use it to perform a cosmological study (Firmani et al. 2006b).
By combining the GRB standard candles with the Type Ia SNe data, 
useful constraints can be placed on a list of cosmological models.
The results are generally consistent with the concordance cosmology 
revealed by WMAP, while the high-$z$ nature of GRBs allows the
data to start to put useful constraints on dynamical dark energy models
(Firmani et al. 2005; Wang \& Dai 2006; Nava et al. 2006; Qin et al.
2006b; Mosquera Cuesta et al. 2006; Su et al. 2006; Li et al. 2006;
Schaefer 2007). 

There are two fundamental difficulties in the GRB cosmology. The first 
one is the calibration problem. Since most GRBs are at high-redshifts
where cosmological effects are important, and since nearby GRBs may
belong to a different population, it is essentially impossible to
calibrate a GRB standard candle using a low-$z$ GRB sample, as has 
been done for Type Ia SNe. Without a calibrated candle, there is a
circularity problem by using a candle determined from one cosmology
to constrain cosmological parameters. The problem could be partially
solved by collecting a sample of GRBs within a redshift bin (e.g.
Lamb et al. 2005b; Ghirlanda et al. 2006; Liang \& Zhang 2006b). In
particular, Liang \& Zhang (2006b) showed that one can well calibrate
the power law indices of various standard candle correlations with
the method. The coefficient cannot be calibrated, but may be 
``marginalized'' within a range of cosmologies. The required redshift bin
is not too narrow, say, $\Delta z \sim 0.3$, so that it may be 
possible to calibrate the GRB candles in the near future when the
sample grows to a large enough size. The second, more fundamental
difficulty is to identify a physically-based standard candle. As 
have been discussed in \S3.3 and \S3.4, the physical origins of 
Ghirlanda, Liang-Zhang and Firmani relations 
are still not identified. More
frustratingly, the Ghirlanda relation is not confirmed by the Swift
XRT data (Sato et al. 2007; Willingale et al. 2006), suggesting 
that the relation is not attached to the jet framework. In the
pre-Swift era, it has been assumed that the relations are generally
valid (e.g. Firmani et al. 2005; Xu et al. 2005a) and simulations
were made to see how large a sample is required in order to achieve
a certain constraint on cosmology. The observations by Swift suggest
that when detecting a burst, the very first thing to do is to check
whether the previously proposed standard candles are still satisfied. 
The growing outliers to the Ghirlanda relation seem to suggest that one
may need to discard it as a useful cosmological tool. The deficiency
of late optical data does not allow a clear test to the Liang-Zhang
relation at the moment. The Firmani relation makes use of prompt
emission data only and is easier to test. It is probably by far the
best GRB standard candle. However, the physical origin of the 
correlation is not understood yet.

\section{Outstanding problems}

Although most have been discussed before, it is informative to
summarize the outstanding GRB problems as of late 2006.

\begin{itemize}
\item {\bf GRB classifications and progenitor systems:} Are there
only two major types of GRBs or there is a third major category
with a distinct progenitor system? Are NS-NS mergers distinct from
BH-NS mergers (or other mergers)? Within the Type II (collapsar-related)
GRBs, are LL-GRBs distinctly different from HL-GRBs? What is the 
very nature of XRFs? Why there are 
two apparent universal tracks for intrinsic optical afterglows?
\item {\bf GRB central engine:} How are relativistic jets launched?
For distinctly different progenitor systems, how could the central 
engines be so similar? In particular, how could a central engine
be restarted to sustain erratic long-term activities to power X-ray
flares for both collapsar-type (Type II) and merger-type (Type I) 
GRBs? Does the central engine also inject energy steadily for a long
time? If so what is the observational evidence? It is worth 
commenting that the latest analysis of
Swift data starts to reveal smoothly decaying components that are not
interpretable within the standard external shock scenarios (Willingale
et al. 2006; Zhang et al. 2007c). Another comment is that magnetic
fields likely play an important role at the central engine (e.g.
Usov 1992; Thompson 1994; M\'esz\'aros \& Rees 1997b), but due
to intrinsic complications, MHD is usually not incorporated in the
central engine models, except for several fruitful first attempts 
(e.g. Proga et al. 2003; Mizuno et al. 2004a,b).
\item {\bf Composition of the GRB outflow:} Are GRB outflows matter
dominated or Poynting-flux dominated? What is the evidence for/against
either possibility? A comment here is that the matter-dominated 
model has been regarded as standard - as long as the data can be 
accommodated within the matter-dominated
model, the Poynting-flux-dominated model is not needed. In reality,
magnetic fields should play an important role at the central engine,
and some tentative evidence of a highly magnetized flow (e.g. bold
gamma-ray polarization, Coburn \& Bogg 2003, cf. Rutledge \& Fox 2004;
Kalemci et al. 2007, the requirement a higher $\epsilon_B$ in the 
reverse shock, Fan et al. 2002; Zhang et al. 2003; Kumar \& Panaitescu 
2003; Fan et al. 2005c; Wei et al. 2006,
as well as the energetics argument for X-ray flares, Fan et al. 2005d)
has been collected. The GRB outflow should be at least hybrid (e.g.
with a moderate $\sigma$ parameter). Studies (e.g. Fan et al. 2004a; 
Zhang \& Kobayashi 2005; Fan et al. 2004b) show that 
the differences with respect to the pure 
hydrodynamical models are not prominent for $\sigma < 1$. In 
high-$\sigma$ regime the dynamical behaviors of the outflow are 
still not fully understood (see Zhang \& Kobayashi 2005; Lyutikov
2006b for preliminary discussions), and detailed MHD simulations are
needed. Due to intrinsic degeneracy of model predictions (e.g. the
reverse shock emission is not significant for both low-$\sigma$
and high-$\sigma$ flows, e.g. Zhang \& Kobayashi 2005), a direct
diagnosis of GRB composition from the data is not an easy task.
\item {\bf GRB prompt emission mechanism and site:} Are prompt 
gamma-rays produced in internal shocks, at the photospheres, or in 
magnetic reconnection regions? Is the emission site ``closer-in''
(near photosphere) or ``further-out'' (near deceleration radius)?
Is the thermal component important in the spectrum? (It is noted
that a thermal component may be also required to fit some of the X-ray
flare spectrum, e.g. Grupe et al. 2006b.) What is the
non-thermal mechanism - synchrotron or Comptonization? Related
questions would be what powers high energy emission (leptonic vs.
hadronic) and whether GRBs are emitters of cosmic rays and high
energy neutrinos. These topics will be discussed in \S6.
\item {\bf GRB jet configuration:} Are GRBs collimated at all
(this question arises after Swift detected a good list of GRBs 
without showing a break in X-rays several months after the triggers,
Willingale et al. 2006; Burrows 2006)? If so, what is the 
collimation angle (this question is raised since some temporal
breaks are not achromatic, so that one could not always simply 
use afterglow breaks to estimate the jet opening angles)? Are jets 
structured (maybe needed to interpret some bursts such as 
GRB 060124, Romano et al. 2006b, and 
GRB 061007, Schady et al. 2006b)? What conclusion could one draw 
regarding the energetics of GRBs and their statistical properties
(the simple picture of standard energy reservoir no longer applies).
\item {\bf Properties and origins of the afterglows:} What are the
origins of the distinct afterglow components (especially the 
shallow decay component)? How much can the external shock model
explain? What is the role of the central engine and the internal
dissipation regions? What is the nature of temporal breaks,
especially the chromatic ones?
Swift observations seem to suggest that what we call
``afterglows'' actually include both the traditional external
component and some other components unrelated to the external
shocks. X-ray flares are a good example of a distinct (late
internal dissipation) origin. Even some smoothly decaying components
may be also related to the central engine or the internal
dissipation regions (e.g. Fan et al. 2006; Zhang et al. 2006c).
The phenomenological two-component modeling of Willingale et al 
(2006) seems to be able to fit most of the X-ray afterglows.
One is driven to consider the physical origins of the fitting.
The most puzzling question is the nature of the afterglow 
temporal breaks, especially those that are not achromatic. 
Extensive data mining and sorting are needed to see whether
some ad hoc scenarios (e.g. Zhang 2007) are indeed needed to 
understand the breaks.
\item {\bf Properties of GRB environment:} What is the 
immediate environment of GRBs? Very early data
collected by Swift and other ground-based telescopes have
allowed a diagnosis of the immediate environment of GRBs.
The data suggest that the GRB immediate environment is a 
constant density medium rather than a stratified stellar wind
(Zhang et al. 2006, 2007b; Molinari et al. 2006; Still et al. 
2005; Blustin et al. 2006). It then more seriously raises the
question why a Type II GRB preferentially lies in a constant
density medium. Other questions include:
Is the ambient density of Type I GRBs lower than that of Type 
II GRBs? Is the ambient medium clumpy (e.g. Dermer 2006)? Is
the ambient medium magnetized (e.g. Li \& Waxman 2006)?
Is there an evolution of medium density with redshift
(Gou et al. 2004)? Are the dust and extinction properties of
GRB host galaxies significantly different from those of
Milky Way or SMC/LMC (Chen et al. 2006)?
\item {\bf Properties of GRB shocks:} Are the electrons
accelerated to a power law distribution? Is the electron
power law index universal or unpredictable (The data seem
to suggest no universality of $p$ among GRBs, e.g. Shen et al. 
2006, cf. Wu et al. 2004; see also Dai \& Cheng 2001 for discussion 
of the $p<2$ afterglows)? What define the shock microphysics
parameters (e.g. $\epsilon_e$, $\epsilon_B$, etc)? Are there
correlations between these parameters (e.g. Medvedev 2006)? 
Do microphysics parameters evolve with time (e.g. Yost et al. 
2003; Ioka et al. 2006; Panaitescu et al. 2006b; 
Fan \& Piran 2006a)? Numerical 
(particle-in-cell) simulations and analytical studies have started 
to answer the fundamental questions about particle acceleration and 
magnetic field generation (e.g. Medvedev \& Loeb 1999; Nishikawa et al. 
2003, 2005; Liang \& Nishimura 2004; Hededal \& Nishikawa 2005; 
Spitkovsky 2005; Kato 2005; Milosavljevic \& Nakar 2006a,b).
\end{itemize}

\section{Outlook}

Although unprecedented information has been collected for GRBs,
there are yet more observational channels that are deemed to be
important to study GRBs, but are so far sparsely covered. These
include the electromagnetic spectrum above 10s of MeV, and
non-electromagnetic signals such as high energy neutrinos and
gravitational waves. These observations are widely expected to
be made in the near future.

\subsection{GRB science with GLAST}

The launch of GLAST (Gehrels \& Michelson 1999) in late 2007 will 
open a new era for GRB studies. The Large Area Telescope (LAT) on
board has a wide energy range from $<20$ MeV to $>300$ GeV. A dedicated
GLAST Burst Monitor (GBM) with energy coverage from $\sim 8$ keV to
$\sim 30$ MeV will promptly localize GRBs and perform spectral analysis
of the bursts. Complemented by other space- and 
ground- based high energy photon detectors (e.g. Milagro, Dingus et 
al. 2004; VERITAS, Horan et al. 2005; AGILE, Pittori \& Tavani 2005; 
HESS, Hinton 2004; MAGIC, Lorenz 2004; CANGAROO-III, Kubo et al. 2004),
GLAST will unveil the last spectral window of GRB observations.
With the overlapping operations of both Swift and GLAST, GRBs will
be studied with a full spectral and temporal coverage for the first
time. 

Observationally, tentative evidence of distinct high-energy
components has been collected in the past. Hurley et al. (1994)
detected long-lasting high energy emission from GRB 940217,
which extended 90 minutes after the trigger and included one
18 GeV photon. Gonzalez et al. (2003) reported the existence of
a distinct high energy component in GRB 941017 which is spectrally
and temporally decoupled from the conventional sub-MeV component.
Atkins et al. (2000, 2003) suggested evidence for TeV emission from 
GRB 970417A by reporting the observation by Milagrito (the prototype 
detector of Milagro) that reveals an excess of events coincident in 
time and space with the burst.

On the theoretical side, the fireball model is not short of mechanisms
to produce these high energy photons. In fact, one could list over a 
dozen of mechanisms to produce high energy photons from a relativistic 
fireball. The challenge is how to identify the correct mechanism at
work. This is also related to the unknown GRB composition as well as
the origin(s) of the prompt emission and afterglow. The following
is an unexhausted list, according to the increasing distance from the 
GRB central engine of the high energy emission site.

\begin{itemize}
\item During fireball acceleration, protons and neutrons may be
decoupled if the fireball entropy is high enough (Derishev et al. 1999;
Bahcall \& M\'esz\'aros 2000). Inelastic collisions between neutron and
proton streams would produce neutrinos and GeV photons (Bahcall \&
M\'esz\'aros 2000). For nearby ($z\sim 0.1$) Type I, neutron-loaded GRBs 
of merger origin, GLAST may be able to detect prompt 100 MeV and 100 GeV 
photon signatures from this process (Razzaque \& M\'esz\'aros 2006a).
For neutron-rich ejecta, beta decay of the free neutrons would also give
unique temporal and spectral signatures that may be used to diagnose the
presence of free neutrons (Razzaque \& M\'esz\'aros 2006b).
\item In internal shocks, if the sub-MeV emission that triggers gamma-ray
detectors is due to synchrotron emission, then a synchrotron self-Compton
(SSC) component naturally extends to high energies. High energy
photons are likely attenuated with low energy photons to produce pairs,
whose secondary emission also contribute to the observed spectrum (e.g. 
M\'esz\'aros et al. 1994; Pilla \& Loeb 1998; Razzaque et al. 2004a; 
Pe'er \& Waxman 2004a, 2005; Takagi \& Kobayashi 2005; 
Pe'er et al. 2005, 2006a).
\item In internal shocks, protons are also accelerated. Their synchrotron
emission or photon-meson interaction would also lead to high energy
photon emissions. Assuming optimistic parameters, these emission
signatures may be detectable (e.g. Totani 1998; Bhattacharjee \& Gupta
2003). However, in a large parameter space (e.g. $\epsilon_e$ not
extremely low), the proton radiation components and the secondary emission
of the leptons produced in photo-meson interactions are not as significant 
as the electron SSC process and therefore not detectable (Fragile et al.
2004; Razzaque \& Zhang 2007; Gupta \& Zhang 2007b).
\item In the external reverse shock, SSC would produce high energy photons
in the GeV range (e.g. M\'esz\'aros et al. 2003; Wang et al. 2001a; 
Granot \& Guetta 2003, cf. Kobayashi et al. 2007).
\item In the external forward shock, SSC at early times also produces 
significant GeV emission that is detectable by GLAST (e.g. M\'esz\'aros
\& Rees 1994; Dermer et al. 2000; 
Zhang \& M\'esz\'aros 2001b, for more general discussion of SSC process in the
external forward shock see Wei \& Lu 1998, 2000a; Panaitescu \& Kumar 2000; Sari
\& Esin 2001; Wu et al 2005b). In particular, Zhang \& M\'esz\'aros (2001b)
showed that for the shock parameter regime commonly inferred from the broadband
afterglow fits, the SSC component is prominent and detectable by GLAST for
GRBs at $z\sim 1$. Due to the slow crossing of the SSC peak energy in the 
GLAST band, GLAST would be able to detect these extended GeV emissions for
hours after the trigger. It is worth commenting that the calculation of
Zhang \& M\'esz\'aros (2001b) was made by assuming a constant energy in
the fireball. The shallow decay phase revealed by Swift XRT may suggest 
substantial energy injection in the early phase (\S3.1.3). If this is the 
case, the SSC signature may be weakened. The presence of X-ray flares would
also cool electrons in the external shock (Wang et al. 2006b; 
Gou et al. 2006). This suggests
a less optimistic prediction of the expected GeV signals due to SSC in the
forward shock region. This model was also used to interpret the distinct hard
component in GRB 941017 (Pe'er \& Waxman 2004b).
\item Photons from the forward and reverse shock regions could be inverse
Compton scattered by electrons in the other regions. These cross IC
processes are important high energy emission contributors (Wang et al.
2001a,b).
\item The prompt sub-MeV photon bath may overlap the external shock region 
(both reverse shock, Beloborodov 2005; Fan et al. 2005e, and forward shock,
Fan et al. 2005e) if the burst duration is long enough. The electrons in 
the shocked region would cool by scattering these prompt gamma-rays and 
produce high energy photons (Beloborodov 2005). The effect is especially 
important in a wind medium where the deceleration radius is small 
(Fan et al. 2005e).
\item Protons in the external shock region would produce high energy photons
through synchrotron emission and photo-meson interaction (B\"ottcher \&
Dermer 1998). The parameter space for this component to dominate is small
(i.e. $\epsilon_e \ll \epsilon_B$, Zhang \& M\'esz\'aros 2001b), and is not 
the preferred parameter space derived from the broad-band afterglow fits.
\item Photons from X-ray flares and probably unobserved UV flares 
would be upscattered by the external shock
electrons to produce GeV-TeV photons (Wang et al. 2006b; Fan \& Piran 2006b).
\item SSC within the X-ray flares would produce high energy photons 
(Wang et al. 2006b).
\item If additional soft photons are available from the GRB progenitor, 
external IC processes would boost soft photons to high
energies. For example, such a process may happen in GRBs associated with
SNe (such as GRB 060218) from which thermal photons 
due to the putative SN shock 
breakout are reprocessed and boosted in energy (Wang \& M\'esz\'aros 2006).
\item TeV photons escaping from GRB fireballs would be attenuated by 
intergalactic infrared background and produce pairs, if the GRB source
is not too close to earth (say $z<0.5$). These pairs would
upscatter the cosmic microwave background and produce GeV photons, which
would be detectable by GLAST if the IGM magnetic field is weak enough (to
avoid significant deflection of pairs before the interaction with CMB 
happens). Such a process would give rise to a delayed high energy emission
following the GRB prompt emission (Plaga 1995; Cheng \& Cheng 1996; Dai \& 
Lu 2002a; Wang et al. 2004; Razzaque et al. 2004a).
\item If a GRB occurred in the past in our galaxy, it is expected that
significant GeV-TeV emissions occur from the GRB remnant (Ioka et al.
2004). 
\item For both prompt and delayed high energy emissions, even if they
are not directly detectable, they could contribute to the gamma-ray
diffuse background. A more careful investigation (Casanova et al. 2007)
suggests that for most optimistic parameters, GRBs are not the dominant
contributor to the diffuse gamma-ray background.
\end{itemize}

Although all the above possibilities have been suggested, it is now high time
to perform a more systematic study of the relative importance of 
various mechanisms. Since early afterglow data have been extensively retrieved
by Swift, one can perform more realistic calculations with the constraints
posed by low energy prompt emission and afterglow data. Such predictions, when
compared with future GLAST data, would give strong constraints on both low
energy and high energy models, narrow down and identify the
physical processes happening in GRBs, and shed light on some 
outstanding problems listed in \S5.

A prospect of GLAST observation is to constrain the bulk Lorentz 
factor of GRBs. This is an important unknown parameter of GRB fireball. 
Due to internal
photon-photon productions, it is expected that there would be a (sharp)
spectral cutoff in the prompt GRB spectrum, which has not been clearly
detected. In the past, the highest energy photons have been used to constrain
the lower limit of GRB fireballs (Woods \& Loeb 1995; Baring \& Harding
1997; Lithwick \& Sari 2001). With the detection of a clear spectral 
cutoff, by combining the variability data, one can give interesting
estimates to the GRB bulk Lorentz factors (Baring 2006). These results
could be compared with the Lorentz factor lower limit derived from the
early X-ray afterglow data (Zhang et al. 2006) and sometimes independent
measurements using the early optical afterglow data (e.g. Molinari et al.
2006). If the $\Gamma$ measurements of a good sample of GRBs become
available, statistical work could be carried out to check how $\Gamma$
is correlated with $E_{iso}$, $L_{iso}$, $E_p$, etc. These correlations
hold the key to identify the correct prompt gamma-ray emission model
(\S3.4). It is worth commenting that gamma-rays become transparent 
from the fireball again at even higher energies (e.g. PeV), and the
opaque window becomes narrower with a higher bulk Lorentz factor
(Razzaque et al. 2004a).

Swift observations led to many surprises. Is it possible to make some
reasonable predictions for GLAST? The chance to make such predictions 
in the pre-GLAST era is better than that in the pre-Swift era, mainly
because we already have detailed information about both the prompt
emission and afterglow in the ``low energy'' regime. The following
is a list of bold, rough predictions for what GLAST would detect for
GRBs.

\begin{itemize}
\item It is almost guaranteed to detect prompt emission in the GLAST
band, with a possible spectral cut-off. The exact location of the cut-off
depends on the properties of the burst. Generally Type-II GRBs would
have higher fluences and Type-I GRBs would have lower fluences, 
mainly because their low energy counterparts are such. 
Type-I GRBs may have a higher cut-off energy
than Type-II GRBs (a harder spectrum and probably a higher $\Gamma$),
but this is not guaranteed. XRFs may not have significant high energy
emission.
\item High energy emission typically lasts longer than the sub-MeV
prompt emission (due to many possible reasons listed above).
The spectrum would have a temporal evolution. Harder photons tend to
be detected at later times when the fireball becomes less compact
for photons.
\item At the low energy regime in the GLAST band, the prompt emission 
lightcurves would have narrower spikes than the the sub-MeV lightcurves, 
a general trend revealed by Swift panchromatic observation (e.g. Romano
et al. 2006b; Page et al. 2006b). However, at higher energies when the
putative IC component takes over, the lightcurves would be more smeared
out with less sharp spikes due to the non-linear IC processes involved.
\item It is possible that GLAST would detect bursts for thousand of seconds.
The long-lasting emission may have a broad temporal bump with flares
overlapping on top of it. The rising and falling indices of the flares
would be less steep (again due to the non-linear IC processes), and the
flare amplitudes would be smaller than those of X-ray flares.
\end{itemize}

More concrete predictions require more detailed study. As suggested by
past experience, surprises and new challenges are also bound to merge in 
the GLAST era.

\subsection{Other future observations}

GRB shocks are ideal sites to accelerate cosmic rays. It has been argued
that they are a good candidate to generate ultra-high energy cosmic rays
(UHECRs, Waxman 1995; Vietri 1995; Migrom \& Usov 1995; 
Vietri et al. 2003; Waxman 2004b). GRBs are also 
emitters of neutrinos of a wide range of energy. Neutrinos are
one of the main agent to launch the relativistic jet
from the central engine. MeV neutrinos from
core collapses would escape. Unless the bursts are close enough, these MeV
thermal neutrinos are however undetectable. Proton-neutron decoupling
during the acceleration phase (Derishev et al. 1999) would produce multi-GeV
neutrinos (Bahcall \& M\'esz\'aros 2000). $pp$ interaction in internal shocks
could produce 30 GeV neutrinos (Paczy\'nski \& Xu 1994). Within the collapsar
scenario, $p\gamma$ and $pp$ interactions in internal shocks within the 
stellar envelope would give multi-TeV neutrinos, regardless of whether
the jet would successfully penetrate through the envelope to make a
successful GRB (M\'esz\'aros \& Waxman 2001; Razzaque et al. 2003, 2004b).
In the internal shocks that produce observable GRB prompt emission, 
$p\gamma$ interactions typically produce $10^{14}$eV neutrinos 
(Waxman \& Bahcall 1997; Rachen \& M\'esz\'aros 1998). For LL-GRBs such 
as GRB 060218, this component typically emits
neutrinos at higher energies (above $10^{17}$ eV), and is probably the
most important GRB emission component in this energy range, thanks to
the very high event rate of LL-GRBs (Gupta \& Zhang 2007a; Murase et al.
2006). X-ray flares of late internal shock origin should be also
accompanied by neutrino emission (Murase \& Nagataki 2006).
In the external shock region, GRBs produce neutrinos with even
higher energies ($\sim$ EeV, Waxman \& Bahcall 2000; Dai \& Lu 2001;
Dermer 2002; Li et al. 2002). A generic upper limit of the extragalactic
neutrino flux has been set up by Waxman \& Bahcall (1999, see also
Bahcall \& Waxman 2001) using the observed UHECR flux. A list of km$^3$ 
neutrino experiments, e.g. ICECUBE (Hill 2006),  ANITA (Barwick et al.
2006), KM3Net (Katz 2006), Auger (Van Elewyck et al. 2006)
are being built and are expected to detect these
possible high energy neutrino signals from GRBs. The detections of
high energy neutrinos from GRBs not only help to constrain GRB models,
but are also valuable for studying neutrino physics (e.g. Li et al. 2005;
Gonzalez-Garcia \& Halzen 2006). 
Several caveats need to be mentioned regarding GRBs as neutrino emitters.
First, in order to maximize the predicted neutrino flux, usually 
a $p=2$ proton spectrum is assumed. Studies of prompt and afterglow
emission suggests that $p$ is typically steeper than 2 for electrons.
If protons also have $p>2$, the predicted neutrino flux would drop.
Second, usually the neutrino spectrum for a burst with typical 
parameters is taken to estimate the diffuse neutrino flux. In principle,
one needs to average over bursts with a wide range of distributions 
of luminosity and other parameters. Such an analysis (Gupta \&
Zhang 2007a) suggests that the predicted diffuse background emission
sensitively depends on some unknown parameters, especially the bulk
Lorentz factor of GRBs. The predicted diffuse neutrino flux level is
therefore rather uncertain. On the other hand, the detection (or tight
upper limit) would present severe constraints on the bulk Lorentz factor
distribution of GRBs. Finally, all the calculations have been performed
under the assumption of the baryonic fireball model. If GRB outflows
are Poynting-flux dominated, and if the prompt emission is due to 
magnetic reconnection rather than shock acceleration, GRBs are not
important contributors to UHECRs and high energy neutrinos.

GRBs are also good candidate gravitational wave (GW) sources. 
The two leading progenitor candidates for GRBs, i.e. mergers of binary 
compact objects (Thorne 1987; Phinney 1991; Kochanek \& Piran 1993;
Kobayashi \& M\'esz\'aros 2003) and stellar core-collapses 
(Rampp et al. 1998; van Putten 2001;
Fryer et al. 2002; Kobayashi \& M\'esz\'aros 2003), have been suggested
as potential GW sources. Fragmentation and subsequent accretion of
a collapsing star (King et al. 2005; Piro \& Pfahl 2006) and 
acceleration of a GRB jet (Sago et al. 2004) would also excite
GW of different wave forms. A coincidence between a GW burst and a GRB
would greatly enhance the statistical significance of the GW signal,
making detections easier (Finn et al. 1999). The GW frequencies of
various phases (in-spiral, merger and ring-down) of both 
types of progenitor
cover the $10-10^3$ Hz band which is relevant to several GW detectors,
such as LIGO (Abramovici et al. 1992), VIRGO (Tournefier et al. 2005), 
GEO600 (Benno et al. 2003) and TAMA300 (Fujimoto et al. 2005). Due to
the intrinsic faintness of the signals, only nearby sources (within
$\sim 200$ Mpc for NS-NS and NS-BH mergers, and within $\sim 30$ Mpc
for collapsars, Kobayashi \& M\'esz\'aros 2003) have strong enough
signals to be detected by LIGO-II. Recent observations of short GRBs
of merger origin suggest a higher event rate than estimated previously
(Nakar et al. 2006a). This is encouraging for GW detections of GRBs.

\section{Conclusion}
\label{sect:conclusion}

Swift has greatly revolutionized our understanding of GRBs. Comparing
with the status of the pre-Swift era (e.g. Zhang \& M\'esz\'aros 2004), 
we have learned a lot about GRB classification (e.g. the nature of 
``short'' GRBs, \S2), GRB physics (e.g. early afterglow properties, 
prompt emission site, etc, \S3) and their cosmological setting (\S4). 
However, new questions and challenges arise (e.g. \S5). In particular,
some pre-Swift pictures (e.g. the nature of afterglow breaks and the
inference about GRB jet configuration and energetics) have to be modified
or even abandoned. X-ray flares open a new era of central engine study.
Time is ripe to perform systematic 
data analyses to peer into the global properties of the
bursts. While one can still gain knowledge from special individual
events (such as GRB 060218 and GRB 060614), for most of the ``normal''
bursts, only global statistical properties can serve to improve our
understanding of GRBs. Swift has collected and will keep collecting
an unprecedented GRB sample for both prompt emission and afterglows. 
Systematical studies of this sample have just commenced (e.g. O'Brien et al.
2006b; Willingale et al. 2006; Zhang et al. 2007b,c; Chincarini et al.
2007; Butler \& Kocevski 2007). 

\medskip
The author thanks D. N. Burrows, G. Chincarini, X. Dai, Z. G. Dai, 
Y. Z. Fan, N. Gehrels, L. J. Gou, N. Gupta, Y. F. Huang, K. Ioka, 
S. Kobayashi, P. Kumar, E. W. Liang, T. Lu, P. M\'esz\'aros, E. Nakar, 
J. Norris, P. O'Brien, A. Panaitescu, A. Pe'er, R. Perna, Y. P. Qin, 
E. Ramirez-Ruiz, S. Razzaque, X. Y. Wang, D. M. Wei, R. Willingale, 
X. F. Wu, D. Xu, and R. Yamazaki for comments on the paper.
This work is supported by NASA under grants NNG06GH62G and NNG05GB67G.

\label{lastpage}

\end{document}